\documentclass[aps,prd,nofootinbib,eqsecnum,10pt,superscriptaddress]{revtex4-2}
\usepackage{graphicx}
\usepackage{amsfonts}
\usepackage{amsmath,amsthm,amssymb}
\usepackage{bm}
\usepackage{here}
\usepackage{xcolor}
\usepackage{subcaption}
\usepackage{caption}
\allowdisplaybreaks[1]

\usepackage{dcolumn}
\usepackage{ulem}
\usepackage{hyperref}

\begin{document}

\newcommand{\be}{\begin{equation}}
\newcommand{\ee}{\end{equation}}
\newcommand{\ba}{\begin{eqnarray}}
\newcommand{\ea}{\end{eqnarray}}
\newcommand{\D}{\partial}
\newcommand{\etal}{{\it et al.}}
\newcommand{\Mpl}{M_{\rm pl}}

\title{Gravitational waves from highly charged binary black holes
in Einstein--Maxwell--dilaton theory}

\author{Ryoto Tomie}
\affiliation{Department of Physics, Faculty of Science,
Tokyo University of Science,
1-3 Kagurazaka, Shinjuku-ku, Tokyo 162-8601, Japan}

\author{Kai Togashi}
\affiliation{Department of Physics, Faculty of Science,
Tokyo University of Science,
1-3 Kagurazaka, Shinjuku-ku, Tokyo 162-8601, Japan}

\author{Kitaro Taniguchi}
\affiliation{Department of Science Communication,
Faculty of Science Division I,
Tokyo University of Science,
1-3 Kagurazaka, Shinjuku-ku, Tokyo 162-8601, Japan}

\author{Ryotaro Kase}
\affiliation{Department of Physics, Faculty of Science,
Tokyo University of Science,
1-3 Kagurazaka, Shinjuku-ku, Tokyo 162-8601, Japan}
\affiliation{Research Center for Space System Innovation,
Research Institute for Science and Technology,
Tokyo University of Science,
2641 Yamazaki, Noda-shi, Chiba 278-8510, Japan}

\date{\today}

\begin{abstract}
We perform fully nonlinear numerical-relativity simulations of electrically charged binary black hole mergers in Einstein--Maxwell--dilaton (EMD) theory, varying the magnitude and relative sign of the input charges and the dilaton coupling $a$. 
In the weak-charge sequence, the dominant gravitational waveforms remain close to the neutral vacuum baseline of general relativity and are visually indistinguishable across the sampled couplings, consistent with earlier EMD simulations. 
For the highly charged binaries with equal-magnitude opposite charges, however, the dominant gravitational strain exhibits visible differences between the EMD cases with $a\neq0$ and the charged Einstein--Maxwell reference at $a=0$. 
At $a=1$, the coordinate-time interval from the same separation to the waveform-based merger is approximately $62\%$ shorter than for the neutral binary. 
In this high-charge sequence at fixed input charges, increasing $a$ strengthens the scalar dressing while leaving the individual dilaton-weighted horizon charges nearly unchanged. 
Within the same sequence, increasing $a$ across the sampled values decreases the dipolar electromagnetic-flux peak, increases the scalar-flux peak, and modestly lengthens the merger interval. 
The sampled like-charged binaries have corresponding merger intervals approximately $15\%$--$20\%$ longer than the neutral baseline and emit predominantly quadrupolar electromagnetic radiation. 
At $a=1$ and $a=\sqrt{3}$, the like-charged mergers leave charged remnants with scalar fields consistent with secondary dilaton hair. 
The opposite-charge remnants instead approach neutral black holes without stationary scalar hair. 
\end{abstract}


\maketitle

\section{Introduction}
\label{introsec}

The detection of gravitational waves (GWs) from binary black hole (BBH) mergers, beginning with GW150914~\cite{LIGOScientific:2016aoc}, has opened a new observational window on gravity in the strong-field and dynamical regime. 
Subsequent observations by the LIGO--Virgo--KAGRA Collaborations~\cite{LIGOScientific:2016sjg, LIGOScientific:2017bnn, LIGOScientific:2017ycc, LIGOScientific:2018mvr, KAGRA:2021vkt} have established BBH coalescences as a common astrophysical process and an important probe of relativistic gravity. 
These observations probe highly nonlinear dynamics in a regime complementary to Solar System tests~\cite{Bertotti:2003rm, Will:2014kxa} and binary-pulsar observations~\cite{Kramer:2006nb, Antoniadis:2012vy, Wex:2014nva}. 
When combined with numerical-relativity predictions~\cite{Pretorius:2005gq, Campanelli:2005dd, Baker:2005vv, Healy:2017psd}, BBH waveforms provide a powerful means of testing general relativity (GR) and constraining deviations predicted by alternative or extended theories of gravity~\cite{Yunes:2013dva, Berti:2015itd}. 

Additional scalar degrees of freedom arise in many extensions of GR and low-energy effective theories~\cite{Kobayashi:2019hrl}. 
Around compact objects, they can support nontrivial field configurations, commonly referred to as scalar hair, and introduce radiative channels in addition to the tensor modes of GR~\cite{Silva:2016smx, Cardoso:2016ryw}. 
Their interactions with the gravitational and matter sectors can modify both the conservative binary dynamics and the dissipative evolution through additional radiation~\cite{Yagi:2016jml}. 
Theories in which a scalar field couples directly to electromagnetism therefore provide a natural setting for studying charged compact binaries beyond GR. 

Einstein--Maxwell--dilaton (EMD) theory provides a prototypical framework for studying scalar--electromagnetic couplings, with early examples arising in supergravity~\cite{Gibbons:1982ih}. 
In four dimensions, the coupling between the Maxwell field and the dilaton is parametrized by a dimensionless constant $a$. 
For $a=0$, the dilaton decouples from the Maxwell field, and the constant-scalar sector reduces to Einstein--Maxwell theory. 
The values $a=1$ and $a=\sqrt{3}$ arise, respectively, in a low-energy effective description of heterotic string theory~\cite{Garfinkle:1990qj} and in four-dimensional Kaluza--Klein theory obtained by reducing five-dimensional vacuum gravity~\cite{Gibbons:1985ac}. 
Static, spherically symmetric charged black hole (BH) solutions for arbitrary $a$ were obtained by Gibbons and Maeda, and the $a=1$ solution was subsequently studied by Garfinkle, Horowitz, and Strominger~\cite{Gibbons:1987ps, Garfinkle:1990qj}. 
These solutions reduce to the Reissner--Nordstr\"om solution in the $a=0$ constant-scalar sector. 
For nonzero $a$, the dilaton generally modifies the horizon geometry and electromagnetic field, and the charge-extremal limit can be singular with vanishing horizon area. 

These properties motivate the study of charged EMD binaries in the strong-field regime. 
For small charge-to-mass ratios, the electromagnetic and scalar sectors are expected to produce perturbative corrections to the vacuum-GR dynamics. 
At larger charges, by contrast, the electromagnetic interaction, the dilaton field, and the associated radiative channels can appreciably affect the inspiral and merger. 
In EMD theory, the electromagnetic field also sources a nontrivial dilaton profile, providing an additional channel through which electric charge can influence the binary dynamics and radiation. 

Although astrophysical BHs are generally expected to carry electric charges far below the extremal scale, charged BBH systems provide a useful theoretical laboratory for understanding how electromagnetic fields can affect compact-binary dynamics. 
Waveform predictions for charged binaries are also required to translate observational agreement with uncharged-BBH models into bounds on charge-to-mass ratios or related parameters~\cite{Bozzola:2020mjx}. 
Several mechanisms have nevertheless been proposed through which BHs may acquire environment-dependent electric charges in magnetized or plasma-rich systems~\cite{Zajacek:2018vsj, Zajacek:2019kla}. 
The highly charged configurations considered here should therefore be viewed primarily as probes of nonlinear EMD dynamics rather than as direct models of typical astrophysical BBHs. 

Charged BBH dynamics has been studied both analytically and through numerical relativity. 
In Einstein--Maxwell theory, previous work has developed constraint-satisfying charged puncture data and performed fully nonlinear simulations of charged BBH inspirals and mergers~\cite{Bozzola:2019aaw, Bozzola:2021elc}. 
In EMD theory, analytical studies have estimated merger and remnant properties and developed post-Newtonian and effective-one-body descriptions of the conservative and radiative binary dynamics~\cite{Jai-akson:2017ldo, Khalil:2018aaj, Julie:2018lfp}. 
Fully nonlinear EMD simulations were performed by Hirschmann et al., who studied quasi-circular BBH mergers with relatively small electric charges and found only a limited influence of the dilaton on the gravitational waveform~\cite{Hirschmann:2017psw}. 

To the best of our knowledge, fully nonlinear numerical-relativity simulations of BBH mergers in the large-charge regime of EMD theory have not yet been reported. 
This leaves an important open problem because the dilaton--electromagnetic interaction can appreciably modify the strong-field dynamics and open additional radiative channels when the black holes carry large charges. 

We address this gap by performing fully nonlinear simulations of charged BBH mergers in EMD theory. 
We construct constraint-satisfying puncture initial data, evolve the coupled gravitational, electromagnetic, and dilaton fields in a $3+1$ formulation, and extract the resulting gravitational, electromagnetic, and scalar radiation. 
By varying the charge magnitude, the relative charge sign, and the dilaton coupling $a$, we examine their effects on the orbital evolution, horizon response, radiative energy fluxes, and gravitational waveform. 
We find differences in the merger dynamics, electromagnetic multipolar structure, and remnant properties among the sampled charge configurations. 
In the highly charged sequence, comparisons with the Einstein--Maxwell reference reveal visible differences in the dominant gravitational strain for nonzero dilaton coupling. 
The coupling also modifies the horizon response and electromagnetic and scalar radiation in a manner that depends on the charge configuration. 

The paper is organized as follows. 
Section~\ref{sec:EMD} reviews the EMD field equations and the relevant static charged BH solutions. 
Section~\ref{sec:formulation} presents the $3+1$ formulation and gauge conditions used in the numerical evolution. 
Section~\ref{sec:Initial_Data_Construction} describes the construction of constraint-satisfying charged puncture initial data. 
Section~\ref{sec:extraction_and_quantities} defines the gravitational, electromagnetic, and scalar radiative quantities, together with the quasi-local quantities evaluated on the apparent horizons. 
Section~\ref{sec:results} presents the charge- and coupling-dependent orbital dynamics, radiation, horizon response, and remnant properties. 
Section~\ref{sec:conclusions} summarizes our findings and outlines future directions. 
Appendix~\ref{app:bssn_constraints} provides the Baumgarte--Shapiro--Shibata--Nakamura (BSSN) evolution system and the constraint-monitoring results. 
Throughout this paper, we use the geometrized unit system, where the speed of light $c$ and the gravitational constant $G$ are equal to 1.

\section{Einstein--Maxwell--dilaton Theory}
\label{sec:EMD}

We consider the EMD theory described by the action
\begin{equation}
S = \frac{1}{16\pi} \int d^4x \sqrt{-g}
\left[
R - 2 (\nabla \phi)^2 - e^{-2 a \phi} F_{\mu\nu} F^{\mu\nu}
\right],
\label{eq:action}
\end{equation}
where $R$ is the Ricci scalar of the spacetime metric $g_{\mu\nu}$ and $\phi$ is the dilaton field. The electromagnetic field strength tensor is defined by $F_{\mu\nu} = \nabla_\mu A_\nu - \nabla_\nu A_\mu$, where $A_\mu$ is the vector potential and $\nabla_\mu$ denotes the covariant derivative compatible with $g_{\mu\nu}$.
The dimensionless coupling constant $a$ controls the strength of the dilaton--electromagnetic coupling. 
For $a=0$, the dilaton decouples from the Maxwell field, and the theory reduces to Einstein--Maxwell gravity supplemented by a free scalar field. 
With the scalar-free initial data adopted in this work, this branch remains in the Einstein--Maxwell sector. 
The values $a=1$ and $a=\sqrt{3}$ correspond to a low-energy effective description of heterotic string theory~\cite{Garfinkle:1990qj} and the Kaluza--Klein value~\cite{Gibbons:1985ac}, respectively. 

Varying the action with respect to $g_{\mu\nu}$, $A_\mu$, and $\phi$ yields the field equations
\begin{align}
	&G_{\mu\nu} = 8 \pi T^{\mathrm{eff}}_{\mu\nu}\,,\qquad 
    T^{\mathrm{eff}}_{\mu\nu} = \frac{1}{4\pi} \left( T^{\phi}_{\mu\nu} + e^{-2a \phi} T^F_{\mu\nu} \right)\,, 
    \label{grav_eom} \\
	&\Box \phi = - \frac{a}{2} e^{-2a \phi} F_{\mu\nu} F^{\mu\nu}\,, \label{scalar_eom} \\
	&\nabla_\mu\left( e^{-2a \phi}F^{\mu\nu} \right) = 0\,. \label{vector_eom}
\end{align}
Here the dilaton and Maxwell stress tensors are
\begin{align}
	T^{\phi}_{\mu\nu} &= \nabla_\mu \phi \nabla_\nu \phi -\frac{1}{2} g_{\mu\nu} \nabla_\alpha \phi \nabla^\alpha \phi\,,
    \label{defTphi}\\
	T^F_{\mu\nu} &= F_{\mu\alpha}F_{\nu}{}^{\alpha}- \frac{1}{4} g_{\mu\nu} F_{\alpha\beta} F^{\alpha\beta}\,.
    \label{defTF}
\end{align}
Equivalently, the Maxwell equation~\eqref{vector_eom} can be written as $\nabla_\mu F^{\mu\nu}=2aF^{\mu\nu}\nabla_\mu\phi$.

Static, spherically symmetric charged BH solutions are known for arbitrary values of the dilaton coupling $a$ in EMD theory. 
These solutions are often referred to as Gibbons--Maeda or GM--GHS solutions~\cite{Gibbons:1987ps, Garfinkle:1990qj}.
As a representative case, we revisit the electrically charged solution for the string value $a=1$ in the areal-radius coordinate form used in Ref.~\cite{Kase:2023kvq}. 
In coordinates $(t,r,\theta,\varphi)$ with $r$ the areal radius, the static and spherically symmetric line element can be written as
\begin{equation}
    ds^2 = - f(r)\, dt^2 + h^{-1}(r)\, dr^2 + r^2 d\Omega^2 ,
\end{equation}
where $f(r)$ and $h(r)$ determine the temporal and radial metric components as $g_{tt}=-f(r)$ and $g^{rr}=h(r)$, respectively, and
$d\Omega^2=d\theta^2+\sin^2\theta\,d\varphi^2$.
For $a=1$, the metric functions, together with the dilaton field $\phi$ and the electric potential $A_0$, take the form
\begin{align}
	f &= 1 - \frac{2M}{r^2} \left( \sqrt{r^2 + r_q^2} - r_q \right)\,, &
	h &= \left( 1+ \frac{r_q^2}{r^2} \right) f\,, \nonumber \\
	\phi &= \phi_0 + \frac{1}{2} \ln \left( \frac{\sqrt{r^2 + r_q^2} - r_q}{\sqrt{r^2 + r_q^2} + r_q} \right)\,, &
	A_0' &= \frac{q r e^{2 \phi_0}}{ \left( \sqrt{r^2 + r_q^2} + r_q \right)^2 \sqrt{r^2 + r_q^2}}\,, 
    \label{GMGHS}
\end{align}
where a prime denotes differentiation with respect to $r$, $M$ is a mass parameter, $\phi_0$ is the asymptotic value of $\phi$ at spatial infinity, and 
\begin{equation}
    r_q = \frac{q^2 e^{2 \phi_0}}{2 M}\,.
\end{equation}
Here, $q$ is the electric charge parameter. 
Compared to the Reissner--Nordstr\"om solution of Einstein--Maxwell theory, this solution has a nontrivial dilaton profile, and the relation between the mass, charge, and horizon structure is modified by the dilaton--electromagnetic coupling.

The secondary nature of the dilaton hair can be seen directly from the asymptotic behavior of the solution in Eq.~\eqref{GMGHS}. 
At large radius, the dilaton field behaves as 
\begin{equation}
\phi=\phi_0-\frac{r_q}{r}+{\cal O}\left(r^{-3}\right)\,.
\end{equation}
Hence, the scalar monopole charge is given by  
\begin{equation}
Q_{\rm S}=-r_q=-\frac{q^2e^{2\phi_0}}{2M}\,.
\end{equation}
The scalar charge is thus fixed by the mass, electric-charge parameter, and asymptotic scalar value, rather than constituting an independent charge, and is therefore secondary hair~\cite{Pacilio:2018gom}. 
Because $Q_{\rm S}$ depends on $q^2$, its sign is unchanged under $q\rightarrow-q$, so black holes carrying electric charges of opposite signs nevertheless carry scalar charges of the same sign. 
In the neutral limit $q=0$, both $r_q$ and $Q_{\rm S}$ vanish, and the solution reduces to a Schwarzschild black hole with a constant dilaton. 
Consequently, at leading order in the weak-field two-body dynamics, the same-sign scalar charges are expected to produce an attractive scalar-mediated interaction, independently of whether the electromagnetic interaction is attractive or repulsive~\cite{Khalil:2018aaj}. 

Subsequent studies have investigated several properties of the GM--GHS family, including the linear perturbations of isolated BHs. 
An early study computed quasinormal modes of the charged dilaton BH and showed that the dilaton breaks the axial--polar isospectrality known for Schwarzschild and Reissner--Nordstr\"om BHs~\cite{Ferrari:2000ep}. 
More recently, work on Gibbons--Maeda BHs in a class of EMD theories has shown that the linear perturbations can be separated and reduced to diagonalized radial equations, and has established mode stability for the class of solutions considered there~\cite{Pope:2024ncb}. 
These results, however, concern isolated BHs in the linear regime. 
Binary dynamics involving charged EMD BHs and fully nonlinear mergers remain much less explored, especially in the large-electric-charge regime considered in this work.

The discussion above illustrates how the dilaton--electromagnetic coupling affects charged BH solutions already in the static sector. 
We now turn to the 3+1 formulation and gauge conditions for the coupled gravitational, electromagnetic, and dilaton fields, keeping the coupling $a$ arbitrary.

\section{Numerical Formulation and Gauge Conditions}
\label{sec:formulation}

In this section, we present the $3+1$ formulation and gauge conditions used to evolve the EMD system.
We first introduce the $3+1$ decomposition of spacetime in the Arnowitt--Deser--Misner (ADM) formalism~\cite{Arnowitt:1959ah} and write the gravitational and dilaton evolution equations in terms of the corresponding $3+1$ variables.
We then define the electromagnetic variables and the effective source terms obtained by projecting the stress-energy tensor onto the spatial hypersurfaces.
Finally, we describe the constraint equations, the extension used to control the Gauss constraint, the BSSN formulation, and the gauge conditions adopted in the numerical evolution.

We foliate spacetime by a family of spacelike hypersurfaces $\Sigma_t$. 
With the signature $(-,+,+,+)$, the ADM line element is written as
\begin{equation}
  ds^2 = -\alpha^2 dt^2 + \gamma_{ij}(dx^i + \beta^i dt)(dx^j + \beta^j dt)\,,
\end{equation}
where $\alpha$ is the lapse function, $\beta^i$ is the shift vector, and $\gamma_{ij}$ is the induced spatial metric on $\Sigma_t$. 
The unit normal vector to $\Sigma_t$ is defined by 
\begin{equation}
  n^\mu = \frac{1}{\alpha} \left( 1, -\beta^i \right)\,,\qquad
  n_\mu = \left( -\alpha, 0, 0, 0 \right).
\end{equation}
The geometric, dilaton, and electromagnetic variables appearing below are decomposed with respect to this foliation.

In terms of the ADM variables, the gravitational and dilaton sectors take a first-order-in-time form.
The evolution equations corresponding to the gravitational~\eqref{grav_eom} and dilaton~\eqref{scalar_eom} field equations are
\begin{align}
\partial_t \gamma_{ij} ={} &-2 \alpha K_{ij} + D_i \beta_j + D_j \beta_i\,,
\\
\partial_t K_{ij} ={} &\alpha \left( R^{(3)}_{ij} + K K_{ij} - 2 K_{ik} K^k{}_j \right) - D_i D_j \alpha - 8 \pi \alpha\left\{  S_{ij}^{\mathrm{eff}} + \frac{1}{2} \gamma_{ij} \left( \rho_H^{\mathrm{eff}} - S^{\mathrm{eff}} \right) \right\} + \mathcal{L}_\beta K_{ij}\,,\\
\partial_t \phi ={} & -\alpha K_\phi +\mathcal{L}_\beta \phi\,,
\\
\partial_t K_\phi ={} & - \alpha \left\{ D_i D^i \phi - K K_\phi + a e^{-2a\phi} \left( D_i A^\perp_j D^i A_\perp^j -D_i A^\perp_j D^j A_\perp^i - E_i E^i \right) \right\} -D^i \phi D_i \alpha + \mathcal{L}_\beta K_\phi\,.
\end{align}
Here, $K_{ij}$ is the extrinsic curvature of $\Sigma_t$, and $K=\gamma^{ij}K_{ij}$. 
The covariant derivative compatible with $\gamma_{ij}$ is denoted by $D_i$, $R^{(3)}_{ij}$ is the corresponding Ricci tensor, and $\mathcal{L}_\beta$ denotes the Lie derivative along the shift vector $\beta^i$. 
The effective source terms $\rho_H^{\mathrm{eff}}$ and $S_{ij}^{\mathrm{eff}}$ are defined below, and $S^{\mathrm{eff}}\equiv\gamma^{ij}S_{ij}^{\mathrm{eff}}$. 
The dilaton sector is described by $\phi$ and its conjugate momentum $K_\phi\equiv -n^\mu\nabla_\mu\phi$. 
The electromagnetic sector is described by the electric field $E^\mu$ measured by the Eulerian observer,
\begin{equation}
E^\mu \equiv F^{\mu\nu} n_\nu\,,\qquad n_\mu E^\mu=0\,,
\end{equation}
and by the $3+1$ decomposition of the vector potential,
\begin{equation}
A^\mu = n^\mu A_\phi + A_\perp^\mu\,,
\qquad
n_\mu A_\perp^\mu = 0\,.
\end{equation}
In the evolution system, the electromagnetic sector is represented by $A_\phi$, the spatial covector $A_i^\perp\equiv\gamma_{i\mu}A_\perp^\mu$, and the electric field $E^i$. 
Their evolution equations are given below after introducing the Lorenz gauge and the damping variable for the Gauss constraint. 

The coupling of the gravitational sector to the dilaton and electromagnetic fields is encoded in the effective energy--momentum tensor $T^{\mathrm{eff}}_{\mu\nu}$ defined in Eq.~\eqref{grav_eom}, with $T^\phi_{\mu\nu}$ and $T^F_{\mu\nu}$ given in Eqs.~\eqref{defTphi} and \eqref{defTF}. 
In the $3+1$ decomposition, the matter source terms appearing in the ADM evolution equations are obtained by projecting $T^{\mathrm{eff}}_{\mu\nu}$ along and orthogonal to the unit normal $n^\mu$. 
Using the projection tensor onto $\Sigma_t$,
$\gamma_{\mu\nu} = g_{\mu\nu} + n_\mu n_\nu$, we define
\begin{align}
\rho_H^{\mathrm{eff}} &\equiv n^\mu n^\nu T_{\mu\nu}^{\mathrm{eff}}\,, \label{eq:rhoH_eff}
\\
J_i^{\mathrm{eff}} &\equiv - \gamma_{i\mu} n_\nu T^{\mu\nu}_{\mathrm{eff}}\,, \label{eq:Ji_eff}
\\
S_{ij}^{\mathrm{eff}} &\equiv \gamma_{i\mu} \gamma_{j\nu} T^{\mu\nu}_{\mathrm{eff}}\,, \label{eq:Sij_eff}
\end{align}
where $\rho_H^{\mathrm{eff}}$, $J_i^{\mathrm{eff}}$, and $S_{ij}^{\mathrm{eff}}$ are the energy density, momentum density, and spatial stress tensor measured by an Eulerian observer, respectively.

For the EMD system considered here, these source terms can be written explicitly in terms of the dilaton variables and the electromagnetic variables defined above as
\begin{align}
\rho_H^{\mathrm{eff}} ={}& \frac{1}{8 \pi} \left( \rho_H^\phi + e^{-2a \phi} \rho_H^F \right)\,, \label{EMD_rho} \\
J_i^{\mathrm{eff}} ={}& \frac{1}{4 \pi} \left( J_i^\phi + e^{-2a \phi} J_i^F \right)\,, \label{EMD_J} \\
S_{ij}^{\mathrm{eff}} ={}& \frac{1}{4 \pi} \left( S_{ij}^\phi + e^{-2a \phi} S_{ij}^F \right)\,,
\end{align}
with
\begin{align}
    \rho_H^\phi ={}& D_i \phi D^i \phi + K_\phi^2\,, &
	\rho_H^F ={}& E_i E^i + D_i A^\perp_j D^i A_\perp^j - D_i A^\perp_j D^j A_\perp^i\,, \\
	J_i^\phi ={}& K_\phi D_i \phi\,, &
	J_i^F ={}& E^j \left( D_i A^\perp_j - D_j A^\perp_i \right)\,, \label{defJi}\\
	S_{ij}^\phi ={}& D_i \phi D_j \phi + \frac{1}{2} \gamma_{ij} \left( K^2_\phi - D_k \phi D^k \phi \right)\,, &
	S_{ij}^F ={}& -E_i E_j + \gamma^{ab} \left( D_i A^\perp_a - D_a A^\perp_i \right) \left( D_j A^\perp_b - D_b A^\perp_j \right) \nonumber\\
    &&&+ \frac{1}{2} \gamma_{ij} \left( E_a E^a - D_a A^\perp_b D^a A_\perp^b + D_a A^\perp_b D^b A_\perp^a \right)\,. 
\end{align}
All quantities in these expressions are evaluated on $\Sigma_t$, and spatial indices are raised and lowered with $\gamma_{ij}$.

In addition to the evolution equations, the ADM formulation gives rise to a set of constraint equations that must be satisfied on each spatial hypersurface. 
The Hamiltonian and momentum constraints are given by
\begin{align}
R^{(3)} + K^2 - K_{ij} K^{ij} =& 16\pi \rho_H^{\mathrm{eff}}\,, \label{Ham} \\
D_j ( K^{ij} - \gamma^{ij} K ) =& 8\pi J^i_{\mathrm{eff}}\,, \label{mom}
\end{align}
where $R^{(3)}=\gamma^{ij}R^{(3)}_{ij}$ is the Ricci scalar associated with $\gamma_{ij}$.
These constraints relate the intrinsic and extrinsic geometry of each spatial hypersurface to the effective source terms. 
In numerical simulations, they are imposed on the initial data and monitored throughout the evolution as checks of numerical accuracy and stability.

The Maxwell equations imply the Gauss constraint,
\begin{equation}
    D_i E^i - 2 a E^i D_i \phi = 0\,, \label{Gauss}
\end{equation}
which must be satisfied on each spatial hypersurface. 
During numerical evolution, violations of this constraint may develop, especially in the highly charged configurations considered below. 
To control such violations and enhance numerical stability, we introduce an auxiliary variable $Z$ associated with the Gauss constraint, following extended Maxwell formulations with hyperbolic constraint damping~\cite{PhysRevD.81.084007, Hirschmann:2017psw}. 
For the electromagnetic vector potential, we impose the Lorenz gauge. 
We therefore use the extended Maxwell evolution system in which deviations from the Gauss constraint are dynamically damped.
This system is given by
\begin{align}
	\partial_t A_\phi ={}& \alpha \left( K A_\phi - D_i A_\perp^i \right) - A_\perp^i D_i \alpha - \alpha Z + \mathcal{L}_\beta A_\phi\,, \label{EMD_Aphi} \\
    \partial_t A_i^\perp ={}& -\alpha E_i-A_\phi D_i\alpha-\alpha D_iA_\phi+\mathcal{L}_\beta A_i^\perp\,,\label{EMD_Ai}\\
	\partial_t E^i ={}& \alpha E^i K + D_j \left\{ \alpha \left( D^i A_\perp^j - D^j A_\perp^i \right) \right\} 
	+ 2 a \alpha \left\{ \left( D^j A_\perp^i - D^i A_\perp^j \right) D_j \phi - E^i K_\phi \right\} + \alpha D^i Z + \mathcal{L}_\beta E^i\,, \\
	\partial_t Z ={}& \alpha \left( D_i E^i - 2 a E^i D_i \phi - \kappa Z \right)+\mathcal{L}_\beta Z\,. \label{EMD_Zeta}
\end{align}

Here, $\kappa>0$ controls the damping of violations of the EMD Gauss constraint. 

Although the coupled EMD system has been written above in terms of ADM variables, we do not evolve the ADM gravitational equations directly.
For numerical stability, we evolve the gravitational sector using the BSSN formulation~\cite{Shibata:1995nv, Baumgarte:1998rp} constructed from the ADM variables $(\gamma_{ij}, K_{ij})$. 
The dilaton and electromagnetic evolution equations are coupled to the BSSN geometry through the lapse, shift, spatial metric, and extrinsic curvature, and they source the BSSN equations through the effective matter terms defined above.
For completeness, the explicit BSSN evolution system and the constraint diagnostics used in the simulations are summarized in Appendix~\ref{app:bssn_constraints}. 

For the spacetime gauge, we adopt the advective moving-puncture gauge, in which the lapse is evolved using the $1+\log$ slicing condition, while the shift is evolved using a hyperbolic Gamma-driver condition ~\cite{Campanelli:2005dd,vanMeter:2006vi}.  The explicit evolution equations and the parameter values used in the simulations are given in Appendix~\ref{app:gauge_conditions}.  These conditions fix the spacetime coordinates without introducing new physical degrees of freedom.  No additional gauge condition is required for the dilaton field.

\section{Initial Data Construction}
\label{sec:Initial_Data_Construction}

We construct BBH initial data for the EMD system by solving the constraint equations derived in Sec.~\ref{sec:formulation}.
On the initial hypersurface, we impose the Hamiltonian, momentum, and Gauss constraints given in Eqs.~\eqref{Ham}, \eqref{mom}, and \eqref{Gauss}, respectively. 

We employ the York--Lichnerowicz conformal decomposition~\cite{JMPA_1944_9_23__37_0, York:1971hw},
\begin{equation}
\gamma_{ij} = \psi^4 \hat{\gamma}_{ij}\,, \qquad
K^{ij} = \psi^{-10} \hat{A}^{ij} + \frac{1}{3}\gamma^{ij}K\,, \qquad
E^i = \psi^{-6} \hat{E}^i\,.
\end{equation}
We assume conformal flatness, $\hat{\gamma}_{ij} = \delta_{ij}$, and maximal slicing, $K=0$, as in standard puncture-based BBH initial data.
Here, $\hat{D}_i$ denotes the covariant derivative associated with the conformal metric $\hat{\gamma}_{ij}$. 
For the choice $\hat{\gamma}_{ij} = \delta_{ij}$, it reduces to the ordinary flat-space derivative, $\hat{D}_i \equiv \partial_i$.

Under this decomposition and the assumptions of conformal flatness and maximal slicing, the constraint equations~\eqref{Ham}, \eqref{mom}, and \eqref{Gauss} take the form
\begin{align}
\hat{D}^2 \psi &= -\frac{\psi^{-7}}{8}\hat{A}_{ij}\hat{A}^{ij}
 - 2\pi \psi^5 \rho_H^{\mathrm{eff}}\,, \label{Ham_York}
\\
\hat{D}_j \hat{A}^{ij} &= 8\pi \psi^{10} J^i_{\mathrm{eff}}\,, \label{mom_York} \\
\hat{D}_i\hat{E}^i &= 2 a \hat{E}^i \hat{D}_i \phi\,, \label{Gauss_York}
\end{align}
where $\rho_H^{\mathrm{eff}}$ and $J^i_{\mathrm{eff}}$ are given by Eqs.~\eqref{EMD_rho} and \eqref{EMD_J}, respectively.

We consider purely electric, nonspinning initial data. 
For the electromagnetic vector potential, we set $A_i^\perp=0$ on the initial slice, so that the magnetic field vanishes initially. 
We also set the initial scalar field to be spatially uniform, $\phi=0$, with vanishing scalar momentum, $K_\phi=0$. 
Under these assumptions, the effective momentum density given in Eq.~\eqref{EMD_J} vanishes, $J^i_{\mathrm{eff}}=0$, and the momentum constraint~\eqref{mom_York} reduces to
\begin{equation}
\hat{D}_j \hat{A}^{ij} = 0\,.
\end{equation}
For two boosted, nonspinning punctures, we adopt the Bowen--York solution of the momentum constraint away from the punctures~\cite{Bowen:1980yu, Bozzola:2019aaw},
\begin{align}
\hat{A}^{ij} &= \hat{A}^{ij}_1 + \hat{A}^{ij}_2\,, \\
\hat{A}^{ij}_A &= \frac{3}{2 r_A^2} \left[P_A^i l_A^j+P_A^j l_A^i
-\left(\delta^{ij}-l_A^i l_A^j\right)P_A^k l_{Ak}\right]\,,
\end{align}

where $A=1,2$ labels the punctures, $r_A=[\delta_{ij}(x^i-C_A^i)(x^j-C_A^j)]^{1/2}$ is the coordinate distance from puncture $A$, $C_A^i$ is its coordinate position, $P_A^i$ is its linear momentum, and $l_A^i=(x^i-C_A^i)/r_A$ is the outward-pointing radial unit vector centered on puncture $A$. 
Spatial indices on $l_A^i$ and $P_A^i$ are raised and lowered with the flat conformal metric $\delta_{ij}$. 

The remaining electromagnetic constraint is the Gauss constraint~\eqref{Gauss_York}.
With $\phi = 0$ imposed on the initial slice, it reduces to
\begin{equation}
\hat{D}_i \hat{E}^i = 0\,,
\end{equation}
away from the punctures.
A multi-puncture Coulomb-type solution is then given by 
\begin{equation}
\hat{E}^i=
\sum_{A=1}^{2}Q_A\frac{x^i-C_A^i}{r_A^3}\,,\label{Coulomb}
\end{equation}
where $Q_A$ denotes the electric charge attributed to the puncture $A$~\cite{Bozzola:2019aaw}.

After the momentum and Gauss constraints have been satisfied, the Hamiltonian constraint~\eqref{Ham_York} remains as the only nontrivial elliptic equation. 
We solve it with the puncture method developed in Ref.~\cite{PhysRevLett.78.3606}. The conformal factor is written as
\begin{equation}
    \psi = 1 + \sum_{A=1}^{2} \frac{m_A}{2r_A} + u\,,
\end{equation}
where $m_A$ are the puncture bare-mass parameters and $u$ is a regular correction field to be solved for numerically on $\mathbb{R}^3$. 
With the choices above, the matter contribution to Eq.~\eqref{Ham_York} is determined by the electric field specified by the Coulomb-type solution given in Eq.~\eqref{Coulomb}.

The total ADM energy of the initial data is defined at the outer spatial infinity by~\cite{Gourgoulhon:2007ADM} 
\begin{equation}
E_{\rm ADM}
=
\frac{1}{16\pi}
\lim_{r\to\infty}
\oint_{S_r}
\delta^{jk}
\left(
\partial_j\gamma_{ik}-\partial_i\gamma_{jk}
\right)n^i\,dS\,,
\label{eq:ADM_energy_def}
\end{equation}
where $S_r$ is a large coordinate sphere in asymptotically Cartesian coordinates, and $n^i$ and $dS$ are its outward unit normal and area element with respect to the flat background metric. 
For conformally flat initial data, this expression reduces to 
\begin{equation}
E_{\rm ADM}
=
-\frac{1}{2\pi}
\lim_{r\to\infty}
\oint_{S_r}\partial_i\psi\,n^i\,dS\,.
\label{eq:ADM_energy_conformal}
\end{equation}
Consequently, if $\psi=1+C/r+{\cal O}(r^{-2})$ and $\partial_r\psi=-C/r^2+{\cal O}(r^{-3})$ at spatial infinity, then
$E_{\rm ADM}=2C$. 
We evaluate the initial ADM energy from the leading asymptotic coefficient of the conformal factor obtained with the
spectral puncture solver~\cite{Ansorg:2004ds}. 
In a frame with vanishing total ADM linear momentum, the ADM mass satisfies $M_{\rm ADM}=E_{\rm ADM}$. 
This global quantity is distinct from the puncture bare-mass parameters, the quasi-local horizon masses and irreducible masses. 

The above procedure yields constraint-satisfying BBH initial data for the EMD system. 
Because $\phi=K_\phi=0$ on the initial hypersurface, the constraint solution at fixed $m_A$, $Q_A$, and $P_A^i$ is independent of the dilaton coupling $a$. 
For charged configurations with $a\neq0$, however, $\phi=0$ is not stationary since the electromagnetic invariant $F_{\mu\nu}F^{\mu\nu}$ sources the dilaton equation~\eqref{scalar_eom}. 
The scalar field therefore develops dynamically around the black holes, producing an early relaxation transient and an initial burst of scalar radiation. 
By contrast, the scalar field remains unsourced for $a=0$ or for neutral initial data. 
In the analysis below, we distinguish the initial relaxation from the subsequent charge- and coupling-dependent inspiral and merger dynamics. 
Initial data incorporating a self-consistent nontrivial dilaton profile, as well as magnetically charged initial data, are left for future work. 

\section{Wave Extraction and Quasi-local Horizon Quantities}
\label{sec:extraction_and_quantities}

We describe the extraction of gravitational, electromagnetic, and scalar radiative quantities, together with the quasi-local horizon quantities used to monitor the black holes during the evolution.

\subsection{Radiative Quantities}
\label{subsec:radiative_quantities}

For the gravitational and electromagnetic radiative fields, we employ the Newman--Penrose (NP) formalism~\cite{Newman:1961qr}. 
These outgoing radiative fields are extracted on coordinate spheres at a sufficiently large radius and decomposed into spin-weighted spherical harmonic modes.

\subsubsection{Gravitational Wave Extraction}

The gravitational radiation is extracted from the Weyl scalar
\begin{equation}
\Psi_4 \equiv -\, C_{\alpha\beta\gamma\delta}\, n_{\rm NP}^\alpha \bar{m}^\beta n_{\rm NP}^\gamma \bar{m}^\delta\,,
\label{eq:Psi4_def}
\end{equation}
where $C_{\alpha\beta\gamma\delta}$ is the Weyl tensor and $(l^\mu,n_{\rm NP}^\mu,m^\mu,\bar{m}^\mu)$ is a null tetrad adapted to the extraction sphere~\cite{Newman:1961qr, Teukolsky:1972my}. 
In the radiation zone, $\Psi_4$ is related to the second time derivatives of the gravitational-wave polarizations $h_+$ and $h_\times$ via
\begin{equation}
\Psi_4 = \ddot{h}_+ - i\, \ddot{h}_\times\,.
\end{equation}

We decompose $\Psi_4$ into spin-weight $-2$ spherical harmonics as
\begin{equation}
  \Psi_4(t,r,\theta,\varphi)
    = \sum_{\ell,m} \Psi_4^{\ell m}(t,r)\, {}_{-2}Y_{\ell m}(\theta,\varphi)\,.
    \label{defpsi4}
\end{equation}
The corresponding gravitational-wave modes of the strain, $h_{\ell m}$, are obtained by double time integration of $\Psi_4^{\ell m}$,
\begin{equation}
  h_{\ell m}(t)
   = \int^t dt' \int^{t'} dt''\, \Psi_4^{\ell m}(t'')\,.
  \label{eq:hlm_from_psi4}
\end{equation}
The double time integration in the above equation was performed using the fixed-frequency integration (FFI) method \cite{Reisswig:2010di}, with a cutoff period of $p_{\rm cut} = 500M$. 
Here $M$ denotes the reference mass scale defined in Sec.~\ref{subsec:parameter_space}. 
\subsubsection{Electromagnetic Wave Extraction}

Electromagnetic radiation is extracted via the NP scalar
\begin{equation}
  \Phi_2 \equiv F_{\alpha\beta}\, \bar{m}^{\alpha} n_{\rm NP}^{\beta}\,,
\label{eq:Phi2_def}
\end{equation}
where $F_{\alpha\beta}$ denotes the Maxwell field strength, and the same Newman--Penrose tetrad as above is used~\cite{Newman:1961qr}. 
The tetrad is normalized according to $l_\mu n_{\rm NP}^\mu=-1$ and $m_\mu\bar{m}^\mu=1$.
In the wave zone, $\Phi_2$ encodes the outgoing radiative degrees of freedom of the electromagnetic field. 

Analogously to the gravitational case, we perform a spin-weight $-1$ decomposition as
\begin{equation}
  \Phi_2(t,r,\theta,\varphi)
    = \sum_{\ell,m} \Phi_2^{\ell m}(t,r)\, {}_{-1}Y_{\ell m}(\theta,\varphi)\,.
    \label{defPhi2}
\end{equation}
\subsubsection{Radiation Zone and Conventions}

All radiative quantities are extracted on a coordinate sphere at a sufficiently large radius. In the wave zone, both $\Psi_4$ and $\Phi_2$ fall off as $r^{-1}$ to leading order. 
We therefore analyze the rescaled quantities $r_{\rm ext}\Psi_4(t,r_{\rm ext})$ and $r_{\rm ext}\Phi_2(t,r_{\rm ext})$ as finite-radius approximations to the corresponding asymptotic radiation fields at future null infinity $\mathcal{I}^+$.
For gravitational radiation, this finite-radius rescaling is commonly adopted in numerical relativity~\cite{Campanelli:2005dd, Baker:2005vv}.

\subsubsection{Radiated Energy Fluxes}

In addition to the waveform quantities, we compute the energy carried away by gravitational, scalar, and electromagnetic radiation. 
In the asymptotic region, the corresponding energy fluxes are evaluated from the leading radial contributions at future null infinity $\mathcal{I}^+$.

For gravitational waves, the energy flux is expressed in terms of the Weyl scalar $\Psi_4$ as~\cite{Teukolsky:1972my}
\begin{equation}
\frac{dE^{\rm GW}}{dt}
= \lim_{R \to \infty} \frac{R^2}{16\pi}
\int d\Omega \left| \int_{-\infty}^t \Psi_4\, d\tilde{t}\, \right|^2\,.
\label{eq:dE_gw}
\end{equation}
In practice, the time integration was performed using the FFI method with a cutoff period of $p_{\rm cut} = 120M$. 
For the scalar channel, the radiated energy flux is computed from the radial component of the scalar energy current,
\begin{equation}
\frac{dE^\phi}{dt}
= + \lim_{R \to \infty} R^2 \int d\Omega\, \mathcal{J}^\phi_R\,,
\label{eq:dE_phi}
\end{equation}
while the electromagnetic contribution is computed analogously from the radial component of the electromagnetic energy current,
\begin{equation}
\frac{dE^F}{dt}
= + \lim_{R \to \infty} R^2 \int d\Omega\, \mathcal{J}^F_R\,.
\label{eq:dE_F}
\end{equation}
Here, $\mathcal{J}^\phi_R=(1/4\pi)\Lambda^i{}_{R}J^\phi_i$ and $\mathcal{J}^F_R=(1/4\pi)\Lambda^i{}_{R}e^{-2a\phi}J^F_i$, where $J_i^\phi$ and $J_i^F$ are defined in Eq.~\eqref{defJi}. 
In the asymptotically Cartesian coordinates used for wave extraction, $\Lambda^i{}_{R}=x^i/R$ projects the spatial current onto the outward radial direction. 
The factor $e^{-2a\phi}$ is included because the electromagnetic contribution to the effective stress tensor is $e^{-2a\phi}T^F_{\mu\nu}/(4\pi)$, as shown in Eq.~\eqref{grav_eom}. 

For the asymptotically flat boundary conditions adopted here, $\phi$ approaches zero as $R\to\infty$, and the leading electromagnetic energy flux can equivalently be expressed in terms of the outgoing NP scalar as~\cite{Bozzola:2021elc} 
\begin{equation}
\frac{dE^F}{dt}
=
\lim_{R\to\infty}
\frac{1}{4\pi}
\sum_{\ell=1}^{\infty}
\sum_{m=-\ell}^{\ell}
\left|
R\Phi_2^{\ell m}
\right|^2\,.
\label{eq:dE_F_Phi2}
\end{equation}
To resolve this flux into multipolar sectors at the finite extraction radius, we define the $m$-summed norm 
\begin{equation}
\mathcal{N}_{\ell}^{\rm EM}(t)
\equiv
\left[
\sum_{m=-\ell}^{\ell}
\left|
r_{\rm ext}\Phi_2^{\ell m}(t,r_{\rm ext})
\right|^2
\right]^{1/2}
\qquad
(\ell\geq1)\,.
\label{eq:em_np_multipole_norm}
\end{equation}
The sum over squared moduli prevents cancellations among modes with different values of $m$ or different complex phases. 
The corresponding finite-radius approximation to the flux carried by the $\ell$ sector is 
\begin{equation}
\mathcal{P}_{\ell}^{\rm EM}(t)
\equiv
\frac{1}{4\pi}
\left[
\mathcal{N}_{\ell}^{\rm EM}(t)
\right]^2\,.
\label{eq:em_np_multipole_flux}
\end{equation}
Accordingly, $\sum_{\ell\geq1}\mathcal{P}_{\ell}^{\rm EM}$ provides a finite-radius approximation to the total electromagnetic flux in Eq.~\eqref{eq:dE_F}. 

\subsection{Quasi-local Horizon Quantities}
\label{subsec:horizon_quantities}

In order to monitor the physical properties of the black holes during the evolution, we compute quasi-local quantities on apparent horizons. 
Once an apparent horizon is found by \texttt{AHFinderDirect}, we evaluate surface integrals on a closed two-dimensional spatial cross section $\mathcal{S}$ of the horizon. 

In particular, we monitor two quasi-local electric charges on each apparent horizon. 
Let $s_i$ denote the outward-pointing spacelike unit normal to a closed two-surface ${\cal S}$ within the spatial hypersurface. 
Adopting the conventions of Ref.~\cite{Bozzola:2019aaw}, the unweighted electric flux is

\begin{equation}
Q_e=\frac{1}{4\pi}\int_{\cal S} E^i s_i\,dA\,.
\label{eq:Qe_def}
\end{equation}

The electric charge associated with the Maxwell equation in EMD theory is instead the dilaton-weighted flux 

\begin{equation}
Q_{\rm EMD}=\frac{1}{4\pi}\int_{\cal S}e^{-2a\phi}E^i s_i\,dA\,.
\label{eq:QEMD_def}
\end{equation}

For $a=0$, the dilaton factor becomes unity and the two definitions coincide. 
For $a\neq0$, however, the dynamically generated dilaton profile can cause $Q_e$ to evolve even when $Q_{\rm EMD}$ remains approximately constant on each horizon. 
This distinction is important for interpreting the horizon data presented below. 

We next define the irreducible mass from the area of each apparent horizon.
Let $A_{\rm H}$ be the area of a spatial cross section of the apparent horizon. 
The corresponding areal radius is defined by
\begin{equation}
  r_{\rm H} \equiv \sqrt{\frac{A_{\rm H}}{4\pi}}\,.
\end{equation}
We then define the irreducible mass as
\begin{equation}
  M_{\rm irr} \equiv \frac{r_{\rm H}}{2}
  = \sqrt{\frac{A_{\rm H}}{16\pi}}\,.
  \label{eq:Mirr}
\end{equation}
For a Schwarzschild black hole, this quantity coincides with the black hole mass. 
Although the horizons in our simulations are dynamical and are not, in general, static or spherically symmetric, $M_{\rm irr}$ provides a useful quasi-local diagnostic based only on the horizon area.
In Sec.~\ref{sec:results}, we use $M_{\rm irr}$ as a remnant-scale diagnostic when identifying the common apparent horizon and determining its first-detection time $t_{\rm CAH}$. 

In order to monitor the scalar field on the horizon, we define the horizon-averaged scalar field
\begin{equation}
  \Phi_{\rm H} \equiv \frac{1}{A_{\rm H}} \int_{\mathcal{S}} \phi \, dA\,.
  \label{eq:PhiH_def}
\end{equation}
This quantity is a signed surface average, so that $|\Phi_{\rm H}|$ denotes the absolute value taken after performing the surface integral.
Since the dilaton is sourced by the electromagnetic field in EMD theory, $\Phi_{\rm H}$ provides a useful diagnostic for tracking the development of the scalar profile around charged black holes. 
All quasi-local quantities defined above are evaluated on the apparent horizons identified during the numerical evolution. 
They allow us to track the quasi-local electric charges, the irreducible mass, and the horizon-averaged scalar field on the individual horizons before merger and on the common horizon afterward. 

\section{Results}
\label{sec:results}

This section examines how the input charge magnitude, the dilaton coupling, and the relative charge sign shape the dynamics and radiation of charged black-hole binaries. 
After introducing the common numerical setup and parameter space, we first vary the charge magnitude at the string-motivated coupling $a=1$. 
We then study the coupling dependence for highly charged binaries carrying equal-magnitude opposite charges and for like-charged binaries with nonzero total charge. 
Finally, we return to the weak-charge regime, where the scalar response admits a perturbative interpretation. 
Together, these comparisons connect the quasi-local horizon response, the electromagnetic multipolar structure, the radiative energy channels, the dominant gravitational waveform, and the properties of the remnant. 

\subsection{Numerical setup and parameter space}
\label{subsec:parameter_space}

We perform the numerical evolutions using the \texttt{EinsteinToolkit}~\cite{EinsteinToolkit:article, EinsteinToolkit:web}. 
Our custom thorn \texttt{EMDEvolve}~\cite{EMDRepository} couples the EMD evolution system described in Sec.~\ref{sec:formulation} to the spacetime evolution, using the \texttt{Scalar} thorn~\cite{Cunha:2017wao} for the scalar variables and the \texttt{Proca} thorn~\cite{Zilhão_2015} as the vector-field infrastructure, both distributed through the Canuda library~\cite{Canuda:zenodo}. 
The puncture initial data are generated with \texttt{TwoPunctures\_EMD}~\cite{EMDRepository}, our extension of \texttt{TwoPunctures}~\cite{Ansorg:2004ds} that incorporates the electromagnetic and dilaton contributions described in Sec.~\ref{sec:Initial_Data_Construction}. 
Apparent horizons are located with \texttt{AHFinderDirect}~\cite{Thornburg:2003sf}, and the quasi-local horizon quantities are evaluated with our extension \texttt{QuasiLocalMeasuresEMD}~\cite{EMDRepository} of \texttt{QuasiLocalMeasures}~\cite{Dreyer:2002mx}. 
The simulation data were analyzed using the Python package \texttt{kuibit}~\cite{kuibit}, with additional functionality implemented as needed.

We compare the simulations using a common input convention. 
All dimensional quantities are normalized by appropriate powers of a common reference mass scale, denoted by $M=1$ in code units.
The puncture data are specified by input bare masses and input charges, whereas the ADM mass and quasi-local horizon quantities are evaluated separately from the resulting numerical data. 
Throughout all simulations, we use the Gauss-constraint damping parameter $\kappa M=1$, where $\kappa$ is defined in Eq.~\eqref{EMD_Zeta}.
Unless otherwise stated, the simulations use the same grid hierarchy, outer-boundary location, and wave-extraction radius $r_{\rm ext}=500M$. 
The initial puncture separation is $10M$ for the neutral sequence, $11M$ for both the weakly charged opposite-charge and like-charged sequences, and $15M$ for the highly charged opposite-charge sequence.
For the charged sequences, these larger separations are chosen to allow the initial field relaxation to settle before the binary reaches $D/M=10$. The merger-time intervals below are measured from this common reference separation. 

We characterize each configuration by the dilaton coupling $a$ and the dimensionless input charge parameters $\lambda_+$ and $\lambda_-$. 
The subscripts $+$ and $-$ label the punctures initially located on the positive and negative sides of the binary axis, respectively. 
The input electric charges are assigned according to 
\begin{equation}
Q_{\pm}^{\rm input}=\lambda_{\pm}m_{\pm}\,,
\label{eq:lambda_input_def}
\end{equation}
where $m_\pm$ are the puncture bare-mass parameters introduced in Sec.~\ref{sec:Initial_Data_Construction}.
For the parameter sets considered here, $m_+/M\simeq m_-/M\simeq0.4859161023$. 
Thus, $\lambda_\pm$ are input charge-to-bare-mass ratios, rather than ratios defined using ADM or horizon masses.
We use $\lambda_\pm=\pm c$ as shorthand for $\lambda_+=+c$ and $\lambda_-=-c$, whereas the neutral case is denoted by $\lambda_\pm=0$.
The input charges $Q_\pm^{\rm input}$ should be distinguished from the quasi-local horizon charges measured during the evolution.

For charged binaries, the $a=0$ simulations provide Einstein--Maxwell reference cases, while the $a=1$ and $a=\sqrt{3}$ simulations probe the effects of the dilaton--electromagnetic coupling. 
For neutral binaries with $\lambda_{\pm}=0$, the Maxwell field vanishes and the dilaton remains unsourced. 
Their evolution therefore reduces to vacuum Einstein gravity independently of the value assigned to $a$, providing a common vacuum-GR baseline for the charged sequences. 

The initial orbital momenta are first computed for the corresponding uncharged binary using \texttt{NRPyPN} ~\cite{Habib:2020dba} and are then rescaled according to the input charge parameters. 
Adapting the approximate prescription used in Einstein--Maxwell binary simulations~\cite{Bozzola:2021elc} to our input-charge convention, we set 
\begin{equation}
p_t=p_t^{\rm NRPyPN}\sqrt{1-\lambda_{+}\lambda_{-}}\,,
\qquad
p_r=p_r^{\rm NRPyPN}\sqrt{1-\lambda_{+}\lambda_{-}}\,.
\label{eq:initial_momentum_charge_rescaling}
\end{equation}
Here, $p_t$ and $p_r$ denote the tangential and radial orbital momentum components, respectively. 
These components determine the puncture momentum vectors $P_A^i$ entering the Bowen--York extrinsic curvature specified in Sec.~\ref{sec:Initial_Data_Construction}. 
The rescaling factor is motivated by the Newtonian balance between gravitational and Coulomb forces. 
It increases the momentum magnitudes for opposite-sign charges, for which the electromagnetic interaction is attractive, and decreases them for like-sign charges, for which it is repulsive. 
We apply the same prescription to all sampled dilaton couplings, thereby providing a common initial-momentum convention for the coupling comparisons below. 

The simulation set summarized in Table~\ref{tab:simulation_parameters} defines four complementary comparisons. 
At fixed $a=1$, the neutral, weakly charged, and highly charged binaries with $\lambda_{\pm}=0$, $\pm0.05$, and $\pm0.7$ isolate the dependence on the input charge magnitude. 
At fixed $\lambda_{\pm}=\pm0.7$ and $\lambda_{\pm}=\pm0.05$, varying $a$ from $0$ to $1$ and $\sqrt{3}$ probes the coupling dependence in the high- and weak-charge regimes, respectively. 
The like-charged sequence with $\lambda_{+}=\lambda_{-}=+0.3$ probes the coupling dependence when the total EMD charge does not cancel. 
The terms ``weak charge'' and ``high charge'' refer only to the two nonzero opposite-charge scales within the present simulation set. 
\begin{table}[h]
\centering
\begin{tabular}{l|c|c}
\hline\hline
Charge configuration
& $(\lambda_+,\lambda_-)$
& Sampled coupling $a$\\
\hline
Neutral
& $(0,0)$
& $0,1,\sqrt{3}$\\
\hline
Opposite charge, weak \hspace{.5cm}
& $(+0.05,-0.05)$
& $0,1,\sqrt{3}$\\
\hline
Like charge
& $(+0.3,+0.3)$
& $0,1,\sqrt{3}$
\\
\hline
Opposite charge, high
& $(+0.7,-0.7)$
& $0,1,\sqrt{3}$\\
\hline\hline
\end{tabular}
\caption{\label{tab:simulation_parameters}
Simulation parameters. 
The input charge parameters $\lambda_{+}$ and $\lambda_{-}$ are defined in Eq.~\eqref{eq:lambda_input_def}. 
Each charge configuration is evolved for the Einstein--Maxwell value $a=0$ and the EMD values $a=1$ and $a=\sqrt{3}$. 
The neutral configurations provide vacuum-GR baselines, while the charged $a=0$ configurations provide Einstein--Maxwell reference cases. 
}
\end{table}
\begin{table}[h]
\centering
\begin{tabular}{l|c|c|c}
\hline\hline
Charge configuration
& $(\lambda_+,\lambda_-)$
& $M_{1,\rm ADM}/M=M_{2,\rm ADM}/M$
& $M_{\rm ADM}/M$\\
\hline
Neutral
& $(0,0)$
& $0.500000$
& $0.98975$ \\
\hline
Opposite charge, weak \hspace{.5cm}
& $(+0.05,-0.05)$
& $0.498666$
& $0.988007$ \\
\hline
Like charge
& $(+0.3,+0.3)$
& $0.498471$ 
& $0.986539$ \\
\hline
Opposite charge, high
& $(+0.7,-0.7)$
& $0.495631$ 
& $0.988775$ \\
\hline\hline
\end{tabular}
\caption{
Initial total and individual ADM masses reported by the puncture solver, normalized by the reference mass scale $M=1$.
The two punctures have equal individual ADM masses to the precision shown. 
For each charged configuration, the reported total and individual masses agree across $a=0$, $1$, and $\sqrt{3}$ to the precision quoted. 
The values refer to the initial data before the dynamical field relaxation, not to horizon-mass measurements.
\label{tab:initial_puncture_masses}
}
\end{table}
Table~\ref{tab:initial_puncture_masses} summarizes the initial total and individual ADM masses reported by the puncture solver. 
Here $M_{\rm ADM}$ is the total ADM mass defined at the outer spatial infinity, while $M_{A,\rm ADM}$ is the ADM mass associated with the inner asymptotically flat end corresponding to puncture $A$. 
The total ADM mass is not, in general, equal to $M_{1,\rm ADM}+M_{2,\rm ADM}$~\cite{Ansorg:2004ds}. 
We organize the results using complementary orbital, quasi-local, and radiative diagnostics. 
The coordinate separation provides a diagnostic of the late orbital evolution, while the near-horizon response is characterized by the horizon-averaged scalar field $\Phi_{\rm H}$, the unweighted Maxwell charge $Q_e$, and the dilaton-weighted EMD charge $Q_{\rm EMD}$ defined in Sec.~\ref{subsec:horizon_quantities}. 
The far-zone response is characterized by the dominant gravitational-wave modes, the gravitational, electromagnetic, and scalar energy fluxes, and the multipolar structure of the electromagnetic Newman--Penrose scalar $\Phi_2$, using the definitions in Sec.~\ref{subsec:radiative_quantities}. 
Taken together, these diagnostics connect the evolution of the orbital motion and near-horizon fields to the radiation emitted during the inspiral, merger, and ringdown. 

\subsection{Charge dependence at fixed coupling \texorpdfstring{$a=1$}{a=1}}
\label{subsec:main_results_a1}

At the string-motivated coupling $a=1$, we examine the dependence on the input charge magnitude by comparing a neutral binary with weakly and highly charged binaries carrying equal-magnitude opposite charges. 
The three configurations are specified by 
\begin{equation}
\lambda_{\pm}=0\,,
\qquad
\lambda_{\pm}=\pm0.05\,,
\qquad
\lambda_{\pm}=\pm0.7\,.
\label{cc1}
\end{equation}
This sequence connects the weak-charge regime, where the dominant gravitational waveform remains close to the vacuum-GR baseline, to the high-charge regime, where the electromagnetic and scalar sectors can substantially modify the binary dynamics and radiation. 
For the horizon diagnostics, we define $t_{\rm CAH}$ as the first coordinate time at which \texttt{AHFinderDirect} detects a common apparent horizon enclosing both individual horizons. 
We identify the common horizon among the tracked horizons
using its remnant-scale irreducible mass and its centroid near
the binary's coordinate center.

The initial data are constructed with $\phi=0$, as described in Sec.~\ref{sec:Initial_Data_Construction}. 
For the charged configurations, however, the electromagnetic invariant $F_{\mu\nu}F^{\mu\nu}$ sources the dilaton equation, and a nontrivial scalar profile develops dynamically. 
The simultaneous evolution of $\Phi_{\rm H}$, $Q_e$, and $Q_{\rm EMD}$ therefore allows us to distinguish the dynamical formation of the dilaton profile from the conservation of the dilaton-weighted EMD charge. 

\begin{figure*}[!htbp]
    \centering
    \begin{subfigure}[b]{0.325\textwidth}
        \centering
        \includegraphics[width=\linewidth]
        {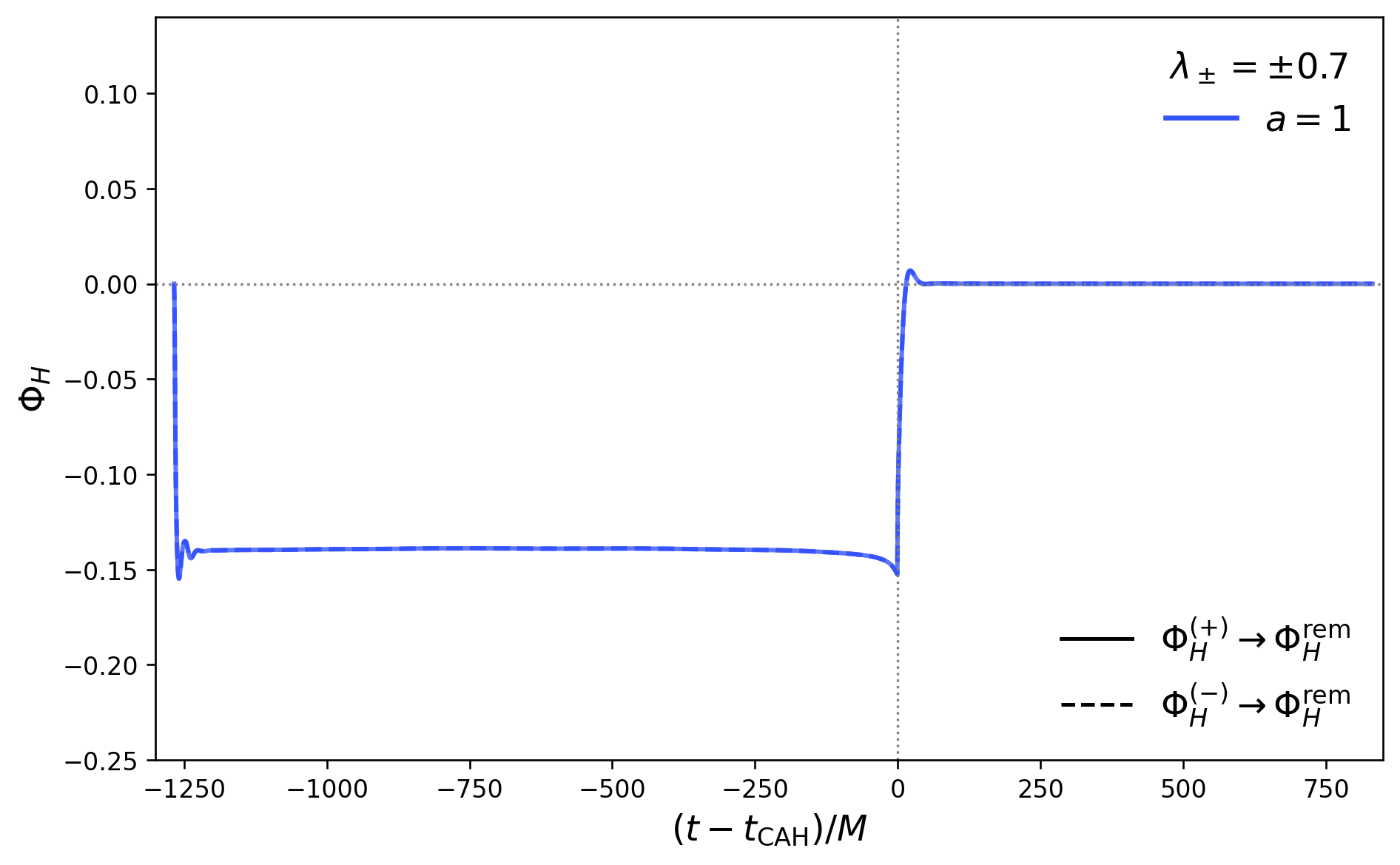}
        \caption{\centering $\Phi_{\rm H}$}
        \label{fig:Phi_H_a1_main}
    \end{subfigure}
    \hfill
    \begin{subfigure}[b]{0.325\textwidth}
        \centering
        \includegraphics[width=\linewidth]
        {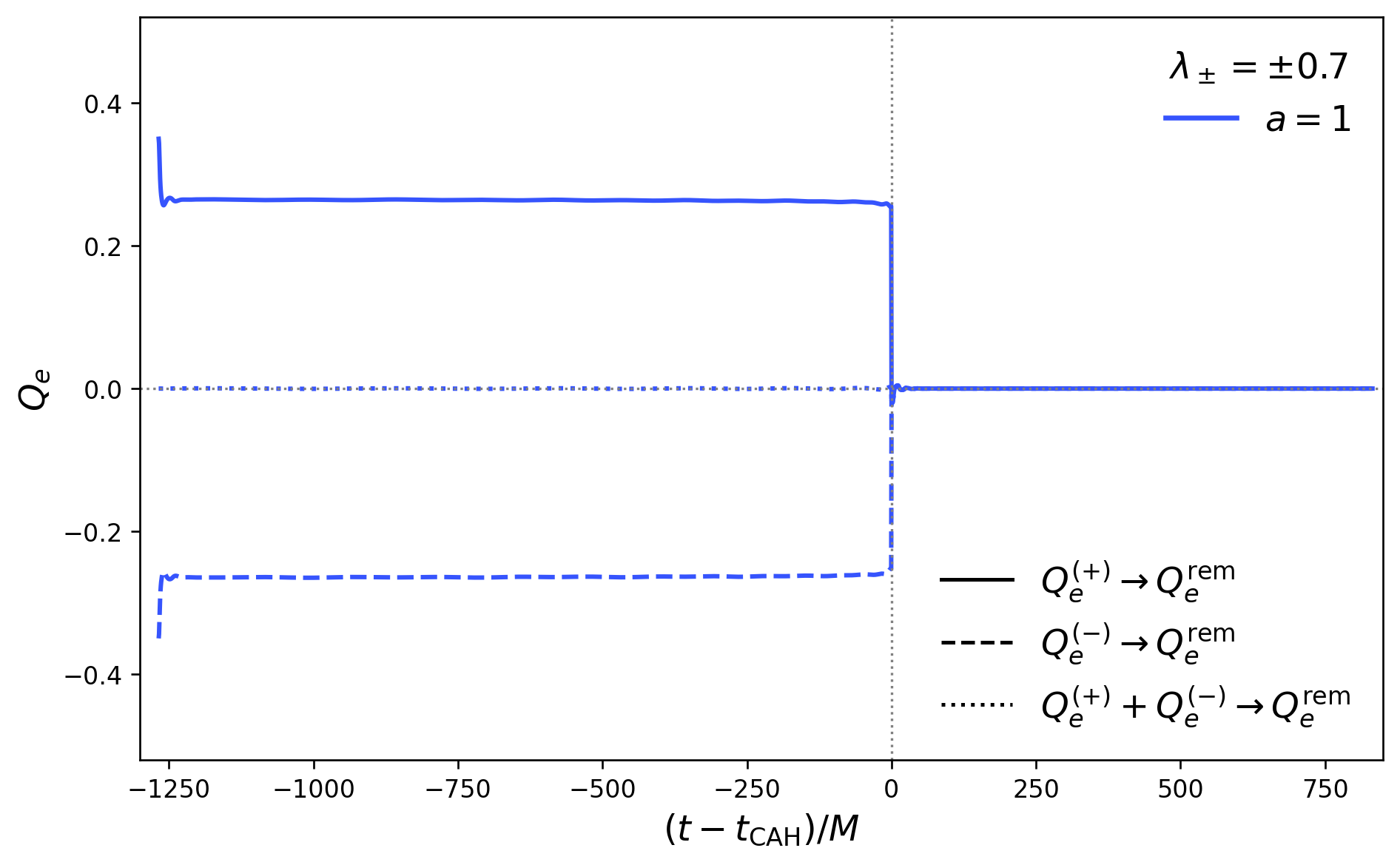}
        \caption{\centering $Q_e$}
        \label{fig:Q_e_a1_main}
    \end{subfigure}
    \hfill
    \begin{subfigure}[b]{0.325\textwidth}
        \centering
        \includegraphics[width=\linewidth]
        {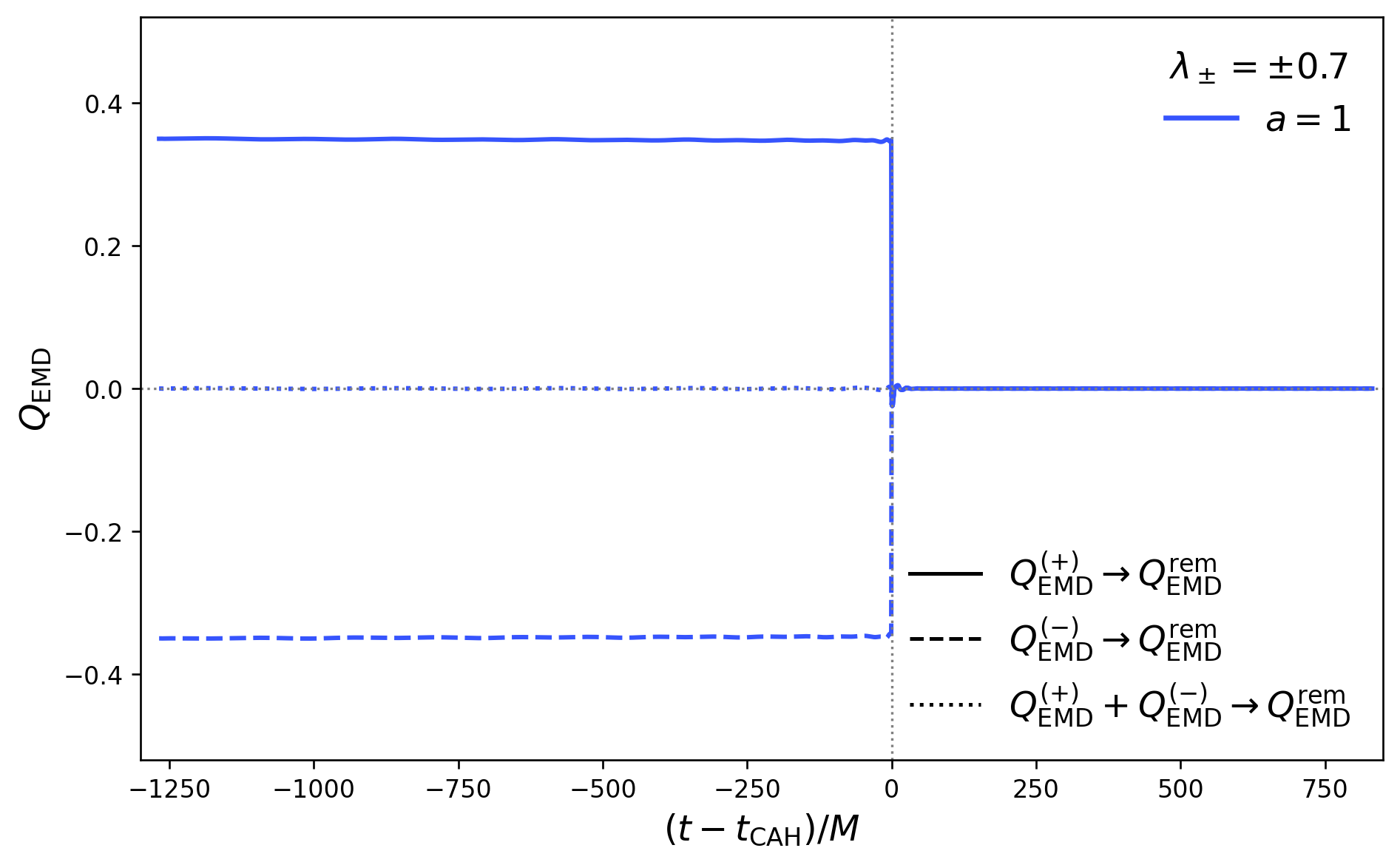}
        \caption{\centering $Q_{\rm EMD}$}
        \label{fig:Q_EMD_a1_main}
    \end{subfigure}
    \caption{
    Evolution of the quasi-local horizon quantities for the highly charged $a=1$ configuration with $\lambda_{\pm}=\pm0.7$. 
    Panel (a) shows the horizon-averaged scalar field $\Phi_{\rm H}$ defined in Eq.~\eqref{eq:PhiH_def}, panel (b) shows the unweighted electric charge $Q_e$ defined in Eq.~\eqref{eq:Qe_def}, and panel (c) shows the dilaton-weighted EMD charge $Q_{\rm EMD}$ defined in Eq.~\eqref{eq:QEMD_def}. 
    The horizontal axis shows $(t-t_{\rm CAH})/M$.  
    Before $t_{\rm CAH}$, the solid and dashed curves denote quantities evaluated on the individual horizons associated with the positive and negative charges, respectively. 
    At and after $t_{\rm CAH}$, the displayed quantities refer to the common horizon. 
    In panels (b) and (c), the dotted curves denote the sum of the two individual horizon charges before $t_{\rm CAH}$. 
    The vertical dotted line marks $t=t_{\rm CAH}$.
    }
    \label{fig:charge_a1_main}
\end{figure*}

In Fig.~\ref{fig:charge_a1_main}, we show the evolution of quasi-local horizon quantities for the highly charged configuration with $a=1$ and $\lambda_{\pm}=\pm0.7$, for which the scalar and electromagnetic responses are most pronounced. 
Because these quantities are evaluated directly on the apparent horizons, their time coordinate is shown as $(t-t_{\rm CAH})/M$ without a propagation-time correction. 
The solid and dashed curves track the two individual horizons before $t_{\rm CAH}$ and the common horizon afterward, while the dotted curves in the charge panels show the net horizon charge. 

Figure~\ref{fig:Phi_H_a1_main} shows that $\Phi_{\rm H}$ rapidly departs from zero during the initial transient and subsequently settles to an approximately constant negative value. 
This adjustment reflects the dynamical formation of a scalar profile around each charged black hole, while the subsequent near constancy of $\Phi_{\rm H}$ indicates a quasi-stationary inspiral regime. 
Concurrently, the magnitude of the unweighted electric charge $|Q_e|$ decreases from its initial value, whereas the quasi-local EMD charge $Q_{\rm EMD}$ associated with each individual horizon remains close to its initial value throughout the inspiral. 
This contrasting behavior follows from the factor $e^{-2a\phi}$ in the electric-flux integral defining $Q_{\rm EMD}$. 
The dynamically generated scalar profile modifies this weighting, allowing $Q_e$ to adjust without a comparable change in $Q_{\rm EMD}$. 

Because the two individual EMD charges have equal magnitudes and opposite signs, their net value remains consistent with zero during the inspiral, and the EMD charge of the common horizon is likewise consistent with zero after $t_{\rm CAH}$. 
Electromagnetic and scalar-field transients can temporarily surround the remnant after merger, but subsequently radiate away. 
The decay of $\Phi_{\rm H}$ toward zero, together with the common-horizon values of $Q_e$ and $Q_{\rm EMD}$ remaining consistent with zero, supports relaxation toward an uncharged black hole without nontrivial secondary dilaton hair. 
This behavior is consistent with the fact that the scalar charge of a stationary EMD black hole is secondary hair fixed by its electromagnetic charge and horizon electric potential~\cite{Pacilio:2018gom}. 
Indeed, in the static sector, the neutral limit of the GM--GHS solution~\eqref{GMGHS} reduces to the Schwarzschild solution. 

The weakly charged configuration with $\lambda_{\pm}=\pm0.05$, which is not shown in Fig.~\ref{fig:charge_a1_main}, exhibits the same qualitative response with substantially smaller changes in the horizon quantities.
Averaging over $-1100\leq(t-t_{\rm CAH})/M\leq-200$ and over the two individual horizons gives representative values of $\Phi_{\rm H}\simeq-6.43\times10^{-4}$, $|Q_e|\simeq2.49\times10^{-2}$, and $|Q_{\rm EMD}|\simeq2.49\times10^{-2}$. 
The small magnitude of the horizon-averaged scalar field indicates
weak scalar dressing.
For the neutral configuration with $\lambda_{\pm}=0$, $\Phi_{\rm H}$, $Q_e$, and $Q_{\rm EMD}$ remain consistent with zero throughout the evolution, as expected for the vacuum-GR baseline. 

For the subsequent comparison of the orbital evolution and radiation, we define a waveform-based radiative merger time. 
The gravitational-wave strain is extracted at $r_{\rm ext}=500M$ and decomposed as 
\begin{equation}
h\equiv h_+-i h_\times
=\sum_{\ell,m}h_{\ell m}\,{}_{-2}Y_{\ell m}\,,
\label{eq:complex_strain_decomposition}
\end{equation}
where $h_{\ell m}$ denotes the corresponding spin-weighted spherical-harmonic mode. 
For the dominant $(\ell,m)=(2,2)$ mode, we define the complex amplitude by 
\begin{equation}
\left|r_{\rm ext}h_{22}\right|
=r_{\rm ext}\sqrt{
\left[{\rm Re}(h_{22})\right]^2+
\left[{\rm Im}(h_{22})\right]^2
}\,.
\label{eq:strain_complex_amplitude}
\end{equation}
We omit the early-time data with $t-r_{\rm ext}<100M$ from the peak search to exclude the initial junk-radiation interval. 
We denote by $t_{\rm peak,ext}^{22}$ the coordinate time of the principal peak of $\left|r_{\rm ext}h_{22}\right|$ in the remaining waveform. 
We then define the corresponding radiative merger time by applying the leading-order propagation-time correction, 
\begin{equation}
t_{\rm merger}^{\rm rad}
\equiv t_{\rm peak,ext}^{22}-r_{\rm ext}\,.
\label{eq:radiative_merger_time}
\end{equation}
For the three $a=1$ configurations considered in this subsection, the resulting values of $t_{\rm merger}^{\rm rad}$ differ from the corresponding common-horizon detection times $t_{\rm CAH}$ by no more than approximately $8M$. 
The two definitions therefore identify nearby stages of the merger, while retaining their distinct waveform-based and horizon-based meanings. 
For radiative quantities evaluated at $r_{\rm ext}$, we use the shifted extraction time 
\begin{equation}
t_{\rm shift}
\equiv t-t_{\rm peak,ext}^{22}
=t-t_{\rm merger}^{\rm rad}-r_{\rm ext}\,.
\label{eq:radiative_time_shift}
\end{equation}
Hereafter, merger-time comparisons use $t_{\rm merger}^{\rm rad}$, whereas quasi-local horizon quantities are referenced to $t_{\rm CAH}$ unless stated otherwise. 

We compare the late orbital evolution from the common coordinate separation $D/M=10$. 
Here, $D$ is measured between the centroids of the two individual horizons, and $t_{D=10}$ denotes the time at which $D/M$ first decreases through $10$ in each simulation. 
The corresponding coordinate-time intervals to the radiative merger are 
\begin{equation}
  \frac{t_{\rm merger}^{\rm rad}-t_{D=10}}{M}
  \simeq
  \begin{cases}
    884.57 & (\lambda_{\pm}=0)\,,\\
    835.47 & (\lambda_{\pm}=\pm0.05)\,,\\
    336.74 & (\lambda_{\pm}=\pm0.7)\,.
  \end{cases}
  \label{eq:merger_time_D10}
\end{equation}
The weakly charged interval is approximately $5.6\%$ shorter than the neutral value. 
By contrast, the highly charged interval is approximately $62\%$ shorter than that of the vacuum-GR baseline, demonstrating a pronounced change in the late orbital evolution at the largest charge magnitude considered here. 

The shorter interval in the highly charged case is qualitatively consistent with the additional electromagnetic attraction between the oppositely charged black holes. 
At a common separation, a Newtonian estimate based on the input charges gives a Coulomb force proportional to $|\lambda_{+}\lambda_{-}|$, with a charge factor $(0.7/0.05)^2=196$ times larger in the high-charge configuration. 
The horizon data in Fig.~\ref{fig:charge_a1_main} confirm that the large, oppositely signed EMD charges persist throughout the inspiral. 
Scalar-mediated interactions and the additional electromagnetic and scalar radiation channels can also influence the evolution, motivating the flux comparison below. 

\begin{figure*}[!htbp]
    \centering
    \begin{subfigure}[b]{0.325\textwidth}
        \centering
        \includegraphics[width=\linewidth]
        {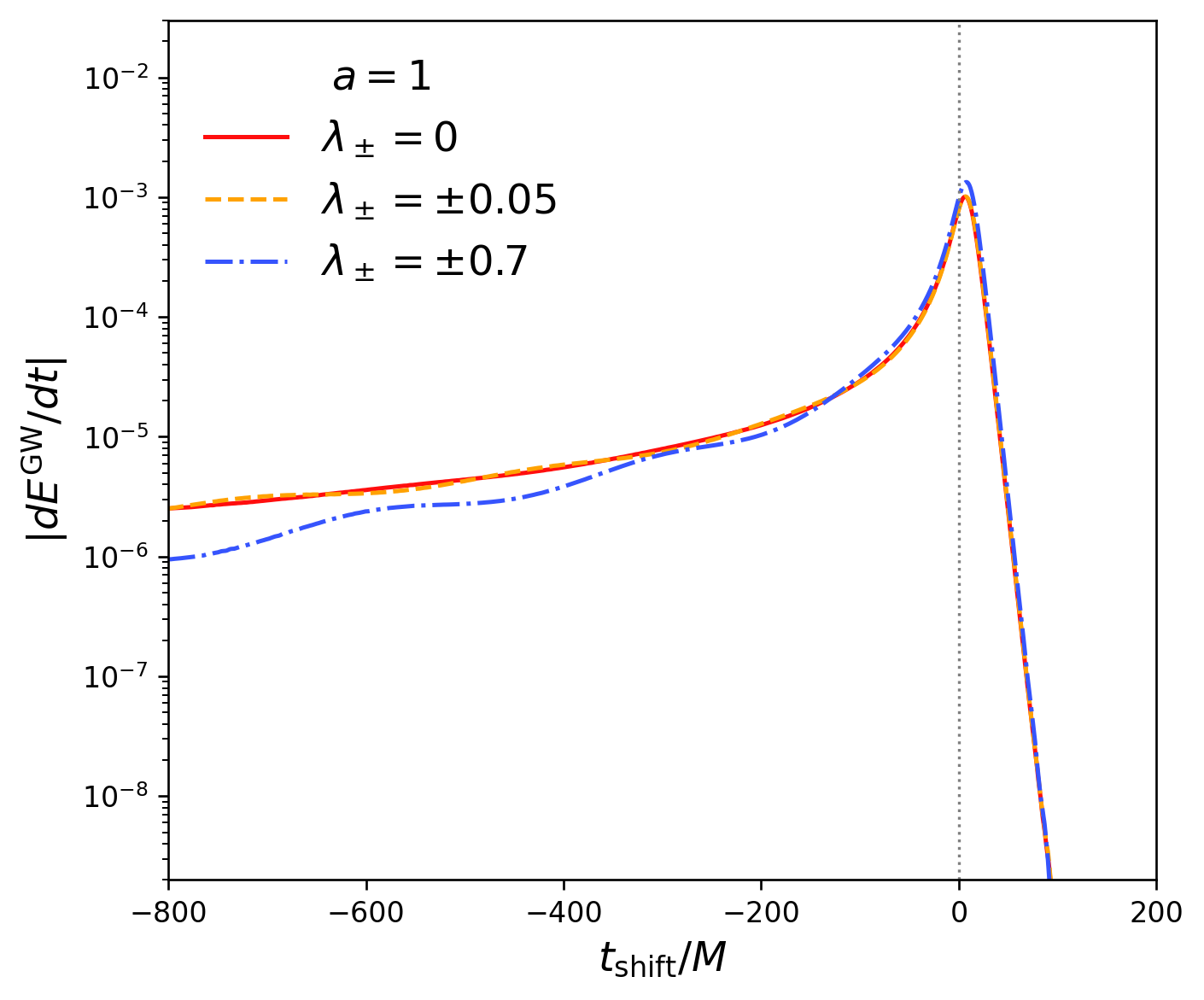}
        \caption{\centering $\left|{dE^{\rm GW}}/{dt}\right|$}
        \label{fig:dEdt_GW_a1}
    \end{subfigure}
    \hfill
    \begin{subfigure}[b]{0.325\textwidth}
        \centering
        \includegraphics[width=\linewidth]
        {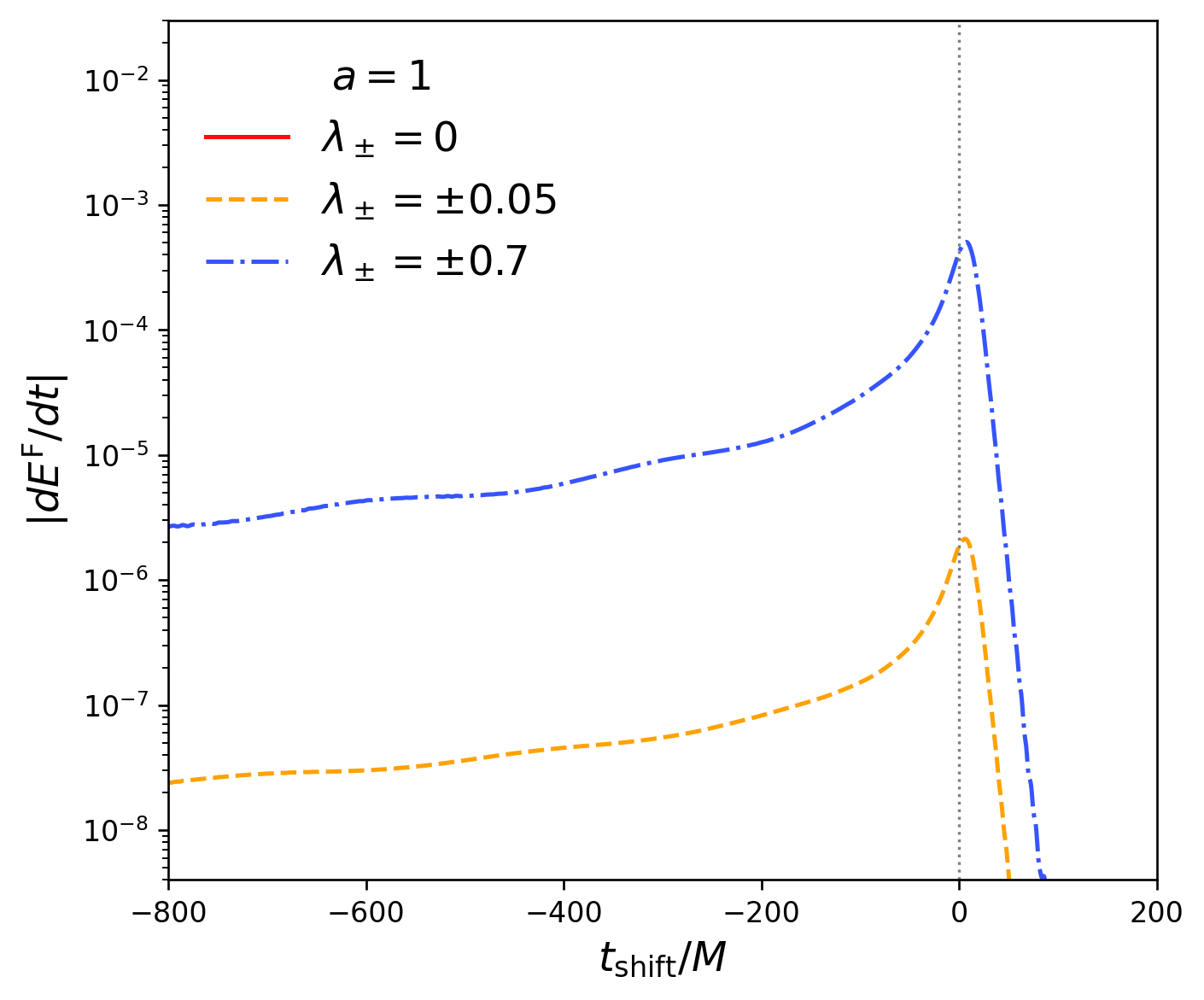}
        \caption{\centering $\left|{dE^{\rm F}}/{dt}\right|$}
        \label{fig:dEdt_EM_a1}
    \end{subfigure}
    \hfill
    \begin{subfigure}[b]{0.325\textwidth}
        \centering
        \includegraphics[width=\linewidth]
        {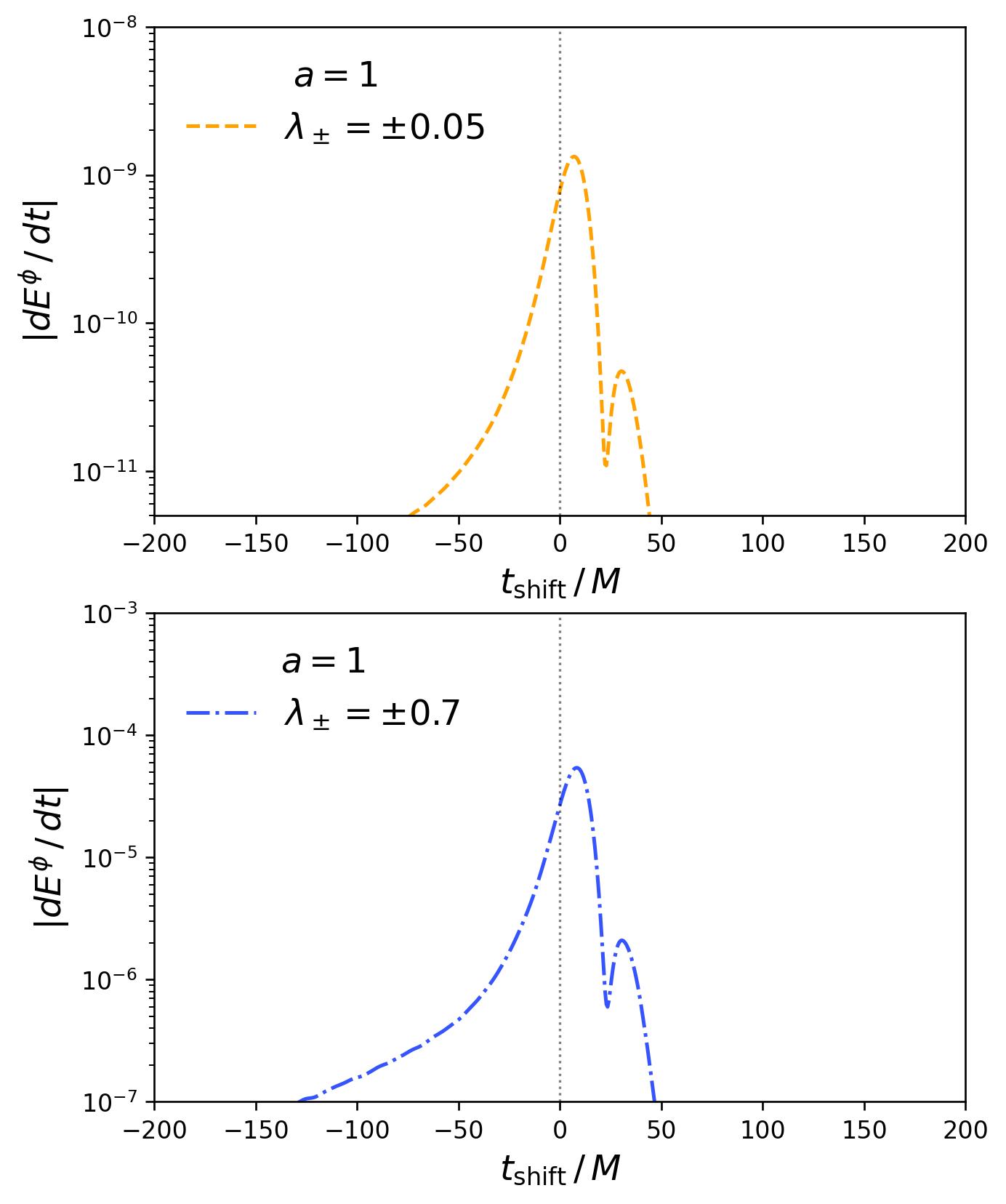}
        \caption{\centering $\left|{dE^\phi}/{dt}\right|$}
        \label{fig:dEdt_Scalar_a1}
    \end{subfigure}
  \caption{
  Magnitudes of the radiative energy fluxes for the $a=1$ cases with $\lambda_{\pm}=0$, $\lambda_{\pm}=\pm0.05$, and $\lambda_{\pm}=\pm0.7$.
    Panels (a), (b), and (c) show the gravitational, electromagnetic, and scalar energy fluxes, respectively.
    The two subpanels in panel (c) display the weak- and high-charge scalar fluxes separately because of their widely different scales.
    The neutral scalar flux is consistent with zero and is therefore omitted.
    For each simulation, the time coordinate is the shifted extraction time $t_{\rm shift}$ defined in Eq.~\eqref{eq:radiative_time_shift}.
    The vertical dotted line marks $t_{\rm shift}=0$, corresponding to the principal peak of $\left|r_{\rm ext}h_{22}\right|$.
    Note that panel (c) uses a different horizontal range. 
  }
  \label{fig:a1_fluxes}
\end{figure*}

Figure~\ref{fig:a1_fluxes} compares the gravitational, electromagnetic, and scalar energy fluxes for the three charge configurations. 
Each simulation is aligned independently using the principal peak of $\left|r_{\rm ext}h_{22}\right|$, and the radiative quantities are plotted against $t_{\rm shift}$ defined in Eq.~\eqref{eq:radiative_time_shift}. 
The peak gravitational-wave flux remains nearly unchanged between the neutral and weakly charged cases, at approximately $1.01\times10^{-3}$. 
In the highly charged case, it increases to approximately $1.34\times10^{-3}$, about $32\%$ above the neutral peak.

The electromagnetic flux exhibits a much stronger charge dependence. 
It remains consistent with zero for the neutral binary, while its peak increases from approximately $2.14\times10^{-6}$ for $\lambda_{\pm}=\pm0.05$ to $5.05\times10^{-4}$ for $\lambda_{\pm}=\pm0.7$, corresponding to a factor of approximately $2.37\times10^2$.
For fixed orbital motion in the weak-charge limit, the leading electromagnetic power scales as $|\lambda_{\pm}|^2$. 
Changing only the input charge magnitude would therefore give a factor of $(0.7/0.05)^2=196$, comparable to the measured peak ratio. 

The scalar flux shows an even steeper dependence on the charge magnitude. 
The neutral binary has a scalar flux consistent with zero because the Maxwell field vanishes and the initially vanishing dilaton remains unsourced. 
For the two charged configurations, the peak scalar flux increases from approximately $1.33\times10^{-9}$ at $|\lambda_{\pm}|=0.05$ to $5.42\times10^{-5}$ at $|\lambda_{\pm}|=0.7$, corresponding to a factor of approximately $4.08\times10^4$.
In the weak-charge expansion, $F_{\mu\nu}={\cal O}(|\lambda_{\pm}|)$, and the electromagnetic invariant sourcing the dilaton is therefore $F_{\mu\nu}F^{\mu\nu}={\cal O}(|\lambda_{\pm}|^2)$. 
The resulting scalar amplitude scales as ${\cal O}(a|\lambda_{\pm}|^2)$, implying a leading scalar-flux scaling of ${\cal O}(a^2|\lambda_{\pm}|^4)$. 
At fixed $a=1$, the charge-only estimate gives $(0.7/0.05)^4\simeq3.84\times10^4$, close to the measured peak-flux ratio. 
Although the highly charged configuration does not lie in a strictly perturbative regime, this numerical proximity shows that the observed hierarchy remains broadly consistent with charge-induced scalar sourcing. 

For the highly charged binary, the electromagnetic and scalar flux peaks are approximately $38\%$ and $4\%$ of the gravitational-wave flux peak, respectively. 
Together with the enhanced electromagnetic attraction between the oppositely charged black holes, these additional channels can contribute to the accelerated late orbital evolution. 

\begin{figure*}[!htbp]
  \centering
  \includegraphics[width=0.48\textwidth]
  {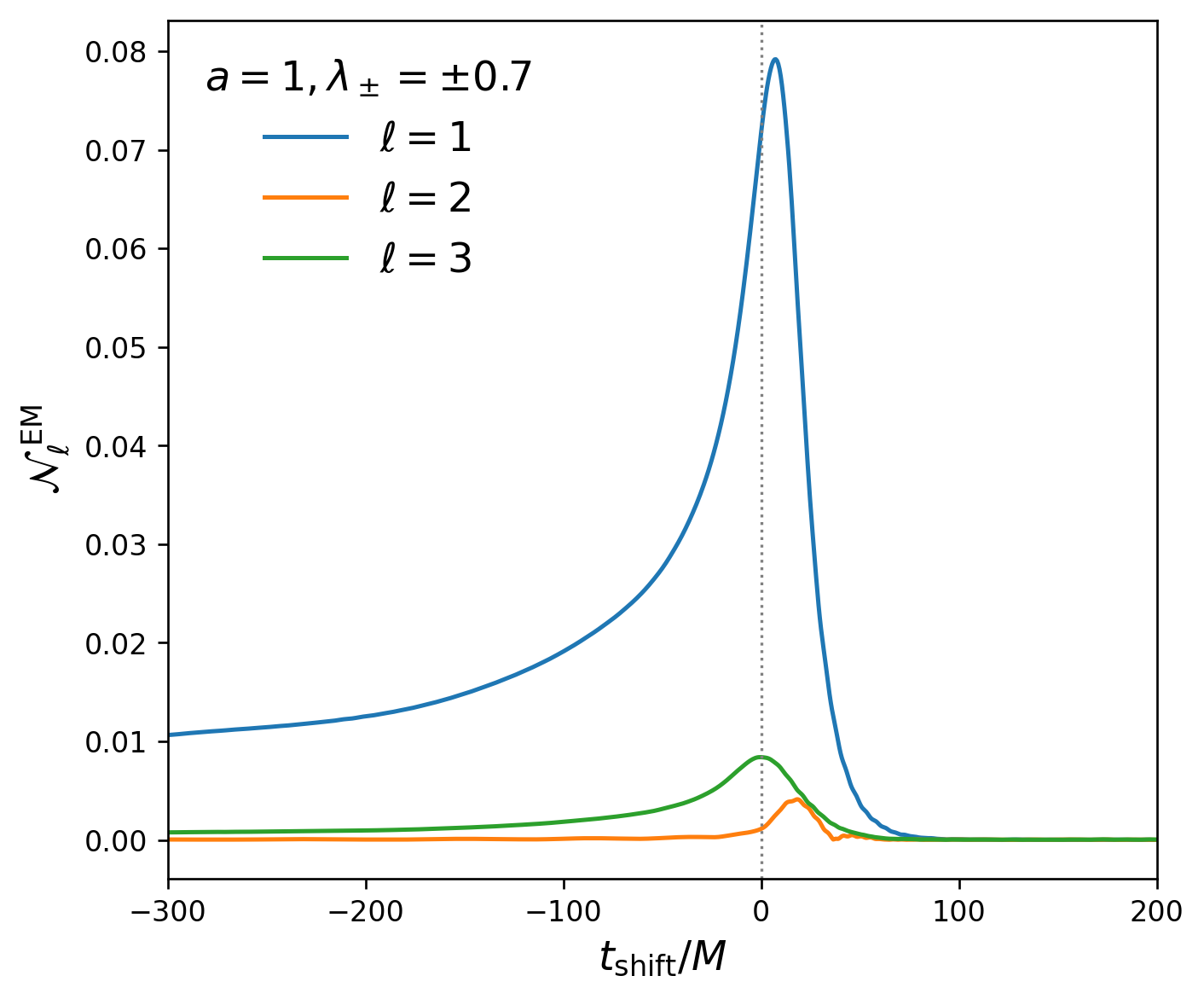}
  \caption{
  Electromagnetic Newman--Penrose multipole norms $\mathcal{N}_{\ell}^{\rm EM}$, defined in Eq.~\eqref{eq:em_np_multipole_norm}, for the highly charged case with $a=1$ and $\lambda_{\pm}=\pm0.7$. 
    The $\ell=1$, $2$, and $3$ sectors are shown. 
    The horizontal axis is the shifted extraction time $t_{\rm shift}$ defined in Eq.~\eqref{eq:radiative_time_shift}.
    The vertical dotted line marks $t_{\rm shift}=0$, corresponding to the principal peak of $\left|r_{\rm ext}h_{22}\right|$.
  }
  \label{fig:rPhi2_modes_a1_charge07}
\end{figure*}

We resolve the electromagnetic radiation into the multipolar sectors defined by $\mathcal{N}_{\ell}^{\rm EM}$ in Eq.~\eqref{eq:em_np_multipole_norm}. 
Because $\Phi_2$ has spin weight $-1$, the radiative decomposition begins at $\ell=1$. 
The corresponding sector fluxes are given by $\mathcal{P}_{\ell}^{\rm EM}$ in Eq.~\eqref{eq:em_np_multipole_flux}. 
Figure~\ref{fig:rPhi2_modes_a1_charge07} compares the $\ell=1$, $2$, and $3$ sectors for the highly charged configuration. 
The electromagnetic radiation is strongly dominated by the $\ell=1$ sector. 
Its flux reaches its maximum at $t_{\rm shift}/M\simeq7.41$, coinciding with the sampled peak time of the total electromagnetic flux in Fig.~\ref{fig:dEdt_EM_a1}. 
At this peak, the $\ell=1$ sector accounts for approximately $98.7\%$ of the total electromagnetic energy flux, quantitatively establishing the predominantly dipolar character of the signal. 

This dipole dominance is expected because the orbital motion of two black holes carrying equal-magnitude opposite charges generates a time-dependent electric dipole moment. 
For an approximately equal-mass binary, the corresponding charge distribution is approximately antisymmetric under spatial inversion. 
In the idealized symmetric limit, this antisymmetry eliminates the even-$\ell$ electric-type multipole moments while allowing the odd-$\ell$ moments. 
Magnetic-type multipoles with even $\ell$ are nevertheless allowed, so the full $\ell=2$ radiative sector need not vanish. 
The displayed hierarchy, with a strongly suppressed $\ell=2$ sector and a larger but still subdominant $\ell=3$ sector, is consistent with this electric-type selection rule. 

\begin{figure}[!htbp]
    \centering
    \begin{subfigure}[b]{0.48\columnwidth}
        \centering
        \includegraphics[width=\linewidth]{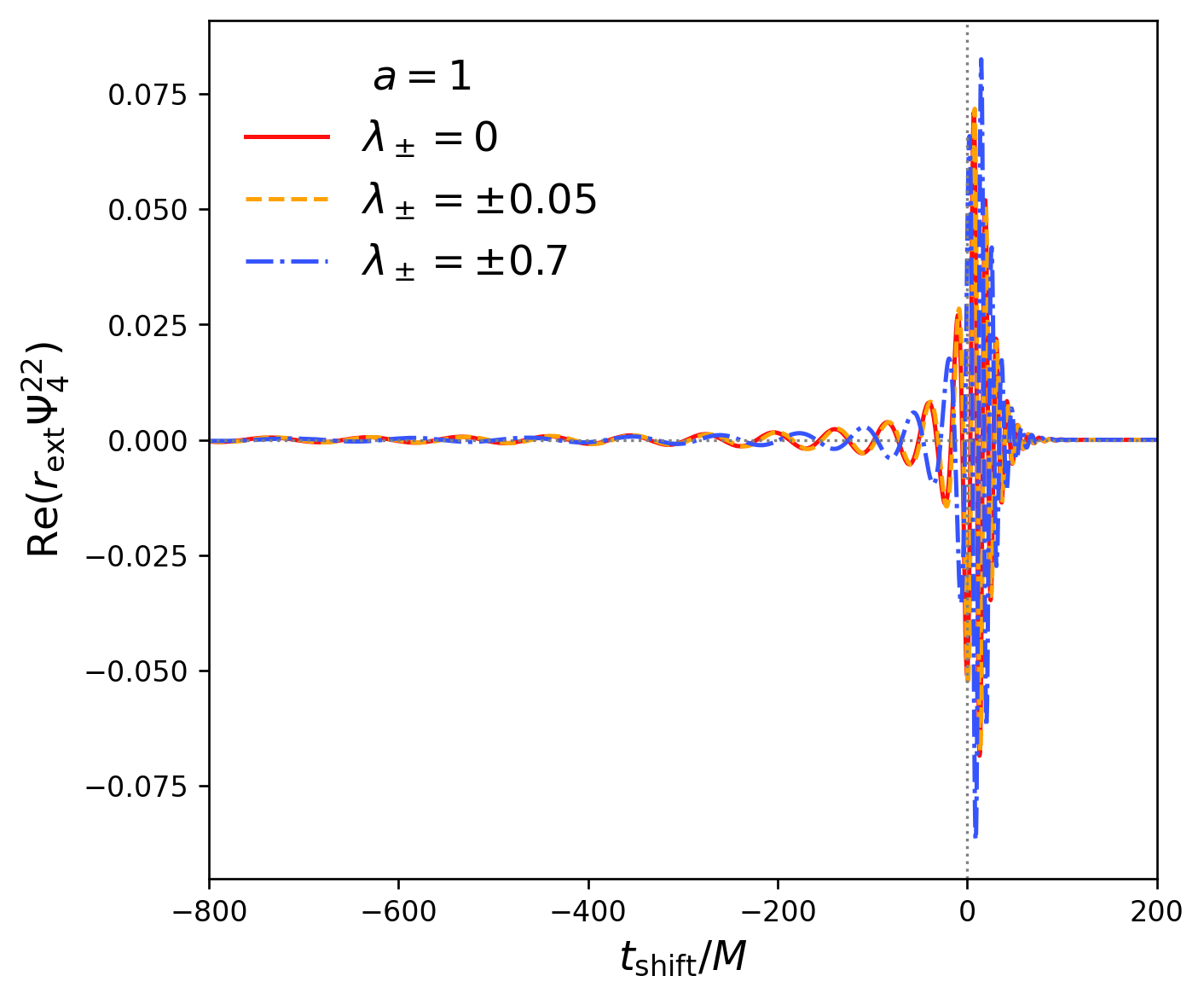}
        \caption{\centering ${\rm Re}(r_{\rm ext}\Psi_4^{22})$}
        \label{fig:rPsi4_a1_main}
    \end{subfigure}
    \hfill
    \begin{subfigure}[b]{0.48\columnwidth}
        \centering
        \includegraphics[width=\linewidth]{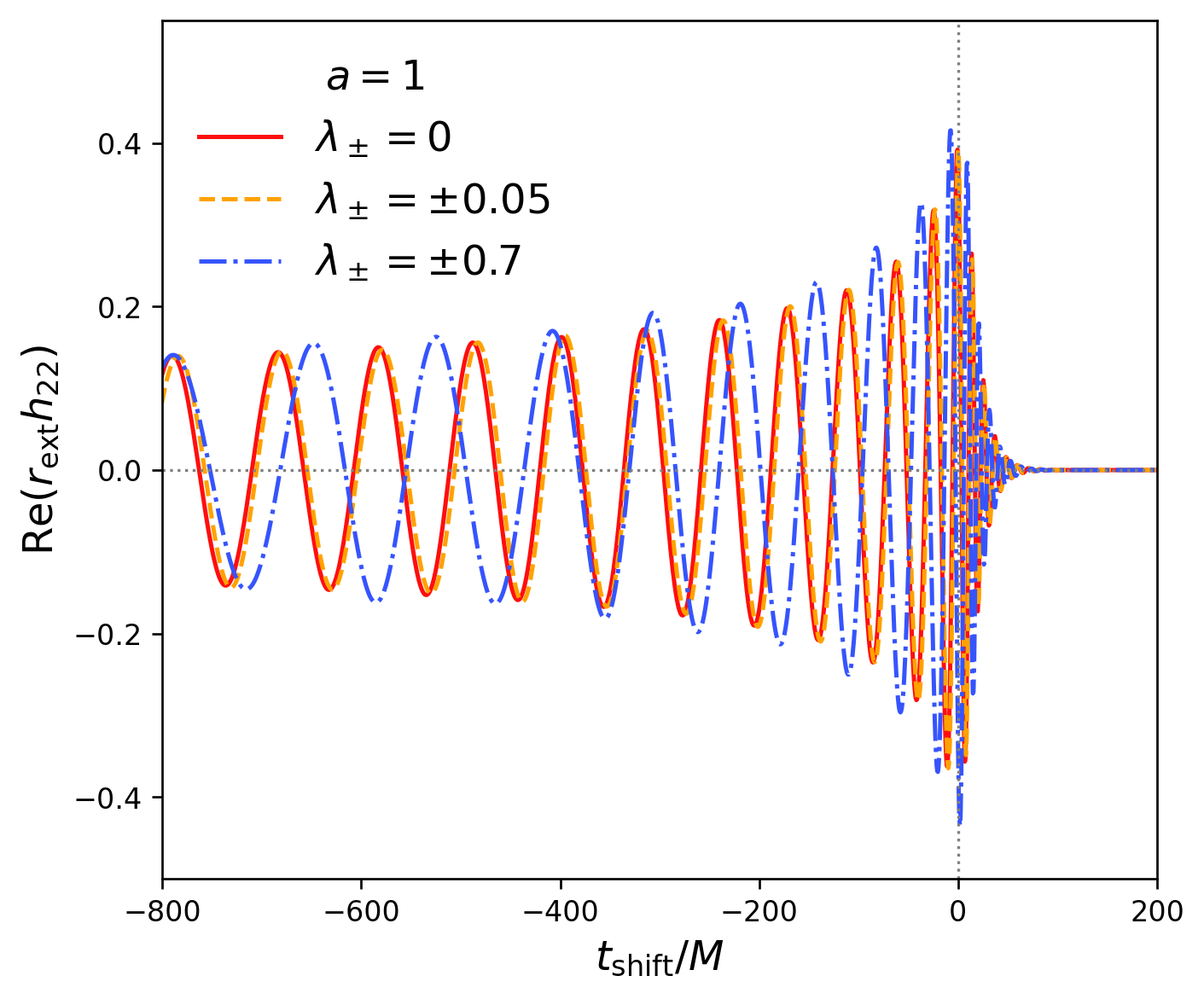}
        \caption{\centering ${\rm Re}(r_{\rm ext} h_{22})$}
        \label{fig:rh_a1_main}
    \end{subfigure}
    \caption{
    Dominant gravitational-wave modes for the $a=1$ cases with $\lambda_{\pm}=0$, $\lambda_{\pm}=\pm0.05$, and $\lambda_{\pm}=\pm0.7$. 
    Panel (a) shows ${\rm Re}(r_{\rm ext}\Psi_4^{22})$, and panel (b) shows the corresponding strain ${\rm Re}(r_{\rm ext}h_{22})$. 
    The horizontal axis is the shifted extraction time $t_{\rm shift}$ defined in Eq.~\eqref{eq:radiative_time_shift}. 
    The vertical dotted line marks $t_{\rm shift}=0$, corresponding to the principal peak of $\left|r_{\rm ext}h_{22}\right|$.
    }
    \label{fig:gw_a1_main}
\end{figure}

Finally, we examine how the charge-dependent orbital and radiative dynamics are reflected in the dominant gravitational waveform. 
Figure~\ref{fig:gw_a1_main} compares ${\rm Re}(r_{\rm ext}\Psi_4^{22})$ and ${\rm Re}(r_{\rm ext}h_{22})$ for the three $a=1$ configurations. 
Both modes use the same time shift for each simulation, determined by the peak of the phase-independent strain amplitude $\left|r_{\rm ext}h_{22}\right|$. 
The comparison therefore highlights waveform structure around merger rather than differences in the time required to reach merger. 
With this strain-peak alignment, the neutral and weakly charged curves nearly overlap in ${\rm Re}(r_{\rm ext}\Psi_4^{22})$, while a small offset in their oscillation timing is visible in ${\rm Re}(r_{\rm ext}h_{22})$. 
The different visual sensitivity of these diagnostics is consistent with the stronger weighting of lower-frequency components in the reconstruction of the strain from $\Psi_4^{22}$. 

The limited departure of the weakly charged waveform from the neutral vacuum-GR baseline is qualitatively consistent with previous EMD binary simulations, which concluded that gravitational signals from the small-charge configurations considered there could be difficult to distinguish from GR binary signals~\cite{Hirschmann:2017psw}. 
By contrast, the highly charged waveform exhibits clearly visible differences from the neutral baseline in both ${\rm Re}(r_{\rm ext}\Psi_4^{22})$ and ${\rm Re}(r_{\rm ext}h_{22})$, particularly during the late inspiral and around merger. 
These differences are consistent with the substantially altered orbital evolution and the enhanced electromagnetic and scalar radiation found above. 

In summary, the charged $a=1$ binaries develop a quasi-stationary dilaton profile while their individual EMD charges remain close to their initial values during the inspiral. 
The weakly charged gravitational waveform remains close to the neutral vacuum-GR baseline. 
By contrast, the highly charged binary reaches the radiative merger after a coordinate-time interval from $D/M=10$ that is approximately $62\%$ shorter than for the neutral binary. 
The persistent opposite charges and enhanced electromagnetic and scalar radiation are consistent with a combination of stronger attractive interactions and additional radiative losses. 
In this highly charged case, the dipole sector carries approximately $99\%$ of the total electromagnetic flux at its peak. 
These dynamical and radiative changes are accompanied by clearly visible modifications of the dominant gravitational waveform. 
After merger, the common-horizon charge is consistent with zero, and the scalar profile decays as the remnant approaches an uncharged black hole without stationary secondary dilaton hair. 

\subsection{Dependence on the dilaton coupling in the high-charge case}
\label{subsec:Dependence_dilaton_coupling}

In order to isolate the dependence on the dilaton coupling at the largest input-charge magnitude considered in this work, we set 
\begin{equation}
\lambda_{\pm}=\pm0.7\,,
\end{equation}
and compare $a=0$, $1$, and $\sqrt{3}$. 
The $a=0$ configuration provides the charged Einstein--Maxwell reference, while $a=1$ and $a=\sqrt{3}$ correspond to the string-motivated and Kaluza--Klein couplings, respectively. 
Because the input charges and the charge-dependent initial-momentum prescription are common to the three simulations, this sequence provides a controlled comparison of how the dilaton coupling modifies the near-horizon fields, orbital evolution, and radiation within the adopted initial-data prescription. 

We first examine the coupling dependence of the quasi-local horizon quantities. 
As in Sec.~\ref{subsec:main_results_a1}, these quantities are displayed relative to the common-apparent-horizon first-detection time $t_{\rm CAH}$, whereas the radiative quantities considered below use the strain-peak-based time coordinate $t_{\rm shift}$. 
\begin{figure*}[!hbp]
    \centering
    \begin{subfigure}[b]{0.32\textwidth}
        \centering
        \includegraphics[width=\linewidth]{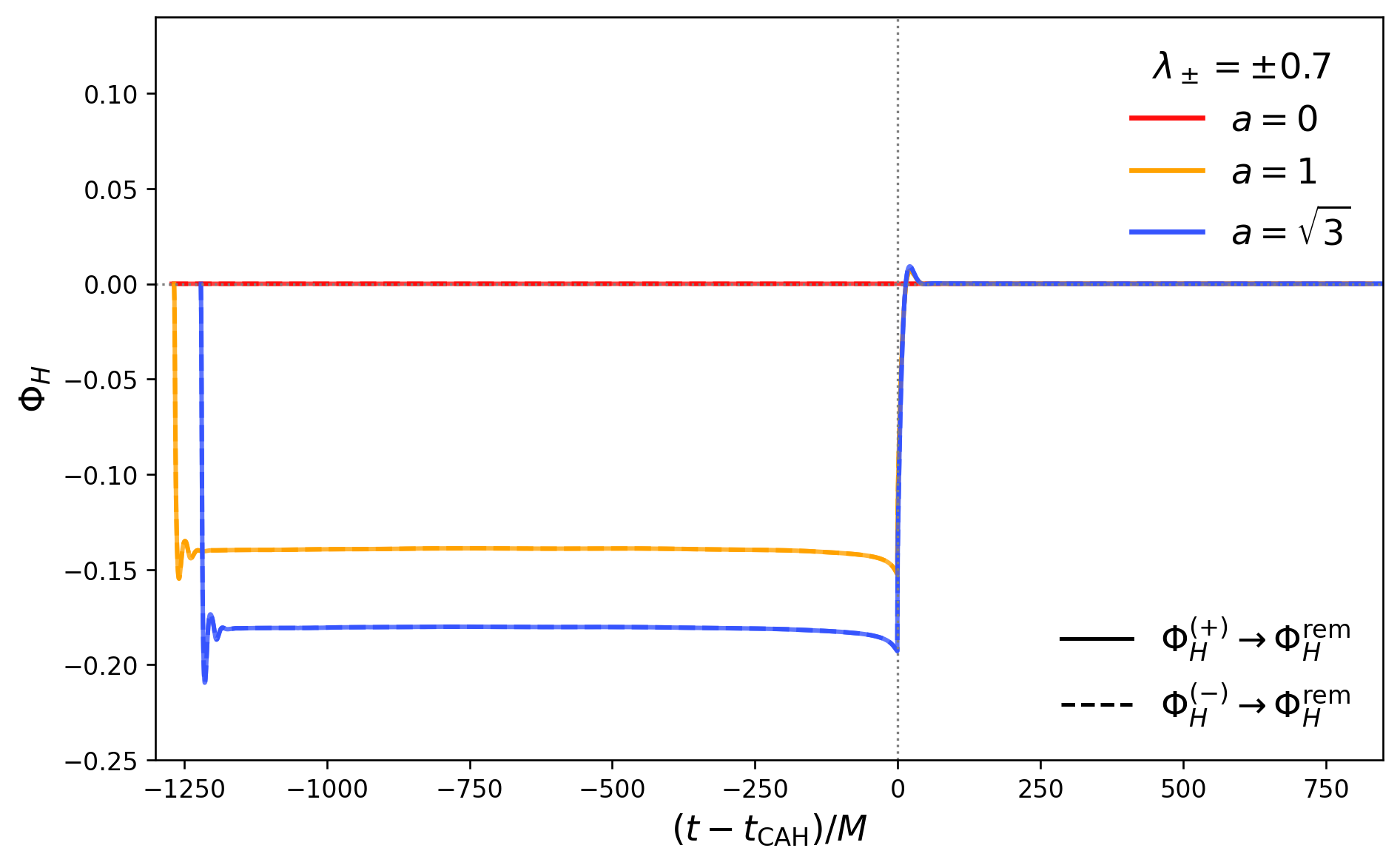}
        \caption{\centering $\Phi_{\rm H}$}
        \label{fig:coupling_PhiH}
    \end{subfigure}
    \hfill
    \begin{subfigure}[b]{0.32\textwidth}
        \centering
        \includegraphics[width=\linewidth]{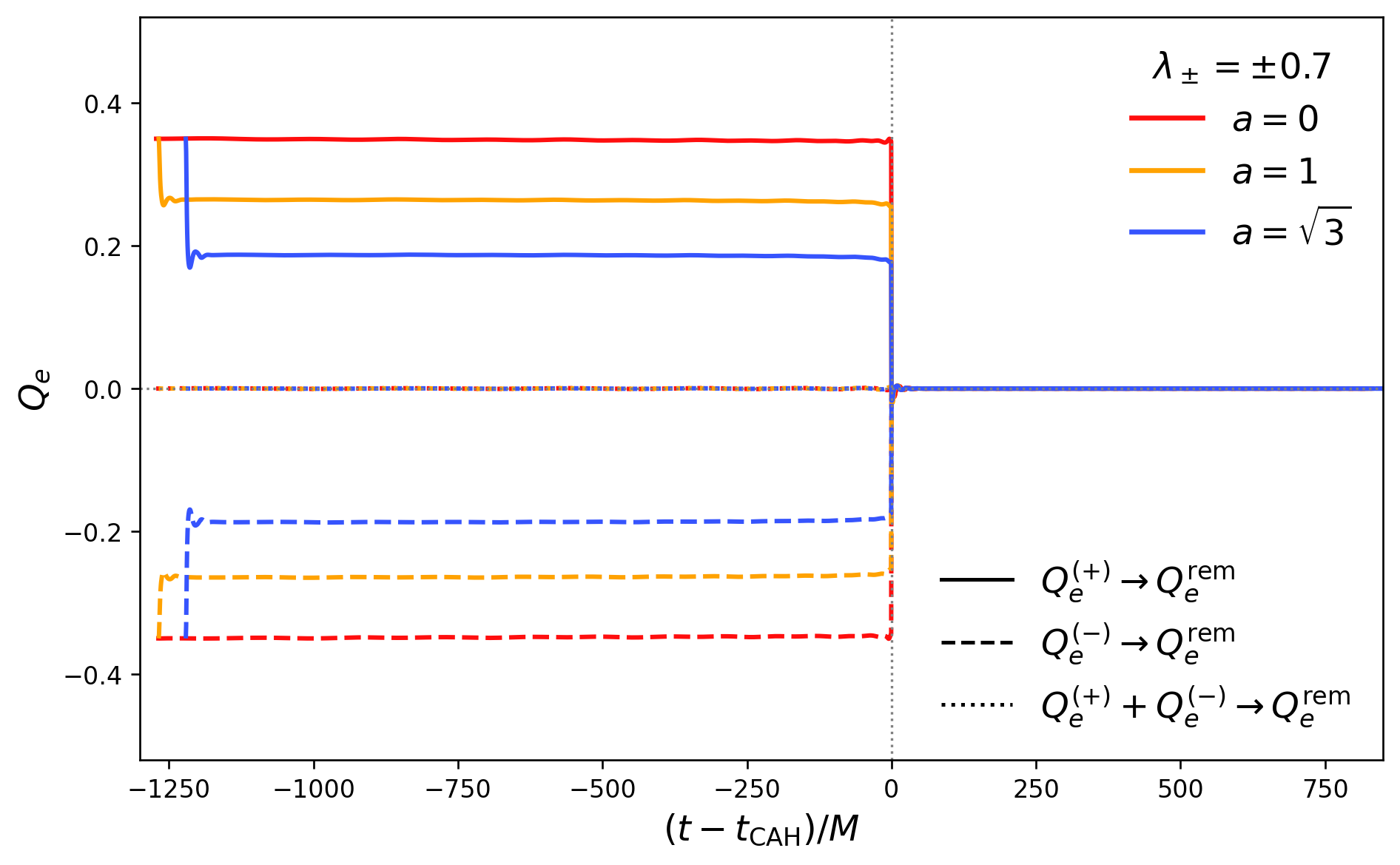}
        \caption{\centering $Q_e$}
        \label{fig:coupling_Qe}
    \end{subfigure}
    \hfill
    \begin{subfigure}[b]{0.32\textwidth}
        \centering
        \includegraphics[width=\linewidth]{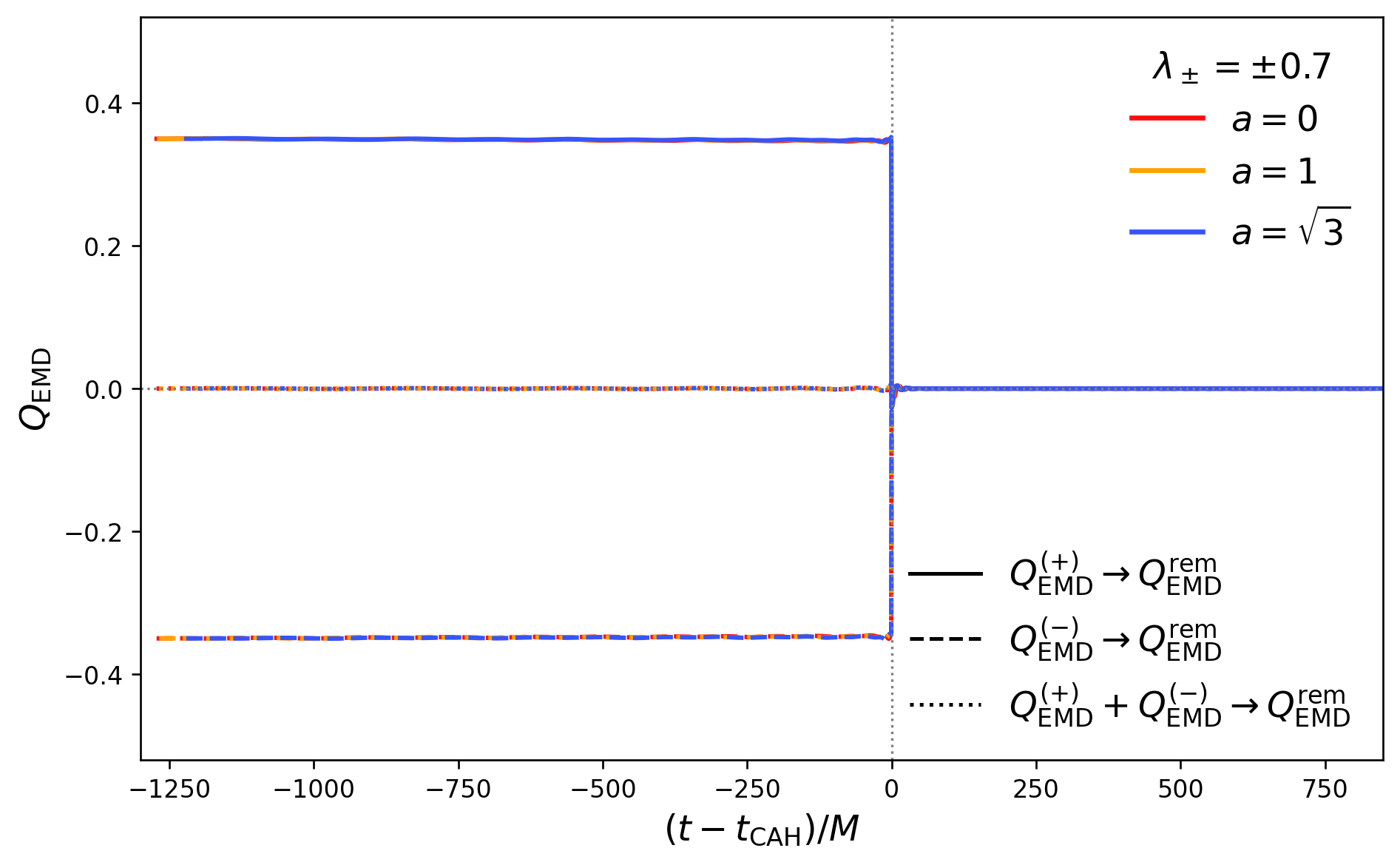}
        \caption{\centering $Q_{\rm EMD}$}
        \label{fig:coupling_QEMD}
    \end{subfigure}
    \caption{
            Evolution of the quasi-local horizon quantities for $a=0$, $1$, and $\sqrt{3}$ at fixed $\lambda_{\pm}=\pm0.7$. 
            Panels (a), (b), and (c) show $\Phi_{\rm H}$, $Q_e$, and $Q_{\rm EMD}$, respectively. 
            The horizontal axis shows $(t-t_{\rm CAH})/M$. 
            Before $t_{\rm CAH}$, the solid and dashed curves denote the individual horizons associated with the positive and negative charges, respectively. 
            At and after $t_{\rm CAH}$, the displayed quantities refer to the common horizon. 
            In panels (b) and (c), the dotted curves show the net horizon charge. 
            The vertical dotted line marks $t=t_{\rm CAH}$. 
    }
    \label{fig:coupling_horizon_quantities}
\end{figure*}
Figure~\ref{fig:coupling_horizon_quantities} shows that $\Phi_{\rm H}$ remains zero for $a=0$, as expected when the dilaton is decoupled and initially vanishes. 
For $a=1$ and $a=\sqrt{3}$, the electromagnetic field dynamically sources a negative scalar profile whose magnitude increases with $a$. 
Representative values of the horizon quantities are summarized in Table~\ref{tab:coupling_horizon_representatives}. 
\begin{table}[h]
\centering
\setlength{\tabcolsep}{8pt}
\begin{tabular}{c|ccc}
$a$ &
$\langle\Phi_{\rm H}\rangle$ &
$\langle|Q_e|\rangle$ &
$\langle|Q_{\rm EMD}|\rangle$ \\
\hline
$0$        & $0$       & $0.3481$ & $0.3481$ \\
$1$        & $-0.1392$ & $0.2637$ & $0.3483$ \\
$\sqrt{3}$ & $-0.1803$ & $0.1867$ & $0.3486$ \\
\end{tabular}
\caption{
Representative horizon quantities for the high-charge sequence. 
Each value is first averaged over $-1100\leq(t-t_{\rm CAH})/M\leq-200$ and then over the two individual horizons. 
}
\label{tab:coupling_horizon_representatives}
\end{table}
The representative values of $\langle|Q_{\rm EMD}|\rangle$ vary by less than approximately $0.2\%$ across the three simulations. 
The corresponding time-averaged net EMD charges lie between approximately $-4.5\times10^{-5}$ and $-3.8\times10^{-5}$, corresponding to only ${\cal O}(10^{-4})$ of the individual charge magnitude. 
Thus, the equal-magnitude opposite charges cancel to high accuracy for all three coupling values. 
By contrast, 
$\langle|Q_e|\rangle$ decreases from $0.3481$ at $a=0$ to $0.2637$ at $a=1$ and $0.1867$ at $a=\sqrt{3}$ as the magnitude of the negative scalar profile increases. 

This behavior follows directly from the dilaton-weighted Gauss law. 
If the scalar field is approximately uniform over an individual horizon, Eqs.~\eqref{eq:Qe_def} and~\eqref{eq:QEMD_def} imply 
\begin{equation}
Q_e\simeq e^{2a\Phi_{\rm H}}Q_{\rm EMD}\,.
\label{eq:horizon_charge_dressing}
\end{equation}
Using the unrounded representative values underlying Table~\ref{tab:coupling_horizon_representatives}, we obtain $\langle|Q_e|\rangle/\langle|Q_{\rm EMD}|\rangle\simeq0.7570$ and $0.5354$ for $a=1$ and $a=\sqrt{3}$, respectively. 
These ratios agree with the corresponding values of $e^{2a\langle\Phi_{\rm H}\rangle}$ to the quoted precision. 
The decrease in $|Q_e|$ therefore does not represent a loss of the conserved EMD charge. 
Instead, it quantitatively reflects the increasing dilaton weighting of the electric flux as $a$ increases. 

After common-horizon formation, $Q_e$ and $Q_{\rm EMD}$ remain consistent with zero for all three couplings, while the scalar transients in the nonzero-coupling cases decay. 
Increasing $a$ therefore strengthens the inspiral scalar dressing without changing the expected neutral, hairless end state discussed in Sec.~\ref{subsec:main_results_a1}. 

The coupling dependence of the late orbital evolution is quantified using the interval from $D=10$ to the radiative merger, following the prescription introduced in Sec.~\ref{subsec:main_results_a1}. 
The resulting intervals are 
\begin{equation}
  \frac{t_{\rm merger}^{\rm rad}-t_{D=10}}{M}
  \simeq
  \begin{cases}
     316.42 & (a=0)\,,\\
     336.74 & (a=1)\,,\\
     350.29 & (a=\sqrt{3})\,.
  \end{cases}
  \label{eq:merger_time_a_D10}
\end{equation}
Across the three sampled couplings, the interval increases monotonically from the Einstein--Maxwell case to the two EMD cases. 
Relative to $a=0$, it is approximately $6.4\%$ longer for $a=1$ and $10.7\%$ longer for $a=\sqrt{3}$. 
The coupling dependence is therefore considerably weaker than the charge dependence found in Sec.~\ref{subsec:main_results_a1}, where increasing the charge magnitude from the neutral case to $|\lambda_{\pm}|=0.7$ reduces the corresponding interval by approximately $62\%$. 

Because the individual values of $|Q_{\rm EMD}|$ vary by less than approximately $0.2\%$ across the coupling sequence, the increase in the merger-time interval cannot be attributed to a reduction of the conserved EMD charge. 
Instead, Fig.~\ref{fig:coupling_horizon_quantities} shows that increasing $a$ strengthens the dilaton dressing of the individual black holes, producing a larger $|\Phi_{\rm H}|$ and a smaller unweighted charge $|Q_e|$ at nearly fixed $|Q_{\rm EMD}|$. 
These coupling-dependent changes in the near-horizon fields can modify both the conservative binary interaction and the radiative energy-loss channels. 
The waveform and flux comparisons below examine how these changes accompany the observed ordering of the merger-time intervals. 

\begin{figure}[!htbp]
    \centering
    \begin{subfigure}[b]{0.48\columnwidth}
        \centering
        \includegraphics[width=\linewidth]
        {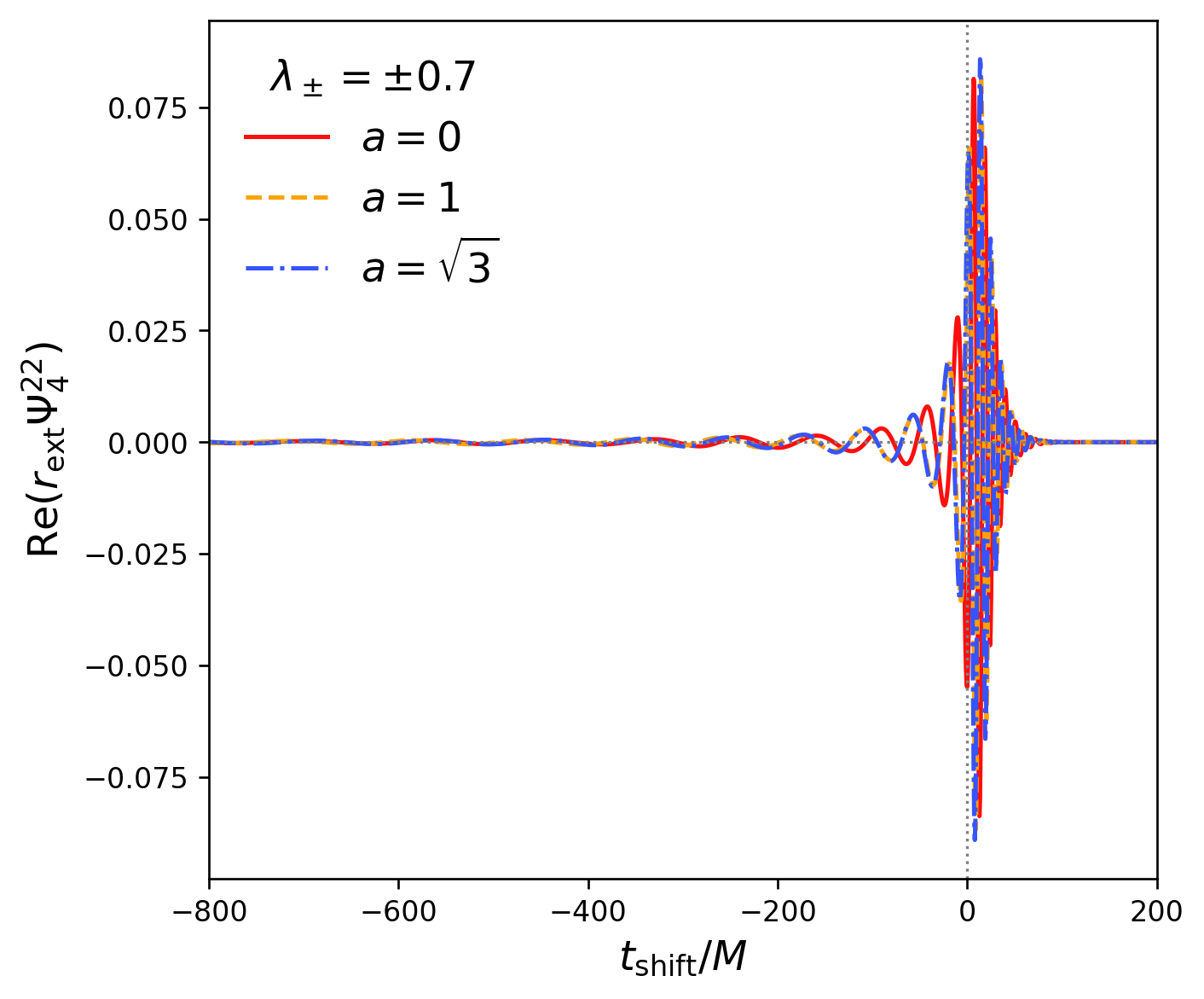}
        \caption{\centering ${\rm Re}(r_{\rm ext}\Psi_4^{22})$}
        \label{fig:rPsi4_dilaton_high_charge}
    \end{subfigure}
    \hfill
    \begin{subfigure}[b]{0.48\columnwidth}
        \centering
        \includegraphics[width=\linewidth]
        {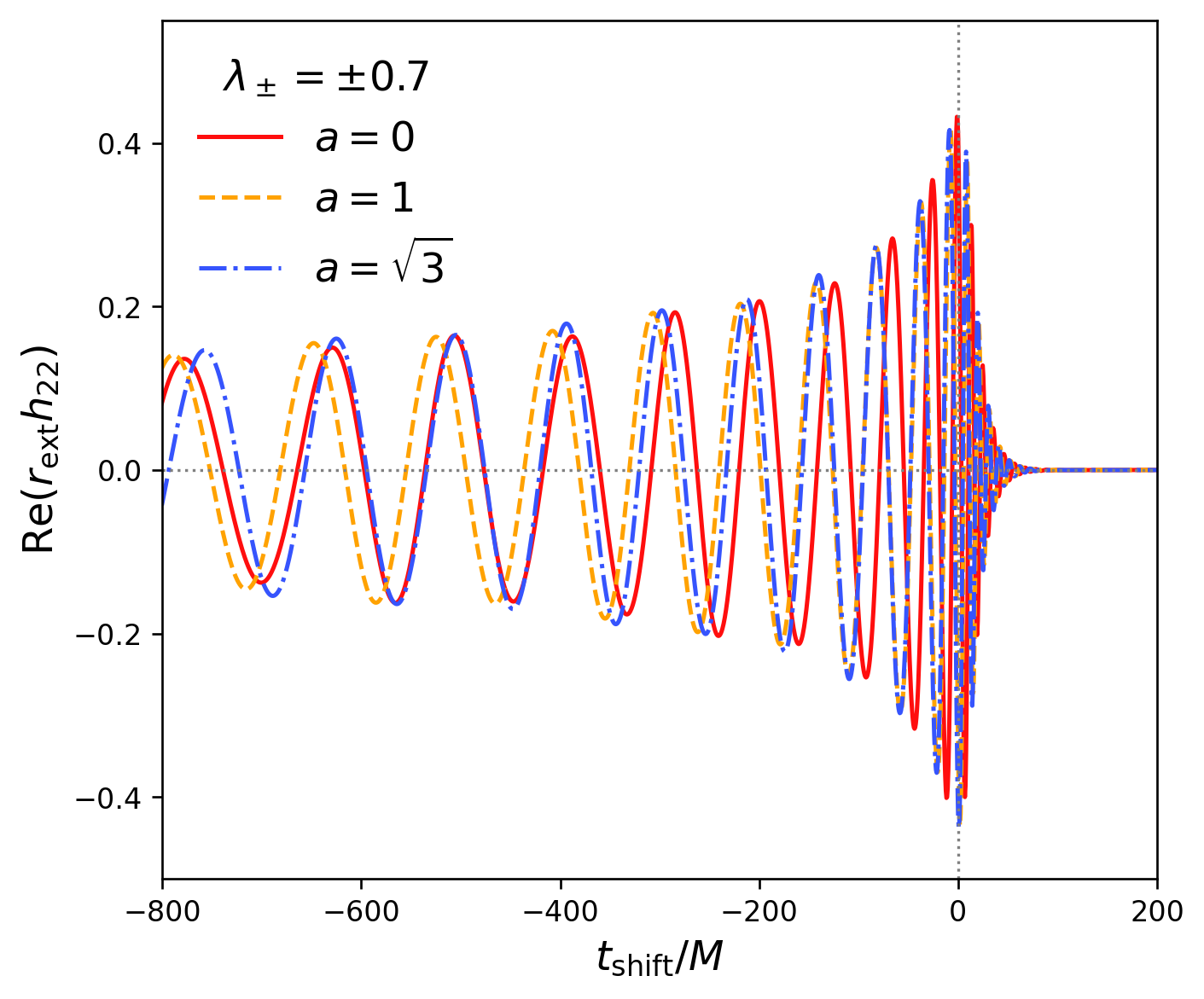}
        \caption{\centering ${\rm Re}(r_{\rm ext}h_{22})$}
        \label{fig:rh_dilaton_high_charge}
    \end{subfigure}
    \caption{
    Dominant gravitational-wave modes for $a=0$, $1$, and $\sqrt{3}$ at fixed $\lambda_{\pm}=\pm0.7$. 
    Panels (a) and (b) show ${\rm Re}(r_{\rm ext}\Psi_4^{22})$ and ${\rm Re}(r_{\rm ext}h_{22})$, respectively. 
    The horizontal axis is the shifted extraction time $t_{\rm shift}$ defined in Eq.~\eqref{eq:radiative_time_shift}. 
    The vertical dotted line marks $t_{\rm shift}=0$, corresponding to the principal peak of $\left|r_{\rm ext}h_{22}\right|$. 
    No additional phase alignment is applied.
    }
    \label{fig:gw_dilaton_high_charge}
\end{figure}

\begin{figure*}[!htbp]
    \centering
    \begin{subfigure}[b]{0.32\textwidth}
        \centering
        \includegraphics[width=\linewidth]
        {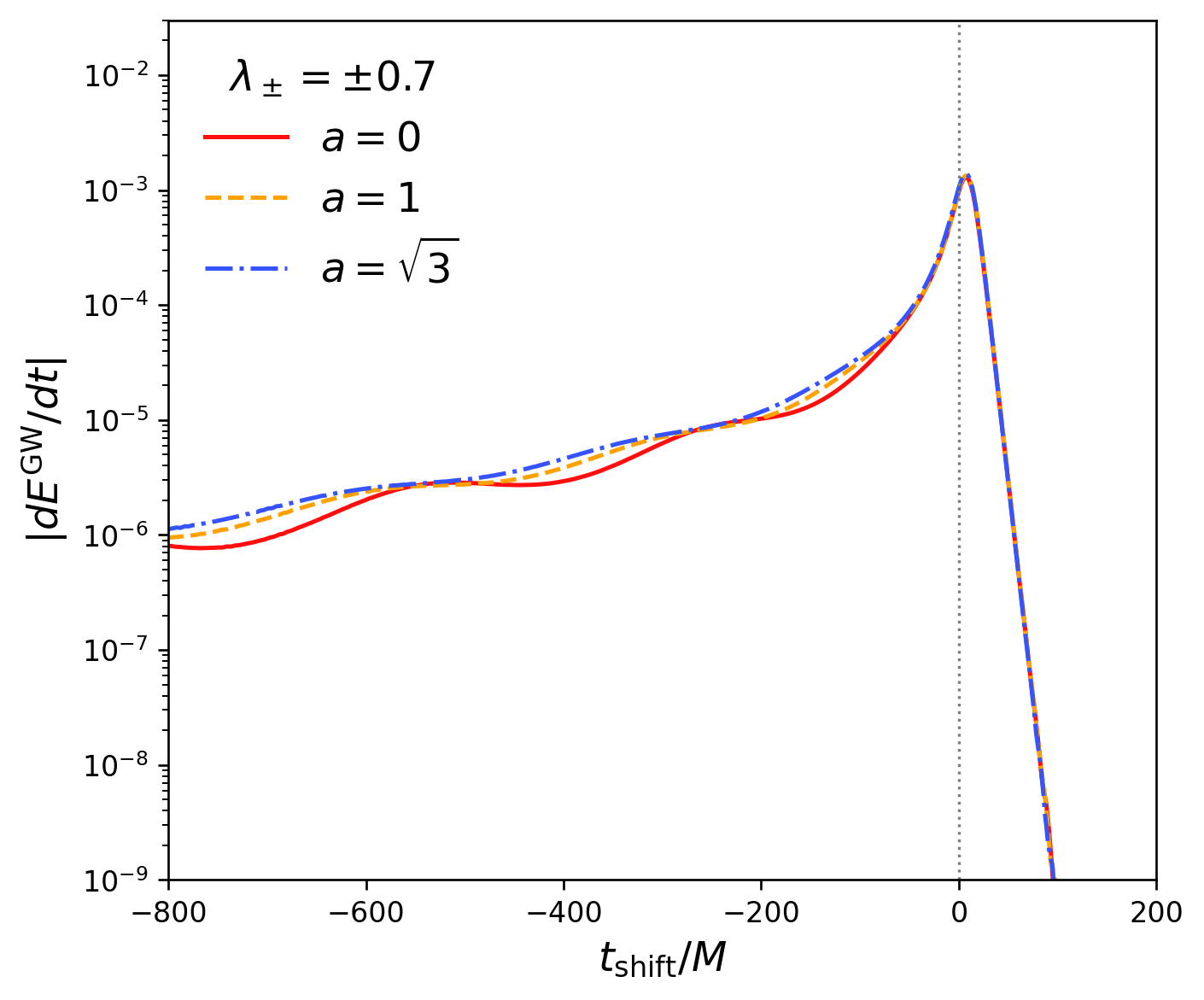}
        \caption{\centering $\left|dE^{\rm GW}/dt\right|$}
        \label{fig:dEdt_GW_dilaton_high_charge}
    \end{subfigure}
    \hfill
    \begin{subfigure}[b]{0.32\textwidth}
        \centering
        \includegraphics[width=\linewidth]
        {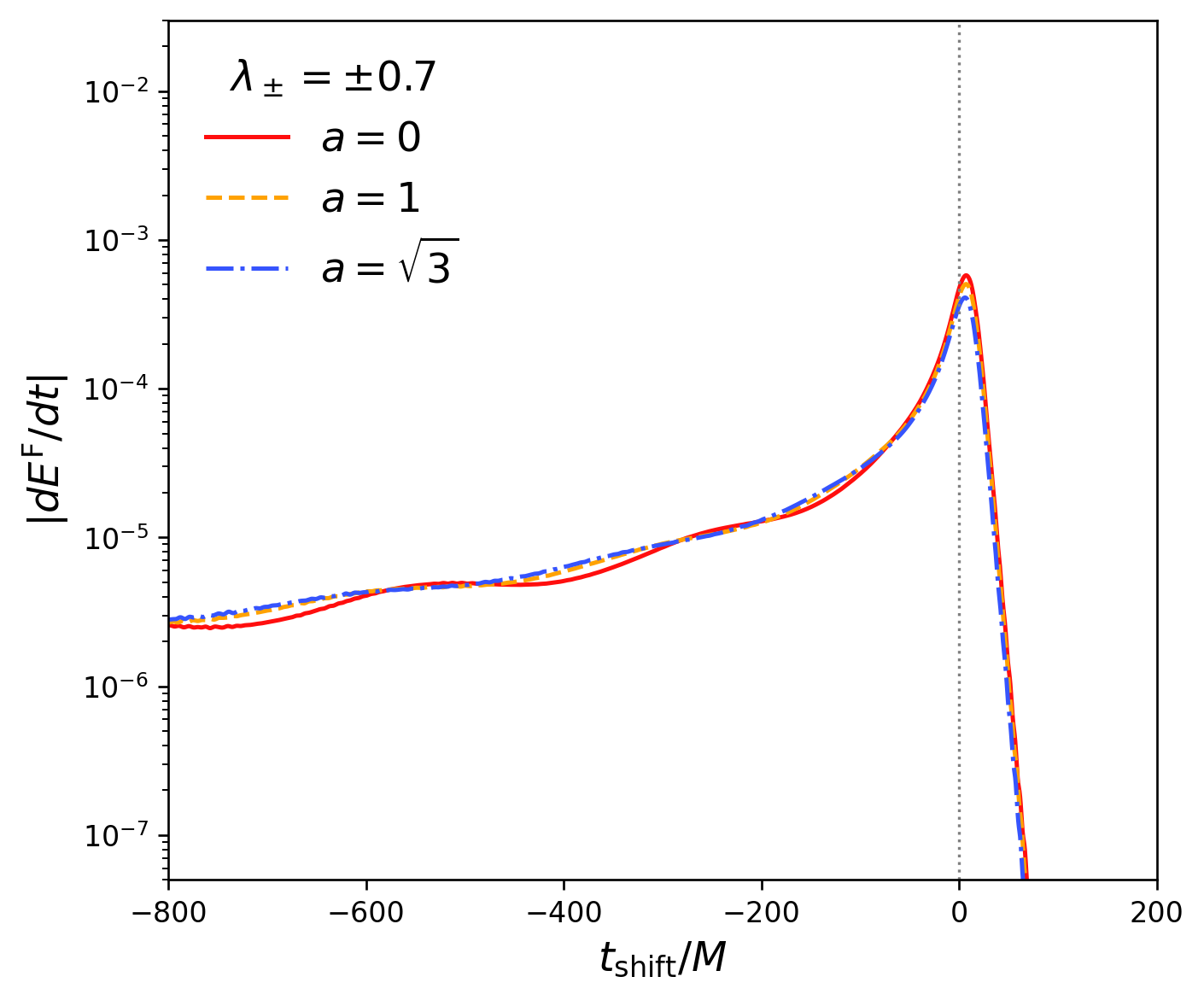}
        \caption{\centering $\left|dE^{\rm F}/dt\right|$}
        \label{fig:dEdt_EM_dilaton_high_charge}
    \end{subfigure}
    \hfill
    \begin{subfigure}[b]{0.32\textwidth}
        \centering
        \includegraphics[width=\linewidth]
        {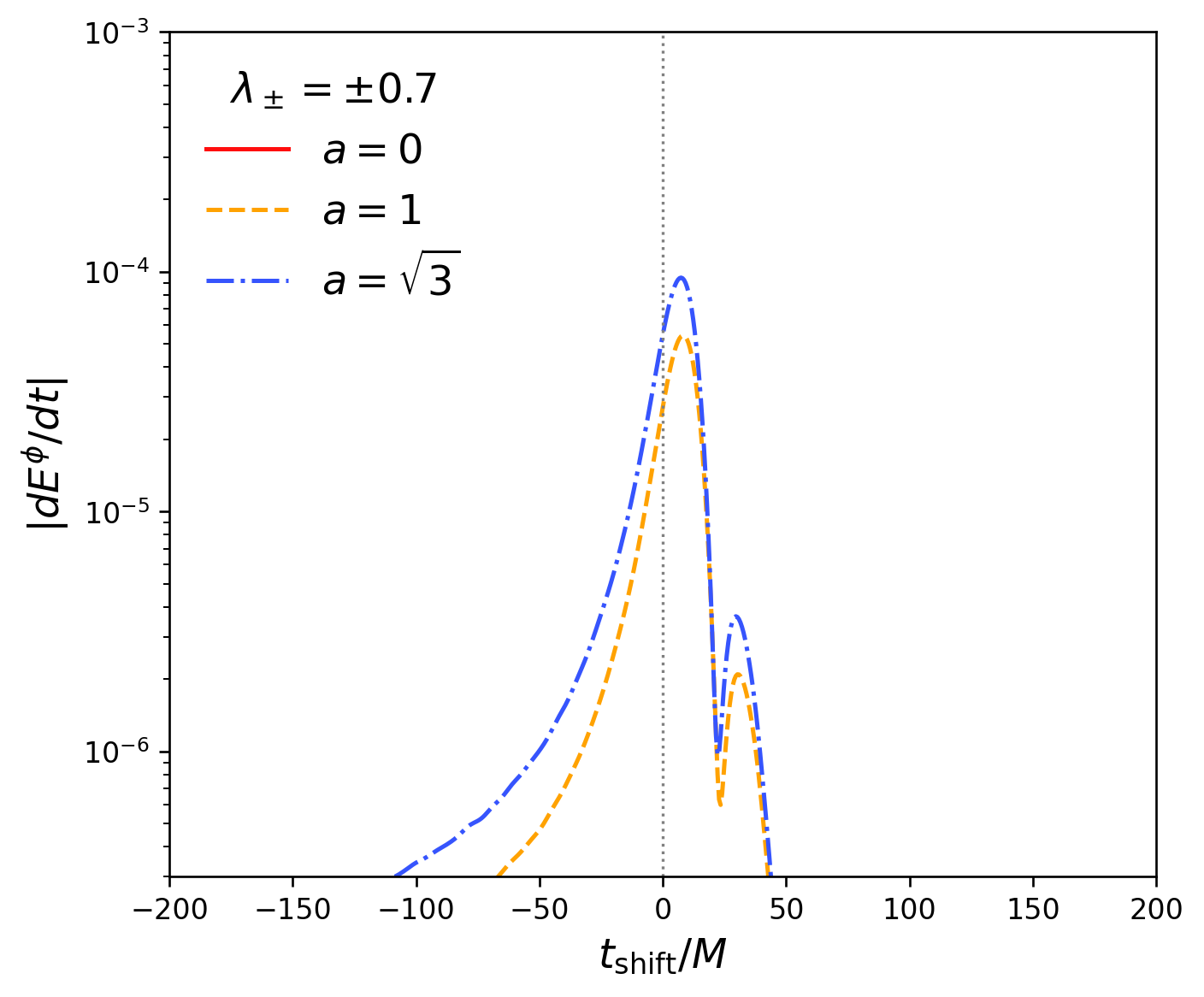}
        \caption{\centering $\left|dE^{\phi}/dt\right|$}
        \label{fig:dEdt_phi_dilaton_high_charge}
    \end{subfigure}
    \caption{
    Magnitudes of the radiative energy fluxes for $a=0$, $1$, and $\sqrt{3}$ at fixed $\lambda_{\pm}=\pm0.7$. 
    Panels (a), (b), and (c) show the gravitational, electromagnetic, and scalar energy fluxes, respectively. 
    The gravitational and electromagnetic fluxes are shown for all three couplings. 
    In panel (c), the scalar flux vanishes for $a=0$ and therefore does not appear on the logarithmic scale. 
    All panels use the shifted extraction time $t_{\rm shift}$ defined in Eq.~\eqref{eq:radiative_time_shift}. 
    The vertical dotted line marks $t_{\rm shift}=0$, corresponding to the principal peak of $\left|r_{\rm ext}h_{22}\right|$. 
    Note that panel (c) uses a different horizontal range.
    }
    \label{fig:energy_flux_dilaton_high_charge}
\end{figure*}

Figure~\ref{fig:gw_dilaton_high_charge} compares ${\rm Re}(r_{\rm ext}\Psi_4^{22})$ and ${\rm Re}(r_{\rm ext}h_{22})$ for $a=0$, $1$, and $\sqrt{3}$ at fixed $\lambda_{\pm}=\pm0.7$. 
Both modes use the same time shift for each simulation, determined by the peak of the phase-independent strain amplitude $\left|r_{\rm ext}h_{22}\right|$. 
The comparison emphasizes waveform structure around merger, while the merger-time intervals are quantified separately in Eq.~\eqref{eq:merger_time_a_D10}. 
The waveform phases are not independently aligned. 
With this strain-peak alignment, the three waveforms have similar overall shapes and comparable amplitudes around merger. 
Both diagnostics show visible offsets in oscillation timing, which are more apparent in the strain. 
This difference in visual sensitivity is consistent with the stronger weighting of lower-frequency components in the reconstruction of the strain from $\Psi_4^{22}$. 
Figure~\ref{fig:energy_flux_dilaton_high_charge} compares the magnitudes of the gravitational, electromagnetic, and scalar energy fluxes for the same high-charge configurations. 
The gravitational fluxes have comparable peak magnitudes and similar near-merger profiles for the three sampled couplings, consistent with the limited waveform differences in Fig.~\ref{fig:gw_dilaton_high_charge}. 
By contrast, the electromagnetic-flux peak decreases systematically from approximately $5.79\times10^{-4}$ at $a=0$ to $5.05\times10^{-4}$ at $a=1$ and $4.10\times10^{-4}$ at $a=\sqrt{3}$.
Relative to the Einstein--Maxwell reference, these values correspond to reductions of approximately $12.6\%$ and $29.1\%$ for $a=1$ and $a=\sqrt{3}$, respectively. 
The scalar channel exhibits the opposite trend. 
Its flux vanishes for $a=0$, while its peak reaches approximately $5.42\times10^{-5}$ for $a=1$ and $9.45\times10^{-5}$ for $a=\sqrt{3}$.
Between the two nonzero couplings, the scalar-flux peak therefore increases by a factor of approximately $1.74$. 

\begin{figure}[!htbp]
  \centering
  \includegraphics[width=0.48\textwidth]
  {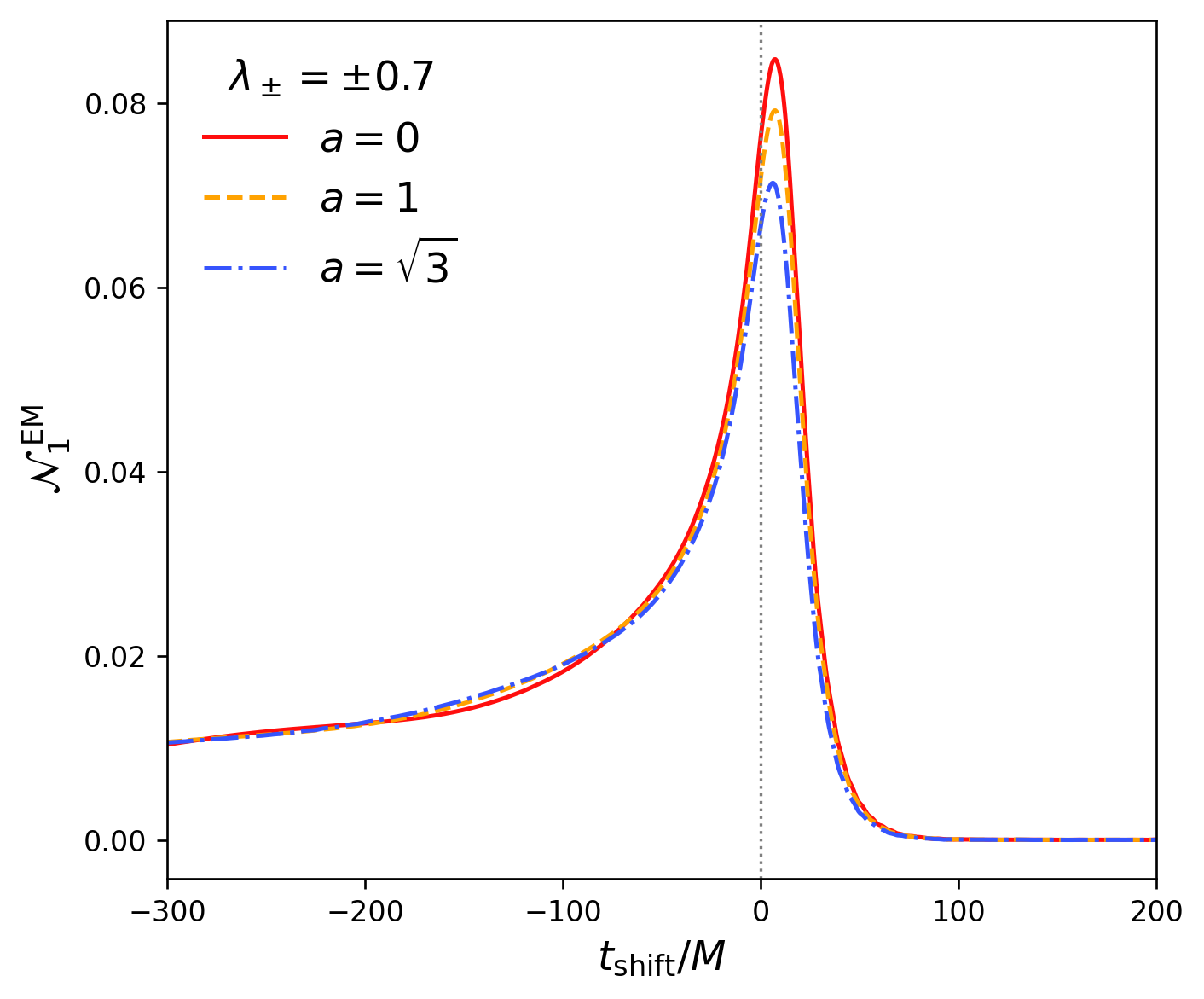}
  \caption{
  Electromagnetic dipole-sector norm $\mathcal{N}_1^{\rm EM}$, defined in Eq.~\eqref{eq:em_np_multipole_norm}, for $a=0$, $1$, and $\sqrt{3}$ at fixed $\lambda_{\pm}=\pm0.7$. 
  The horizontal axis is the shifted extraction time $t_{\rm shift}$ defined in Eq.~\eqref{eq:radiative_time_shift}. 
  The vertical dotted line marks $t_{\rm shift}=0$, corresponding to the principal peak of $\left|r_{\rm ext}h_{22}\right|$. 
  }
  \label{fig:coupling_em_dipole}
\end{figure}

As $a$ increases at nearly fixed $|Q_{\rm EMD}|$, the enhanced scalar radiation accompanies stronger scalar dressing, whereas the smaller electromagnetic-flux peak accompanies the decrease in the unweighted horizon charge $|Q_e|$ shown in Fig.~\ref{fig:coupling_horizon_quantities}. 
Figure~\ref{fig:coupling_em_dipole} compares the dipole-sector norm $\mathcal{N}_1^{\rm EM}$ defined in Eq.~\eqref{eq:em_np_multipole_norm} for the three coupling values. 

The maximum of $\mathcal{N}_1^{\rm EM}$ decreases from approximately $0.0847$ at $a=0$ to $0.0792$ at $a=1$ and $0.0713$ at $a=\sqrt{3}$.
Because the dipole-sector flux is $(4\pi)^{-1}(\mathcal{N}_1^{\rm EM})^2$, these values correspond to peak-flux reductions of approximately $12.7\%$ and $29.2\%$ relative to the Einstein--Maxwell reference. 
For each coupling, the dipole-sector and total electromagnetic fluxes reach their sampled maxima at the same time, and the dipole accounts for approximately $99\%$ of the total peak. 
The nearly unchanged dipole fraction therefore shows that the reduction of the total electromagnetic-flux peak is governed primarily by a weaker dipole rather than by a substantial shift of power toward higher multipoles. 

The decrease in the dipolar electromagnetic-flux peak accompanies the increase in the interval from $D=10$ to the radiative merger, suggesting that reduced electromagnetic dissipation can contribute to the longer evolution at larger $a$. 
The larger scalar-flux peak points to an opposing dissipative tendency near merger. 
The observed merger-time ordering is therefore consistent with the combined influence of these competing radiative responses and coupling-dependent changes in the conservative binary interaction. 

In summary, increasing $a$ across the sampled values at fixed $\lambda_{\pm}=\pm0.7$ strengthens the scalar dressing while leaving the individual quasi-local EMD charges nearly unchanged. 
The increasingly negative near-horizon scalar field corresponds to a smaller unweighted charge $|Q_e|$ through the dilaton-weighted Gauss law. 
The electromagnetic-flux peak decreases predominantly through a weaker dipole, whereas the scalar-flux peak increases. 
The interval from $D=10$ to the radiative merger grows with $a$ and is approximately $10.7\%$ longer for $a=\sqrt{3}$ than for the Einstein--Maxwell reference. 
After strain-peak alignment, the dominant gravitational waveforms exhibit visible differences in oscillation timing, while retaining similar overall shapes and comparable merger amplitudes. 
For all three couplings, the remnant approaches a neutral black hole without stationary secondary dilaton hair. 

\subsection{Dependence on the dilaton coupling for like-charged binaries}
\label{subsec:like_charged}

The nonzero-charge configurations considered in Secs.~\ref{subsec:main_results_a1} and~\ref{subsec:Dependence_dilaton_coupling} carry charges of equal magnitude and opposite sign and are therefore globally neutral. 
To examine the coupling dependence when the total charge does not cancel, we now consider a like-charged sequence with 
\begin{equation}
\lambda_+=\lambda_-=+0.3\,.
\end{equation}
Each puncture carries the input charge $Q_\pm^{\rm input}=0.3m_\pm$, corresponding to $Q_\pm^{\rm input}/M\simeq0.145775$ and a total input charge $Q_{\rm tot}^{\rm input}/M\simeq0.291550$.
We compare the Einstein--Maxwell case $a=0$ with the EMD cases $a=1$ and $a=\sqrt{3}$. 
Unlike the globally neutral sequences, the common horizon can inherit a nonzero EMD charge and, for $a\neq0$, support charge-induced secondary dilaton hair. 
Because the input charge magnitude differs from $|\lambda_{\pm}|=0.7$ in Sec.~\ref{subsec:Dependence_dilaton_coupling}, the present sequence probes the coupling dependence of like-charged binaries rather than isolating the effect of the relative charge sign at fixed magnitude. 

\begin{figure*}[!htbp]
    \centering
    \begin{subfigure}[b]{0.326\textwidth}
        \centering
        \includegraphics[width=\linewidth]{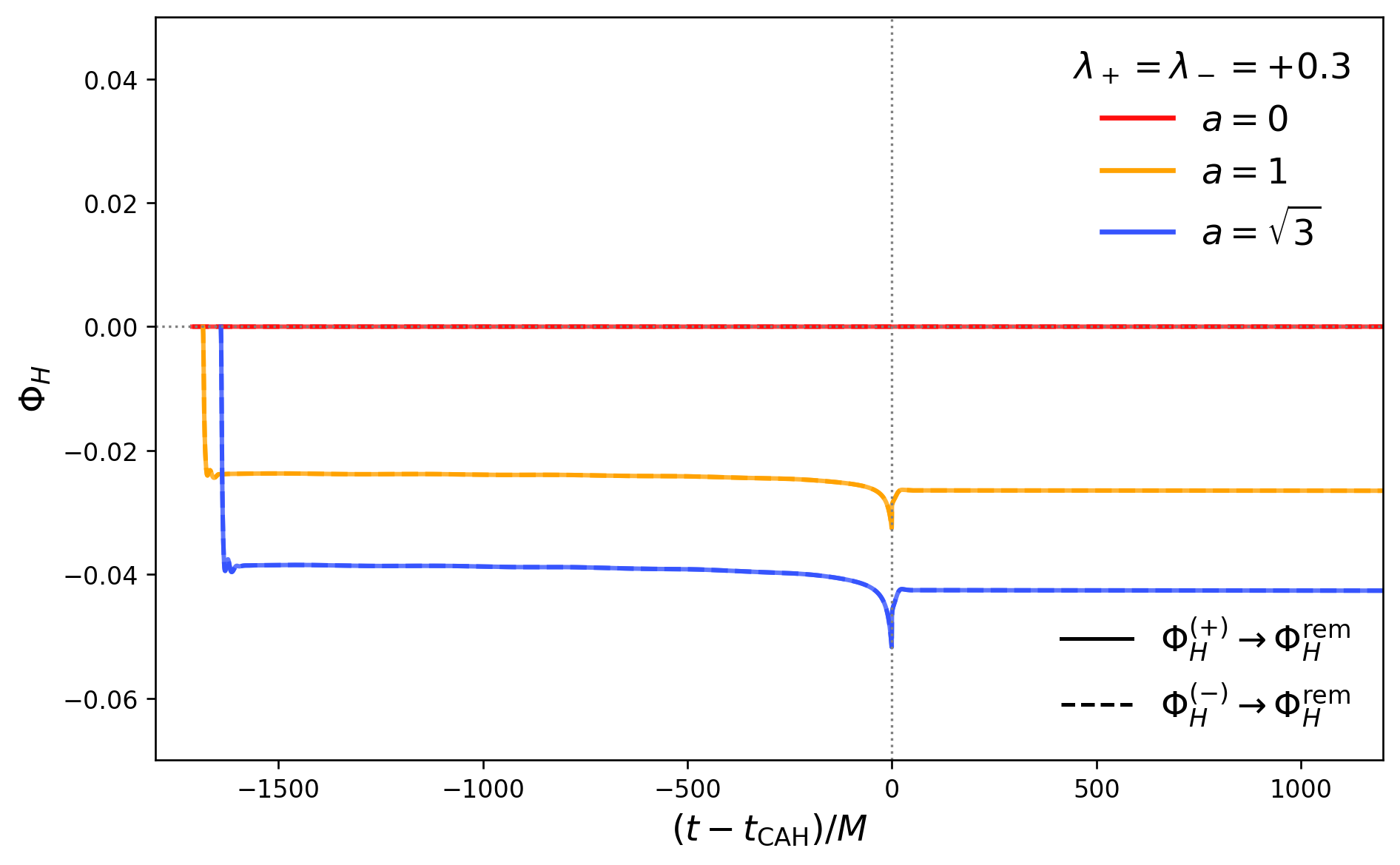}
        \caption{\centering $\Phi_{\rm H}$}
        \label{fig:sec6d_PhiH}
    \end{subfigure}
    \hfill
    \begin{subfigure}[b]{0.326\textwidth}
        \centering
        \includegraphics[width=\linewidth]{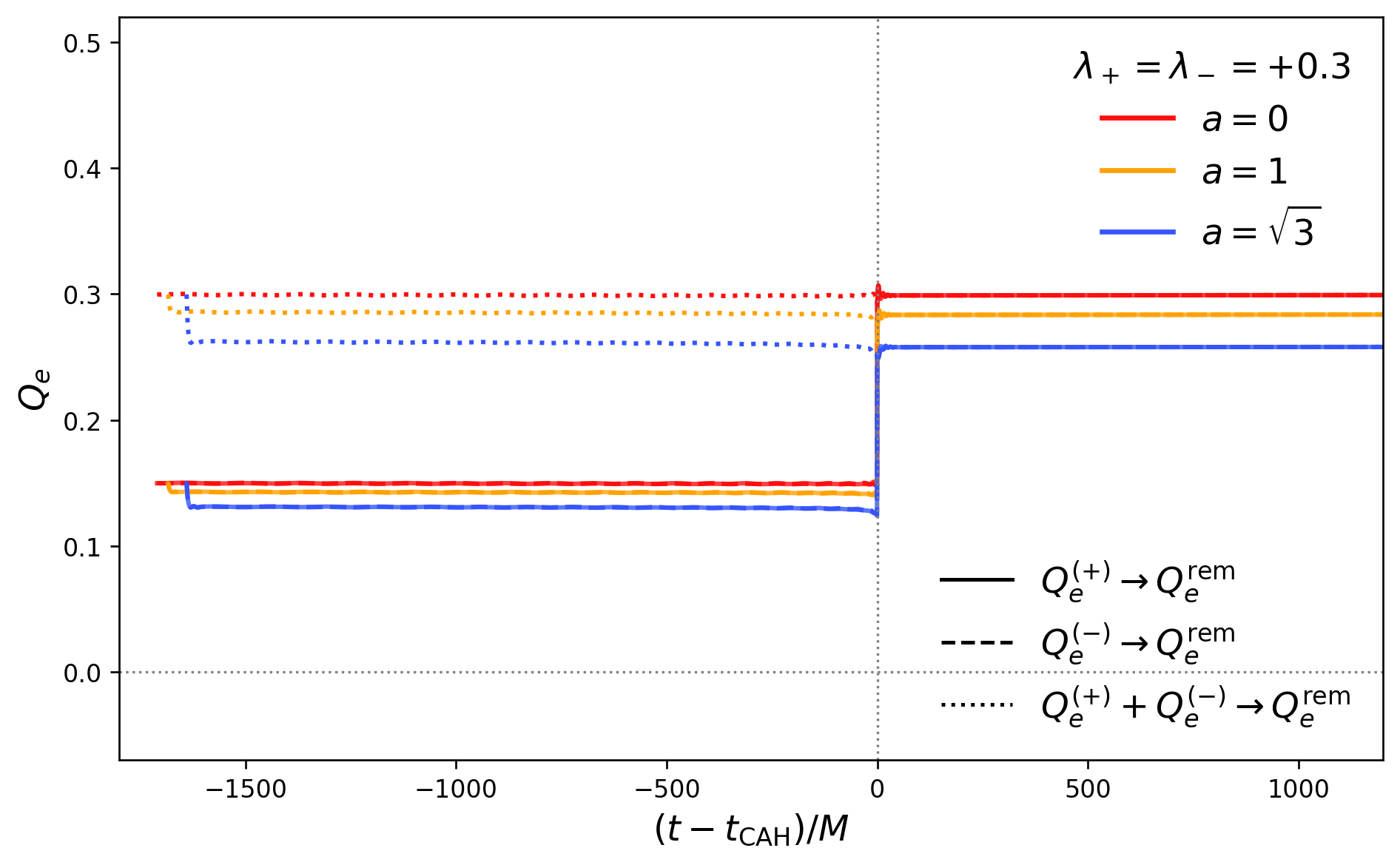}
        \caption{\centering $Q_e$}
        \label{fig:sec6d_Qe}
    \end{subfigure}
    \hfill
    \begin{subfigure}[b]{0.326\textwidth}
        \centering
        \includegraphics[width=\linewidth]{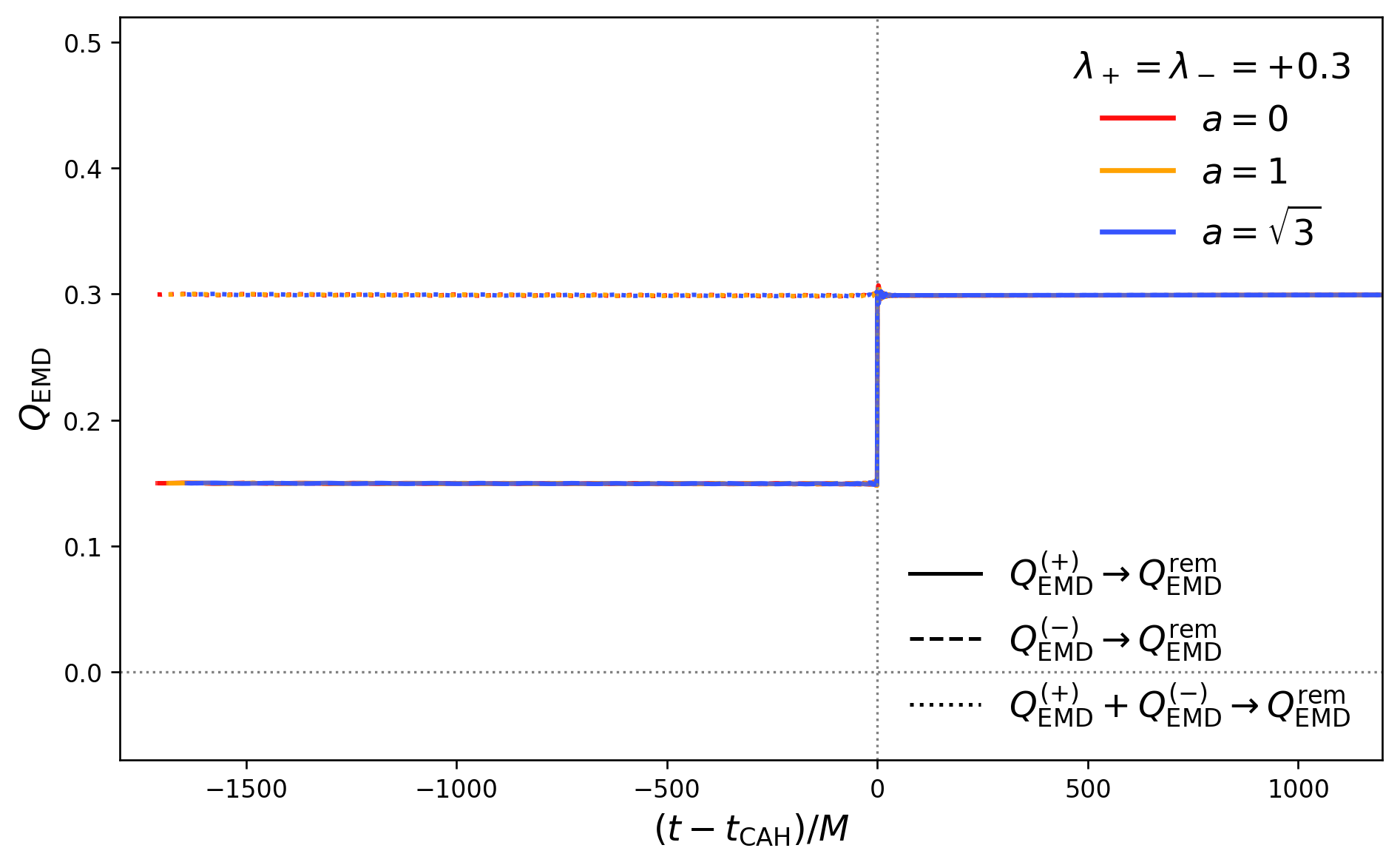}
        \caption{\centering $Q_{\rm EMD}$}
        \label{fig:sec6d_QEMD}
    \end{subfigure}
    \caption{
        Evolution of the quasi-local horizon quantities for $a=0$, $1$, and $\sqrt{3}$ in the like-charged sequence with $\lambda_+=\lambda_-=+0.3$. 
        Panels (a), (b), and (c) show the horizon-averaged scalar field $\Phi_{\rm H}$, the unweighted electric charge $Q_e$, and the dilaton-weighted charge $Q_{\rm EMD}$, respectively. 
        Solid and dashed curves denote the horizons originating from the two punctures before common-horizon formation. 
        After common-horizon formation, both curves trace the corresponding common-horizon value and therefore overlap. 
        In panels (b) and (c), the dotted curves show the sum of the two individual charges before common-horizon formation and the remnant charge afterward. 
        The horizontal axis shows $(t-t_{\rm CAH})/M$, and the vertical dotted line marks $t=t_{\rm CAH}$. 
    }
    \label{fig:sec6d_horizon}
\end{figure*}

Figure~\ref{fig:sec6d_horizon} shows the quasi-local horizon quantities for this sequence. 
Following the convention established in Sec.~\ref{subsec:main_results_a1}, the horizon data are displayed as functions of $(t-t_{\rm CAH})/M$. 
After the initial transient, the two individual horizons have nearly identical values of each quantity. 
Representative values are summarized in Table~\ref{tab:sec6d_horizon_representatives}. 
The time-averaged total EMD charges, $Q_{\rm EMD}^{(+)}+Q_{\rm EMD}^{(-)}$, are approximately $0.2991M$, $0.2991M$, and $0.2992M$ for $a=0$, $1$, and $\sqrt{3}$, respectively. 
The total horizon EMD charge connects smoothly to the common-horizon value after merger.
The remnant therefore retains a nonzero EMD charge, unlike the neutral remnants of the opposite-charge sequences. 
For $a=1$ and $a=\sqrt{3}$, $\Phi_{\rm H}$ remains nonzero on the common horizon throughout the displayed postmerger interval, consistent with a charged remnant carrying secondary dilaton hair. 
\begin{table}[h]
\centering
\setlength{\tabcolsep}{8pt}
\begin{tabular}{c|ccc}
$a$ &
$\langle\Phi_{\rm H}\rangle$ &
$\langle|Q_e|\rangle$ &
$\langle|Q_{\rm EMD}|\rangle$ \\
\hline
$0$        & $0$       & $0.1496$ & $0.1496$ \\
$1$        & $-0.0241$ & $0.1425$ & $0.1496$ \\
$\sqrt{3}$ & $-0.0390$ & $0.1307$ & $0.1496$ \\
\end{tabular}
\caption{
Representative horizon quantities for the like-charged sequence. 
Each value is first averaged over $-1400\leq(t-t_{\rm CAH})/M\leq-200$ and then over the two individual horizons. 
}
\label{tab:sec6d_horizon_representatives}
\end{table}
The coupling dependence of $Q_e$ is consistent with the charge-dressing relation in Eq.~\eqref{eq:horizon_charge_dressing}. 
Using the unrounded representative values underlying Table~\ref{tab:sec6d_horizon_representatives}, we obtain $\langle|Q_e|\rangle/\langle|Q_{\rm EMD}|\rangle\simeq0.953$ and $0.874$ for $a=1$ and $a=\sqrt{3}$, respectively. 
These ratios agree with the corresponding values of $e^{2a\langle\Phi_{\rm H}\rangle}$ at the quoted precision. 
The decrease in $|Q_e|$ therefore reflects the increasing dilaton weighting of the electric flux rather than a loss of the conserved EMD charge. 

For the like-charged sequence, we quantify the coupling dependence of the late orbital evolution using the interval from $D=10$ to the radiative merger, following the prescription introduced in Sec.~\ref{subsec:main_results_a1}. 
The resulting intervals are 
\begin{equation}
 \frac{t_{\rm merger}^{\rm rad}-t_{D=10}}{M}
 \simeq
 \begin{cases}
  1058.73 & (a=0)\,,\\
  1041.35 & (a=1)\,,\\
  1013.95 & (a=\sqrt{3})\,.
 \end{cases}
 \label{eq:sec6d_merger_intervals}
\end{equation}
For comparison, the neutral binary in Eq.~\eqref{eq:merger_time_D10} has $(t_{\rm merger}^{\rm rad}-t_{D=10})/M\simeq884.57$. 
The corresponding intervals for the like-charged binaries are approximately $15\%$--$20\%$ longer than the neutral value. 
This delay is qualitatively consistent with the repulsive electromagnetic interaction reducing the net inward attraction and slowing the late orbital evolution. 
The substantially shorter intervals of the oppositely charged high-charge sequence in Eq.~\eqref{eq:merger_time_a_D10} are likewise consistent with the electromagnetic interaction becoming attractive when the two charges have opposite signs. 

Within the like-charged sequence, however, the interval decreases as $a$ increases. 
Relative to the Einstein--Maxwell case, it is approximately $1.64\%$ shorter for $a=1$ and $4.23\%$ shorter for $a=\sqrt{3}$. 
This coupling dependence is opposite to that found for the oppositely charged high-charge sequence in Sec.~\ref{subsec:Dependence_dilaton_coupling}. 
As discussed in Sec.~\ref{sec:EMD}, the same-sign dilaton charges of electrically charged black holes contribute an attractive scalar-mediated interaction in the leading weak-field dynamics. 
The increase in $|\langle\Phi_{\rm H}\rangle|$ with $a$ in Table~\ref{tab:sec6d_horizon_representatives} indicates a stronger scalar response at larger coupling. 
The shorter merger intervals at larger $a$ are consistent with this scalar attraction partially offsetting the electromagnetic repulsion. 

\begin{figure*}[!htbp]
  \centering
    \begin{subfigure}[b]{0.48\columnwidth}
    \centering
    \includegraphics[width=\linewidth]
    {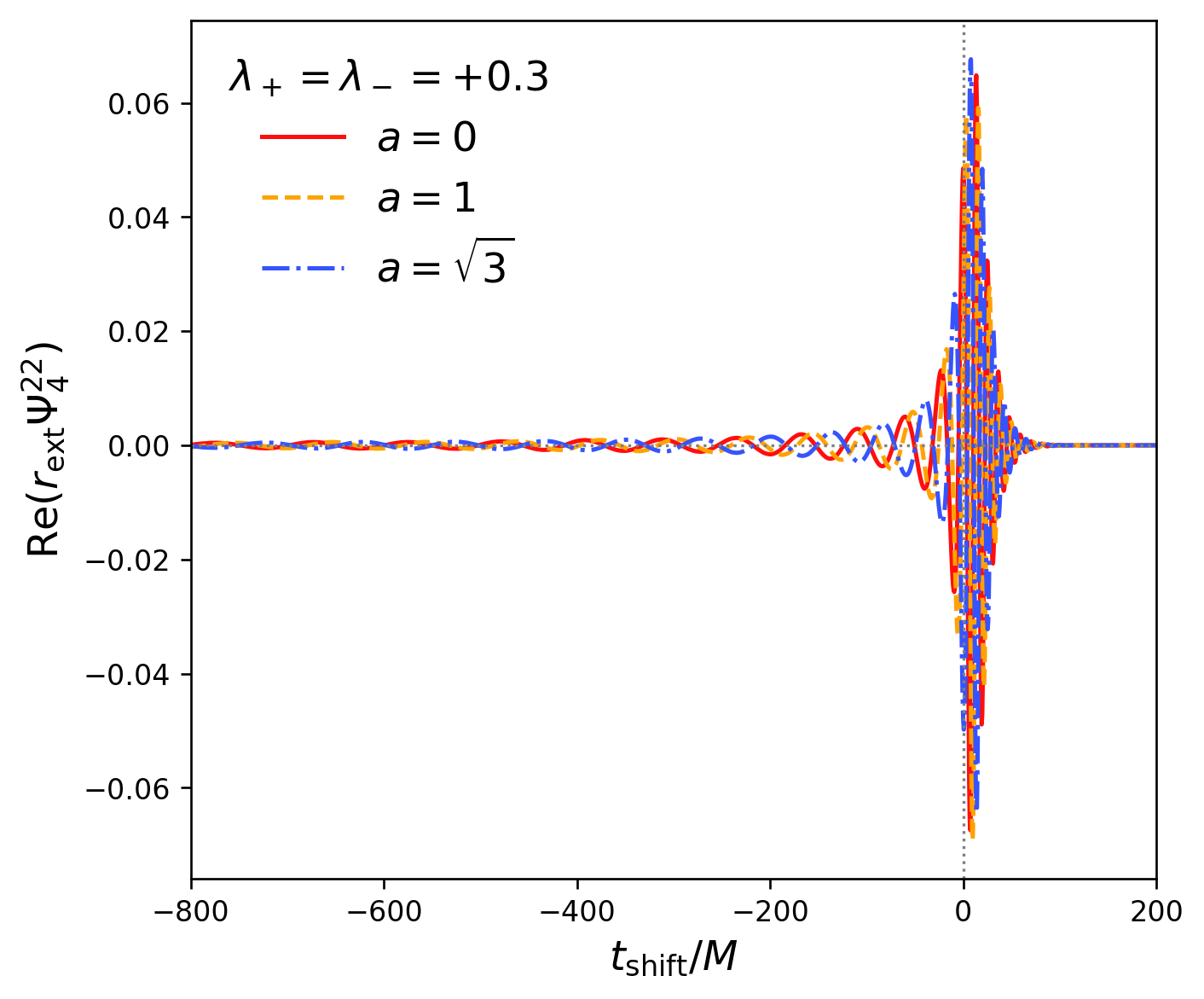}
    \caption{\centering ${\rm Re}(r_{\rm ext}\Psi_4^{22})$}
    \end{subfigure}
    \hfill
    \begin{subfigure}[b]{0.48\columnwidth}
    \includegraphics[width=\linewidth]
    {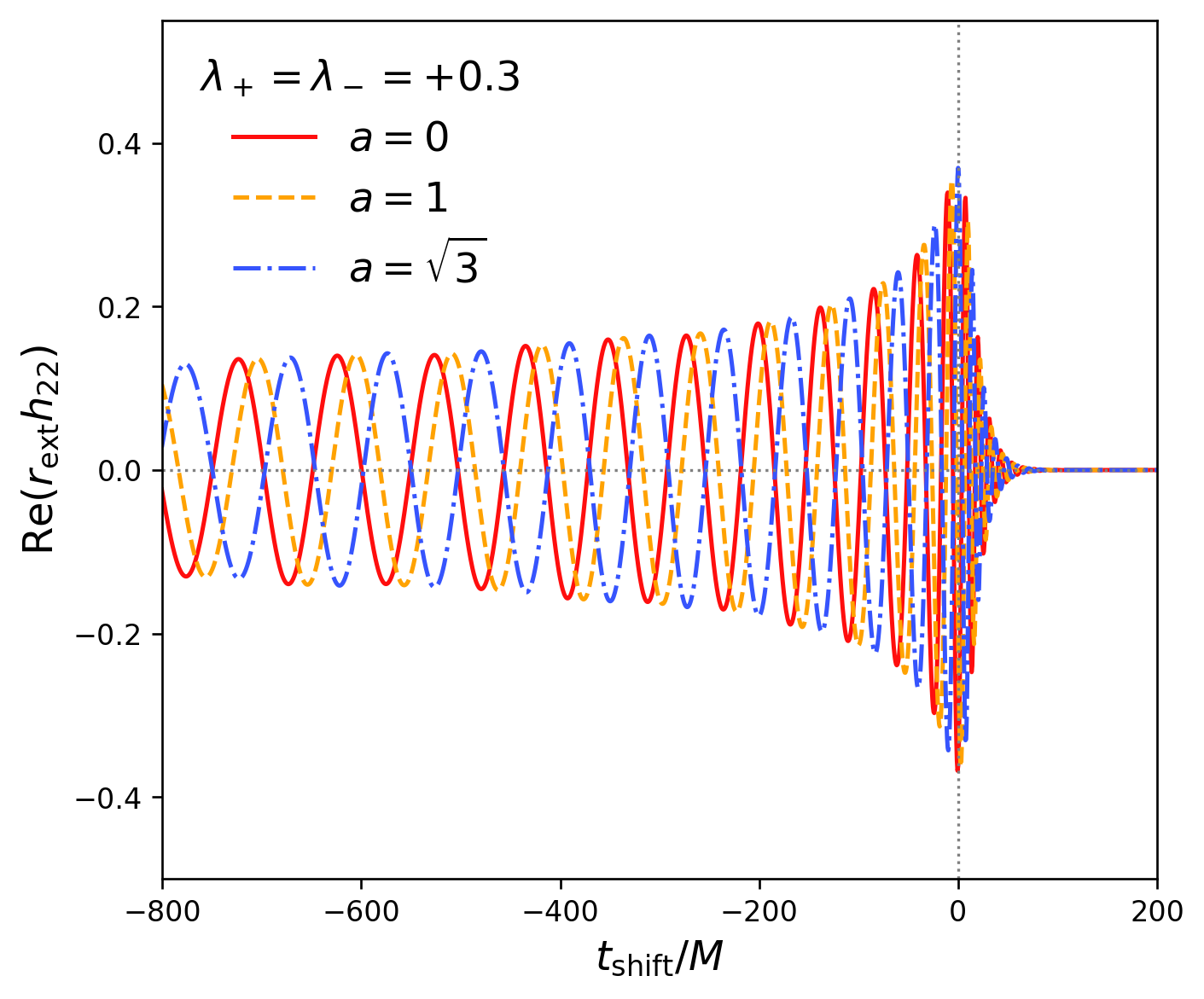}
    \caption{\centering ${\rm Re}(r_{\rm ext}h_{22})$}
    \end{subfigure}
    \\
  \caption{
  Dominant gravitational-wave signals for $a=0$, $1$, and $\sqrt{3}$ at fixed $\lambda_{+}=\lambda_{-}=+0.3$.
  Panels (a) and (b) show ${\rm Re}(r_{\rm ext}\Psi_4^{22})$ and ${\rm Re}(r_{\rm ext}h_{22})$, respectively.
  The horizontal axis is the strain-peak-aligned extraction time $t_{\rm shift}$ defined in Eq.~\eqref{eq:radiative_time_shift}.
  The vertical dotted line marks $t_{\rm shift}=0$, corresponding to the principal peak of $\left|r_{\rm ext}h_{22}\right|$. 
  No additional phase alignment is applied.
  }
  \label{fig:sec6d_gw}
\end{figure*}

Figure~\ref{fig:sec6d_gw} compares the dominant gravitational-wave signals for the three coupling values. 
The curves use the strain-peak-aligned extraction time $t_{\rm shift}$ defined in Eq.~\eqref{eq:radiative_time_shift}, without an additional phase alignment. 
The comparison therefore concerns waveform structure around merger, rather than the intervals from $D=10$ to merger quantified in Eq.~\eqref{eq:sec6d_merger_intervals}. 

The three waveforms show similar overall shapes and amplitude evolution through the late inspiral, merger, and ringdown. 
The peak complex strain amplitudes are $0.36804$, $0.36843$, and $0.36953$ for $a=0$, $1$, and $\sqrt{3}$, respectively, varying by approximately $0.41\%$ across the sequence. 
Visible offsets in the timing of the oscillations occur in both observables and are more readily seen in ${\rm Re}(r_{\rm ext}h_{22})$. 

\begin{figure*}[!htbp]
  \centering
    \begin{subfigure}[b]{0.32\textwidth}
    \centering
    \includegraphics[width=\linewidth]
    {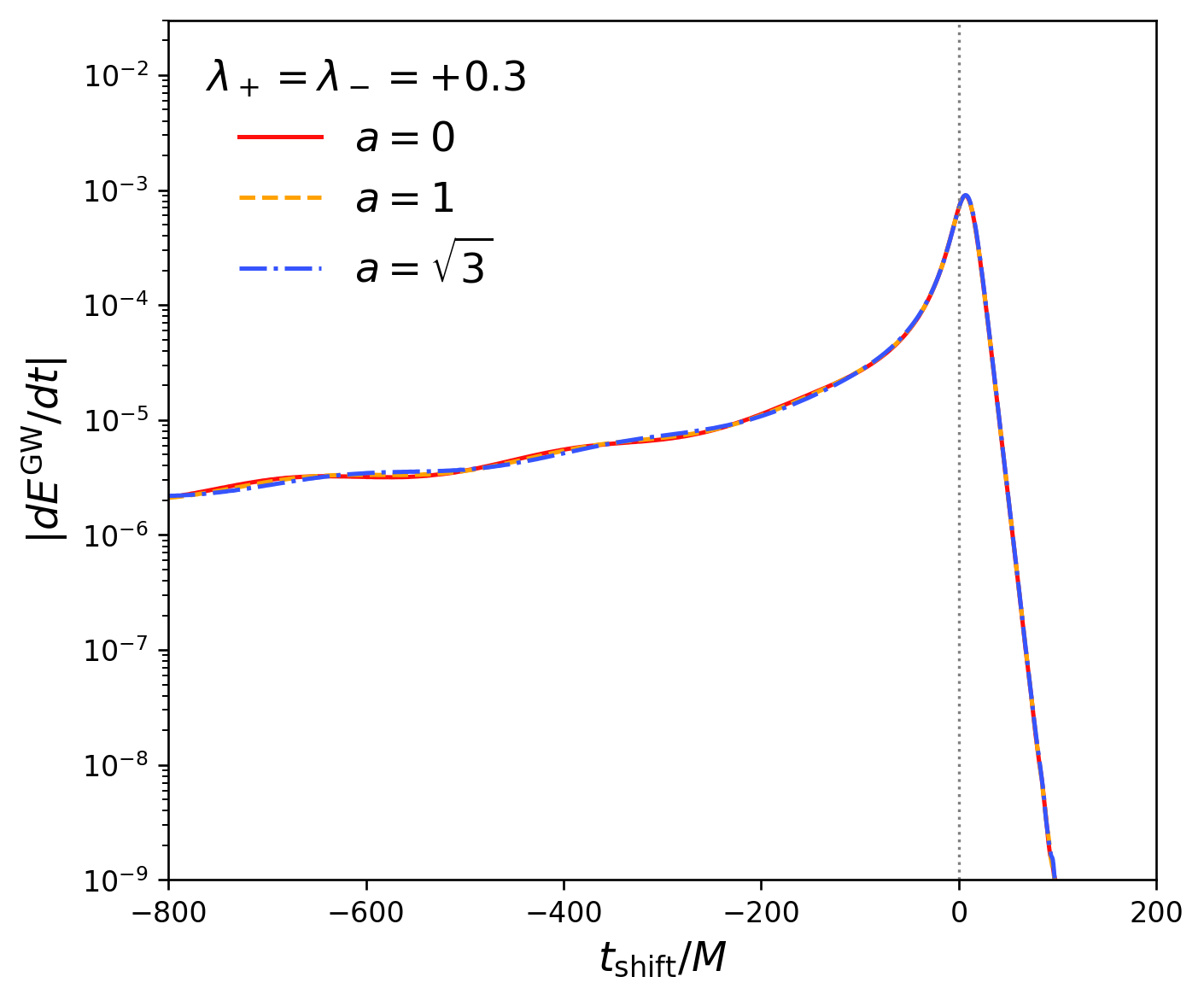}
    \caption{\centering $\left|dE^{\rm GW}/dt\right|$}
    \end{subfigure}
    \hfill
    \begin{subfigure}[b]{0.32\textwidth}
    \centering
    \includegraphics[width=\linewidth]
    {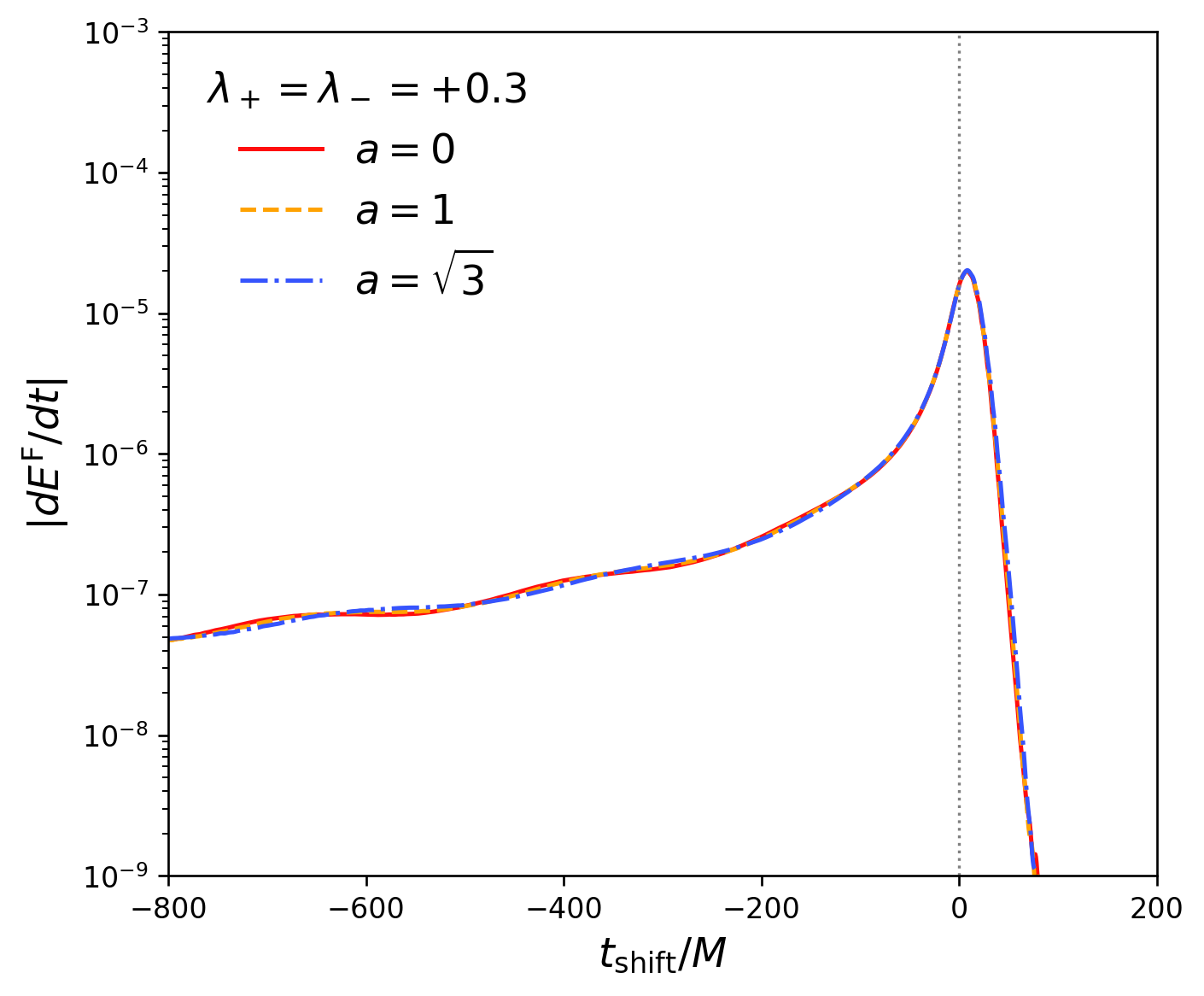}
    \caption{\centering $\left|dE^{\rm F}/dt\right|$}
    \end{subfigure} 
    \hfill
    \begin{subfigure}[b]{0.32\textwidth}
    \centering
    \includegraphics[width=\linewidth]
    {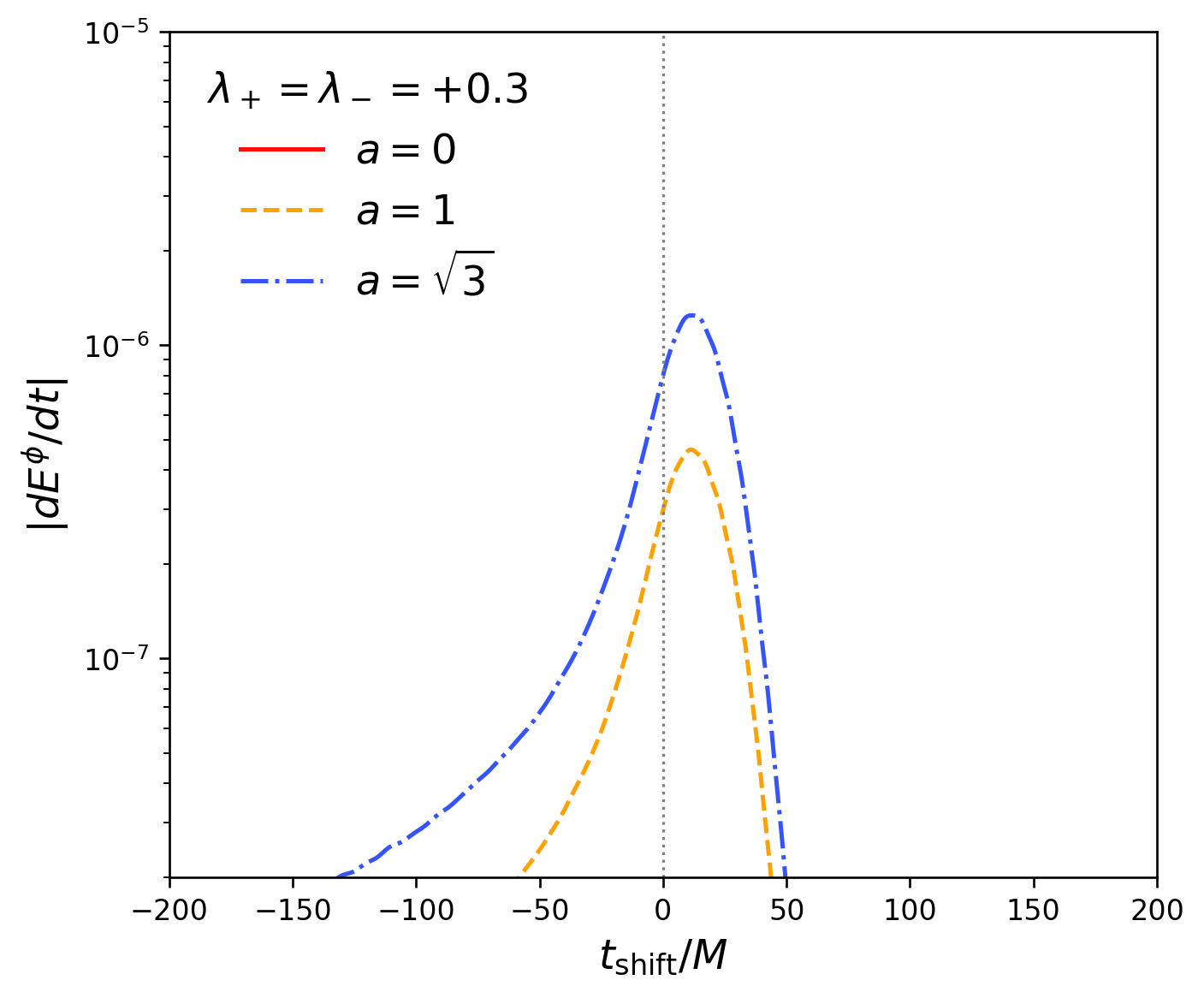}
    \caption{\centering $\left|dE^{\rm \phi}/dt\right|$}
    \end{subfigure} 
  \caption{
  Magnitudes of the radiative energy fluxes for $a=0$, $1$, and $\sqrt{3}$ at fixed $\lambda_{+}=\lambda_{-}=+0.3$.
  Panels (a), (b), and (c) show the gravitational, electromagnetic, and scalar energy fluxes, respectively.
  All panels use the strain-peak-aligned extraction time $t_{\rm shift}$ defined in Eq.~\eqref{eq:radiative_time_shift}.
  The scalar flux for $a=0$ is consistent with zero and therefore does not appear on the logarithmic scale. 
  Note that panel (c) uses a different horizontal range.
  }
  \label{fig:sec6d_fluxes}
\end{figure*}

The corresponding gravitational, electromagnetic, and scalar energy fluxes are shown in Fig.~\ref{fig:sec6d_fluxes}. 
All three panels use the same strain-peak-aligned extraction time as the gravitational waveforms. 
The gravitational- and electromagnetic-flux peaks vary by less than $1\%$ and $2\%$, respectively, across the three sampled couplings. 
By contrast, the scalar flux is consistent with zero for $a=0$, while its peak increases from approximately $4.65\times10^{-7}$ for $a=1$ to $1.25\times10^{-6}$ for $a=\sqrt{3}$. 
The latter is approximately $2.68$ times the former. 
This enhancement accompanies the increase in $|\langle\Phi_{\rm H}\rangle|$ found in Table~\ref{tab:sec6d_horizon_representatives}, indicating a stronger scalar response both on the horizons and in the radiation. 
Scalar radiation provides an additional channel for orbital-energy loss, consistent with the shortening of the late orbital evolution as $a$ increases. 

\begin{figure}[!htbp]
  \centering
  \includegraphics[width=0.48\columnwidth]
  {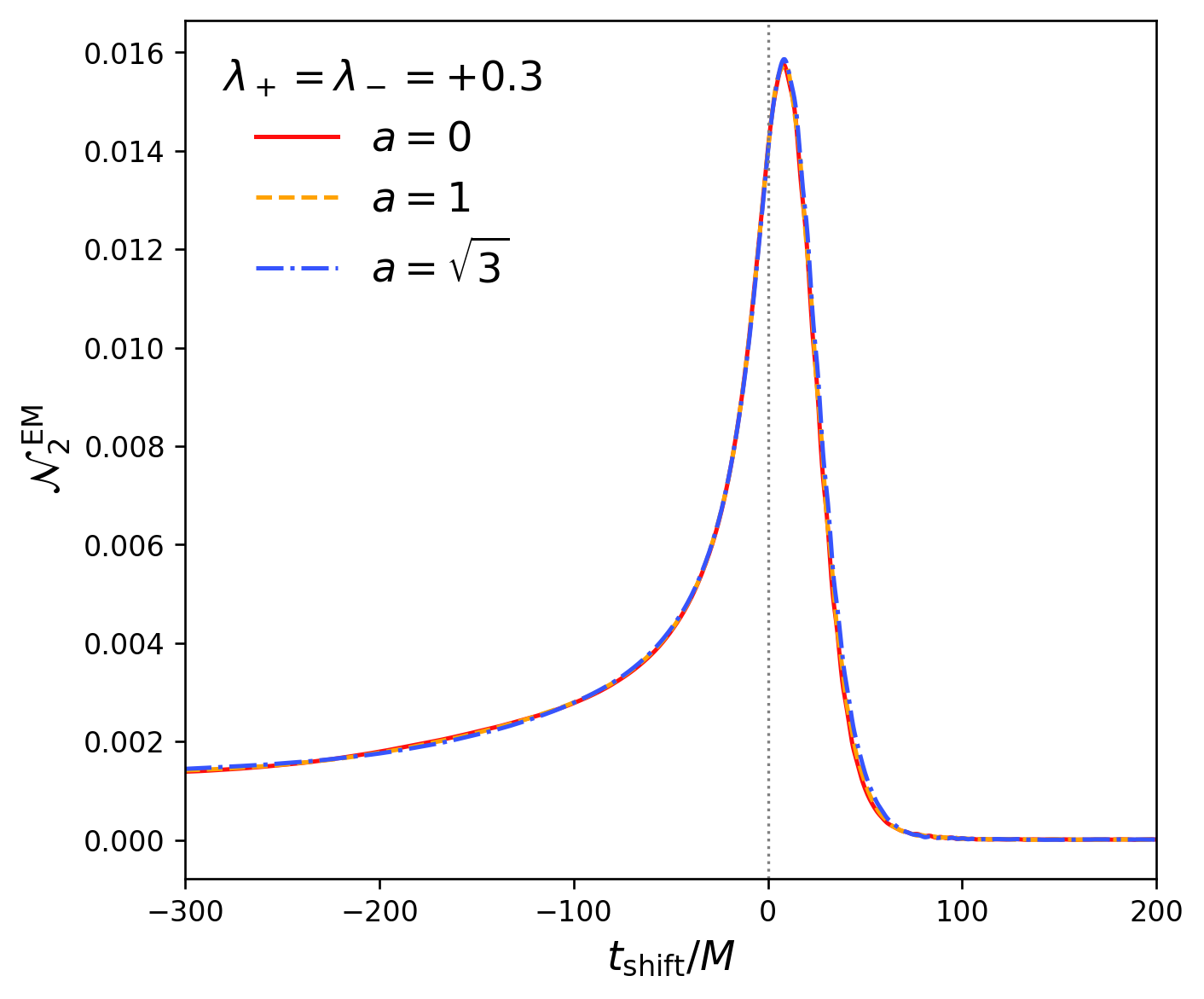}
  \caption{
  Quadrupolar electromagnetic Newman--Penrose norm $\mathcal{N}_2^{\rm EM}$, defined in Eq.~\eqref{eq:em_np_multipole_norm}, for $a=0$, $1$, and $\sqrt{3}$ in the like-charged sequence with $\lambda_{+}=\lambda_{-}=+0.3$. 
  The outgoing Newman--Penrose scalar has spin weight $-1$. 
  The horizontal axis is the strain-peak-aligned extraction time $t_{\rm shift}$ defined in Eq.~\eqref{eq:radiative_time_shift}. 
  The vertical dotted line marks $t_{\rm shift}=0$, corresponding to the principal peak of $\left|r_{\rm ext}h_{22}\right|$.
  The near coincidence of the three curves shows that the dominant quadrupolar amplitude depends only weakly on $a$ across the sampled sequence.
  \label{fig:sec6d_em_multipoles}
  }
\end{figure}

The multipolar structure of the electromagnetic radiation provides a sharper distinction from the oppositely charged sequence. 
Figure~\ref{fig:sec6d_em_multipoles} compares the quadrupolar norm $\mathcal{N}_2^{\rm EM}$ defined in Eq.~\eqref{eq:em_np_multipole_norm} for the three coupling values. 

The peak values of $\mathcal{N}_2^{\rm EM}$ lie between approximately $0.01574$ and $0.01586$ across the three couplings. 
The $\ell=1$ and $\ell=3$ peak norms are both ${\cal O}(10^{-3})$. 
The quadrupolar flux $\mathcal{P}_2^{\rm EM}=(4\pi)^{-1}(\mathcal{N}_2^{\rm EM})^2$ accounts for approximately $99\%$ of the total electromagnetic flux at its peak for all three couplings. 
The weak coupling dependence of the total electromagnetic-flux peak therefore reflects that of the dominant quadrupolar sector. 

This quadrupolar hierarchy follows naturally from the approximate inversion symmetry of an equal-mass binary carrying equal charges. 
In the center-of-mass frame, the two equally charged black holes occupy approximately opposite positions, so their contributions to the leading electric dipole moment cancel. 
The electric charge distribution is approximately invariant under spatial inversion, suppressing the odd-$\ell$ electric-type source multipoles while allowing the time-dependent electric quadrupole. 
The small residual $\ell=1$ and $\ell=3$ signals can contain symmetry-allowed magnetic-type radiation, together with weak electric-type contributions generated by small departures from exact binary symmetry. 
For equal-magnitude opposite charges, by contrast, the charge distribution is approximately antisymmetric under spatial inversion, allowing a nonzero electric dipole and explaining the dipole-dominated signal found in Sec.~\ref{subsec:main_results_a1}. 

For approximately equal charge-to-mass ratios, the leading electric quadrupole is proportional to the mass quadrupole. 
The individual EMD charges are nearly independent of $a$, while the dominant gravitational waveform and the gravitational-flux peak show only modest coupling dependence. 
These features provide a natural leading-order explanation for the nearly coincident electromagnetic quadrupolar signals. 
The lower horizon charge $Q_e$ at larger $a$ does not imply an equally reduced far-zone quadrupole, because it reflects local dilaton dressing rather than a loss of conserved EMD charge. 
This behavior contrasts with the opposite-charge sequence at $\lambda_{\pm}=\pm0.7$ in Fig.~\ref{fig:coupling_em_dipole}, where the dominant dipole norm decreases by approximately $15.8\%$ from $a=0$ to $a=\sqrt{3}$. 
The dilaton coupling therefore produces different radiative responses in these two charge configurations, rather than a uniform rescaling of the electromagnetic signal. 

In summary, the like-charged binaries retain a nonzero EMD charge after merger and, for $a\neq0$, a persistent horizon scalar field consistent with secondary dilaton hair. 
Their intervals from $D=10$ to the radiative merger are approximately $15\%$--$20\%$ longer than the neutral baseline, consistent with electromagnetic repulsion. 
Within this sequence, increasing $a$ strengthens the scalar response and shortens the merger interval by approximately $4.23\%$ from $a=0$ to $a=\sqrt{3}$. 
The electromagnetic radiation remains strongly quadrupole dominated, while the dominant gravitational-wave amplitude and electromagnetic-flux peak vary only weakly with $a$.

\subsection{Dependence on the dilaton coupling in the weak-charge case}
\label{subsec:main_results_low_charge}

We finally compare the Einstein--Maxwell case $a=0$ with the EMD cases $a=1$ and $a=\sqrt{3}$ at fixed $\lambda_{\pm}=\pm0.05$. 
This sequence tests whether the dilaton coupling leaves a discernible imprint on the dominant gravitational waveform in the weak-charge regime. 

\begin{figure}[!htbp]
\centering
\begin{subfigure}{0.48\linewidth}
\centering
\includegraphics[width=\linewidth]{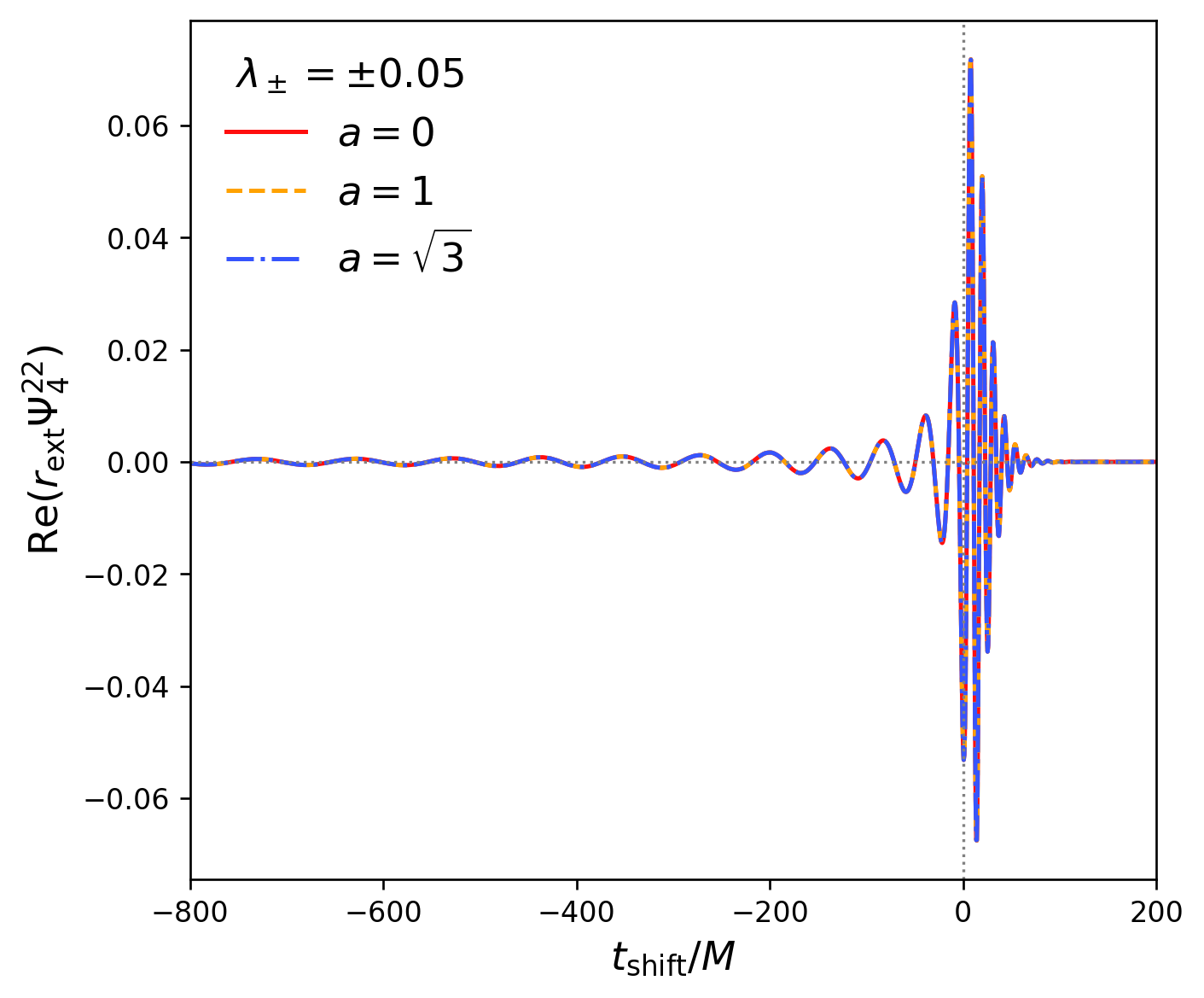}
\caption{\centering
${\rm Re}(r_{\rm ext}\Psi_4^{22})$.
}
\label{fig:low_charge_psi4}
\end{subfigure}
\hfill
\begin{subfigure}{0.48\linewidth}
\centering
\includegraphics[width=\linewidth]{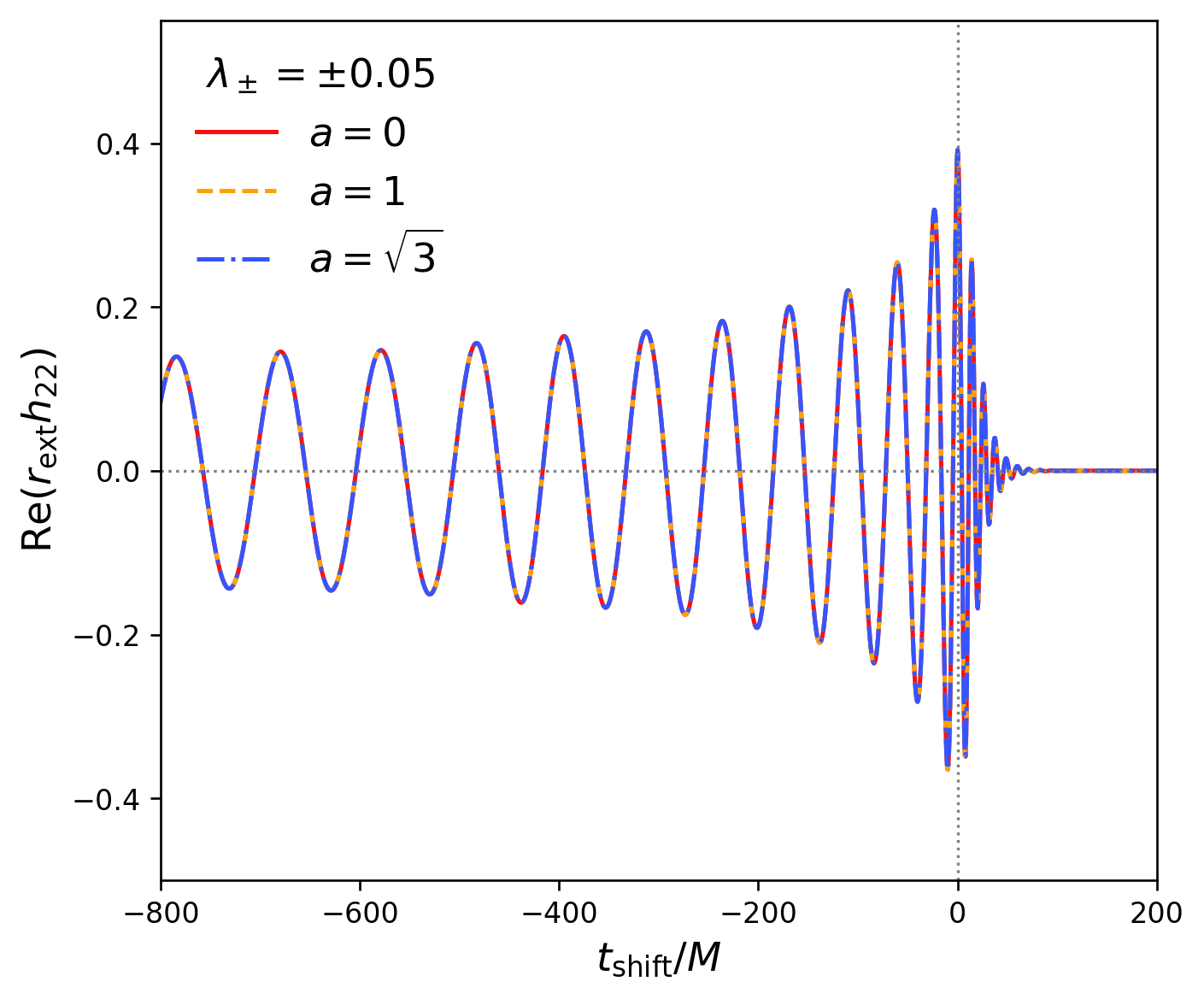}
\caption{\centering
${\rm Re}(r_{\rm ext}h_{22})$.
}
\label{fig:low_charge_gw}
\end{subfigure}
\caption{
Dominant gravitational-wave signals for $a=0$, $1$, and $\sqrt{3}$ in the weak-charge sequence with $\lambda_{\pm}=\pm0.05$. 
Panels (a) and (b) show ${\rm Re}(r_{\rm ext}\Psi_4^{22})$ and ${\rm Re}(r_{\rm ext}h_{22})$, respectively. 
The signals are extracted at $r_{\rm ext}=500M$ and are shown as functions of the strain-peak-aligned extraction time $t_{\rm shift}$ defined in Eq.~\eqref{eq:radiative_time_shift}. 
The vertical dotted line marks $t_{\rm shift}=0$, corresponding to the principal peak of $\left|r_{\rm ext}h_{22}\right|$. 
}
\label{fig:low_charge_waveforms}
\end{figure}
\begin{figure*}[!htbp]
    \centering
    \begin{subfigure}[b]{0.32\textwidth}
        \centering
        \includegraphics[width=\linewidth]
        {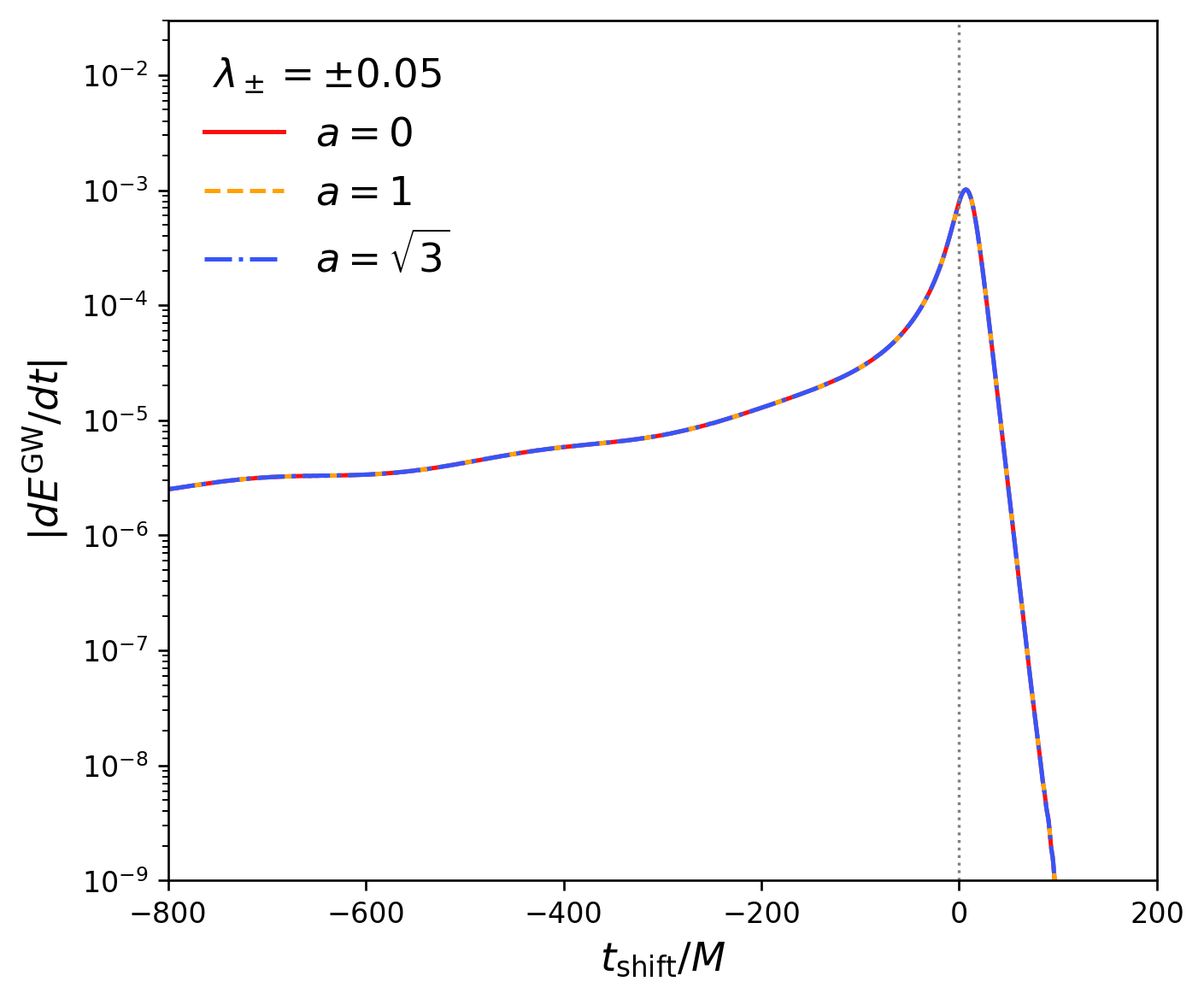}
        \caption{\centering
        $\left|dE^{\rm GW}/dt\right|$
        }
        \label{fig:dEdt_GW_dilaton_weak_charge}
    \end{subfigure}
    \hfill
    \begin{subfigure}[b]{0.32\textwidth}
        \centering
        \includegraphics[width=\linewidth]
        {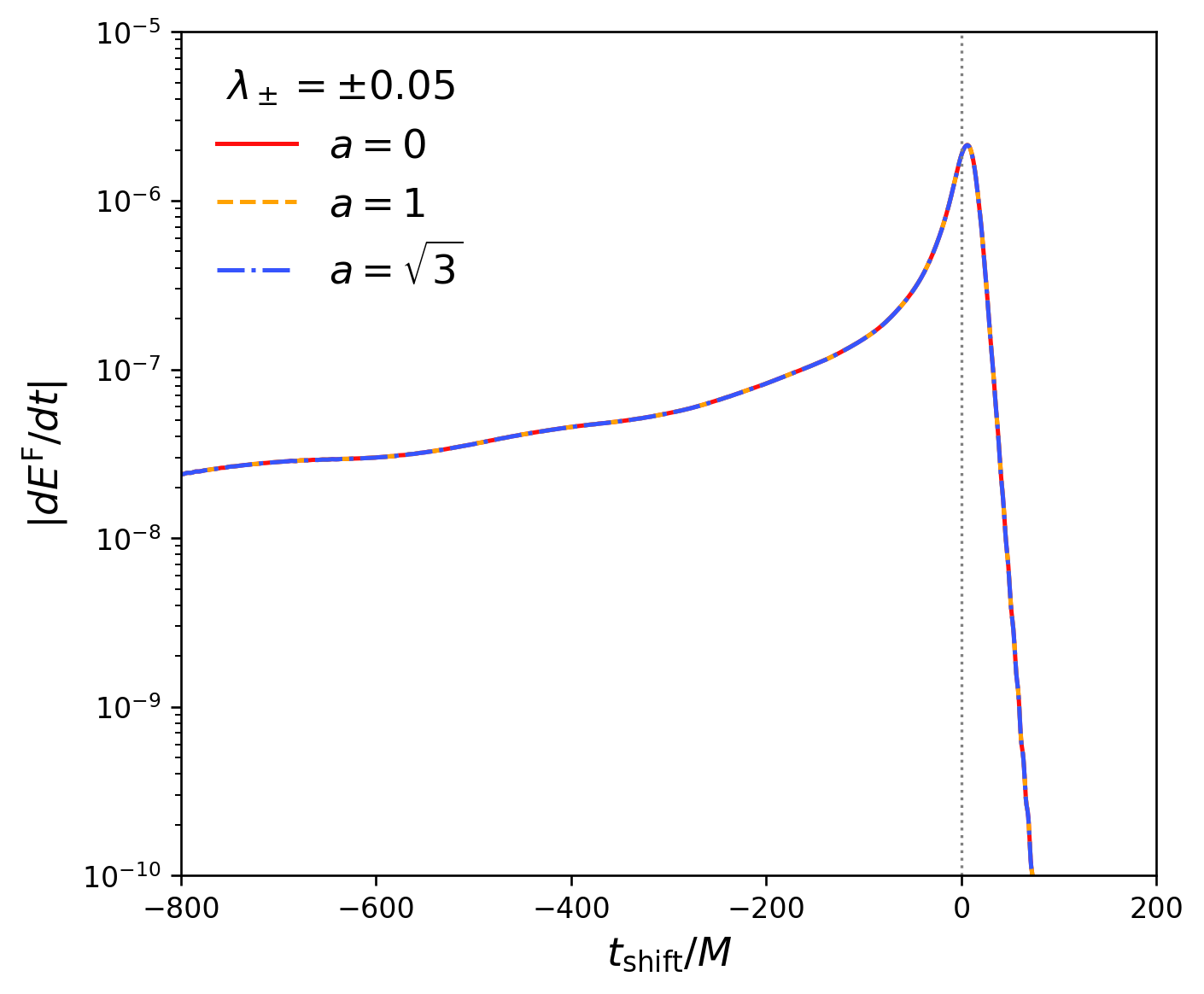}
        \caption{\centering
        $\left|dE^{\rm F}/dt\right|$
        }
        \label{fig:dEdt_EM_dilaton_weak_charge}
    \end{subfigure}
    \hfill
    \begin{subfigure}[b]{0.32\textwidth}
        \centering
        \includegraphics[width=\linewidth]
        {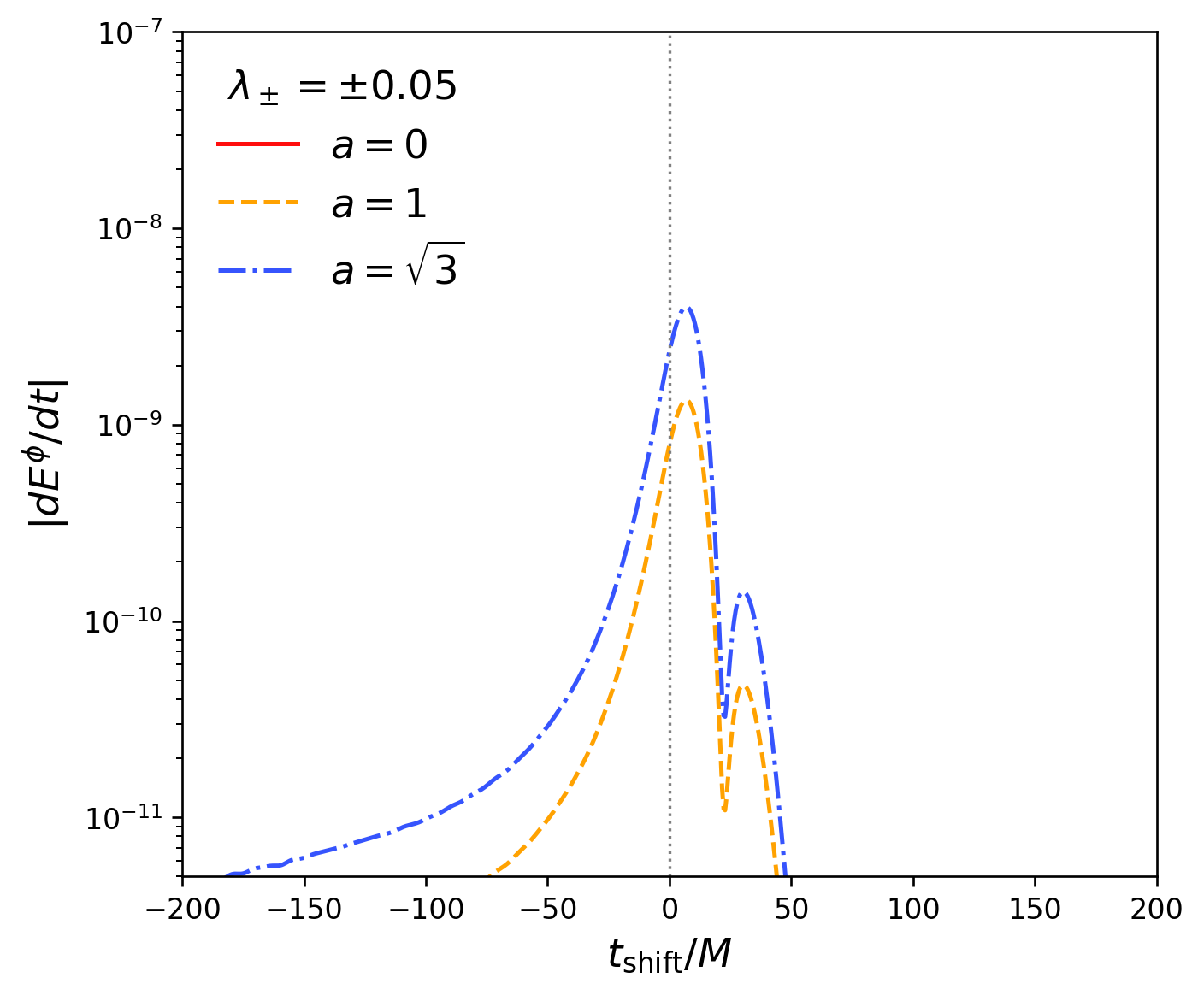}
        \caption{\centering
        $\left|dE^\phi/dt\right|$
        }
        \label{fig:dEdt_phi_dilaton_weak_charge}
    \end{subfigure}
    \caption{
    Magnitudes of the gravitational, electromagnetic, and scalar energy fluxes for $a=0$, $1$, and $\sqrt{3}$ in the weak-charge sequence with $\lambda_{\pm}=\pm0.05$. 
    Panels (a), (b), and (c) show the gravitational, electromagnetic, and scalar fluxes, respectively. 
    The fluxes are extracted at $r_{\rm ext}=500M$ and are shown as functions of the strain-peak-aligned extraction time $t_{\rm shift}$ defined in Eq.~\eqref{eq:radiative_time_shift}. 
    The vertical dotted line marks $t_{\rm shift}=0$, corresponding to the principal peak of $\left|r_{\rm ext}h_{22}\right|$.  
    The scalar flux for $a=0$ is consistent with zero and therefore does not appear on the logarithmic scale. 
    Note that panel (c) uses a different horizontal range.
    }
    \label{fig:energy_flux_dilaton_weak_charge}
\end{figure*}
Figure~\ref{fig:low_charge_waveforms} compares the strain-peak-aligned $(\ell,m)=(2,2)$ gravitational-wave signals for the three couplings. 
In both ${\rm Re}(r_{\rm ext}\Psi_4^{22})$ and ${\rm Re}(r_{\rm ext}h_{22})$, the three curves are visually indistinguishable throughout the displayed interval. 
The peak complex strain amplitudes agree to four significant figures, with $\max\left|r_{\rm ext}h_{22}\right|\simeq0.3920$ for all three couplings. 
The dilaton-coupling dependence of the dominant gravitational waveform is therefore not appreciably resolved at the scale displayed. 

Figure~\ref{fig:energy_flux_dilaton_weak_charge} shows the corresponding gravitational, electromagnetic, and scalar energy fluxes. The gravitational- and electromagnetic-flux curves nearly overlap across the three couplings, with peak variations below $0.001\%$ and $0.3\%$, respectively. 
The scalar flux vanishes for $a=0$, while its peak reaches approximately $1.33\times10^{-9}$ for $a=1$ and $3.97\times10^{-9}$ for $a=\sqrt{3}$. 
The ratio of approximately $2.99$ agrees closely with the factor of three predicted by the leading weak-charge scaling $dE^\phi/dt\propto a^2|\lambda_{\pm}|^4$ discussed in Sec.~\ref{subsec:main_results_a1}. 
Despite this clear coupling dependence, the scalar-flux peaks remain far below the gravitational- and electromagnetic-flux peaks. 

The weak-charge sequence therefore retains the expected coupling dependence in the scalar sector without an appreciable change in the dominant gravitational waveform. 
This result is qualitatively consistent with Hirschmann et al., who found only a limited influence of the dilaton coupling on the gravitational waveforms for $q_e=10^{-3}$~\cite{Hirschmann:2017psw}. 

\section{Conclusions}
\label{sec:conclusions}

We have performed fully nonlinear numerical-relativity simulations of electrically charged BBH mergers in EMD theory. 
Sections~\ref{sec:EMD}--\ref{sec:extraction_and_quantities} presented the EMD theory, the $3+1$ evolution system, the construction of constraint-satisfying charged puncture initial data, and the radiative and quasi-local diagnostics used in our analysis. 
The initial data combine a Bowen--York extrinsic curvature with a Coulomb-type electric field and initially vanishing scalar variables $\phi=K_\phi=0$. 
Consequently, at fixed puncture parameters, the initial constraint solution is independent of the dilaton coupling, while charged configurations with $a\neq0$ subsequently develop a nontrivial scalar profile through an early relaxation. 
Using this framework, we varied the magnitude and relative sign of the input charges and the dilaton coupling to investigate their effects on the late inspiral, merger, radiation, and postmerger remnant. 

In Sec.~\ref{subsec:main_results_a1}, we examined the charge dependence at the fixed, string-motivated coupling $a=1$ by comparing the neutral binary with oppositely charged binaries having $\lambda_{\pm}=\pm0.05$ and $\pm0.7$. 
In the charged evolutions, the scalar profile and the unweighted horizon charge $Q_e$ undergo an initial adjustment, whereas the individual dilaton-weighted charges $Q_{\rm EMD}$ remain nearly constant from the beginning of the evolution to just before merger. 
As quantified in Eq.~\eqref{eq:merger_time_D10}, the intervals from $D=10$ to the radiative merger differ by approximately $5.6\%$
between the neutral and weakly charged binaries, whereas the corresponding interval for the highly charged binary is approximately $62\%$ shorter than in the neutral case. 
This pronounced shortening is consistent with the additional electromagnetic attraction between the oppositely charged black holes, whose individual charges remain large until merger. 

The radiative diagnostics show the same transition from weak to pronounced charge dependence. 
At fixed $a=1$, the electromagnetic- and scalar-flux peak ratios are comparable to estimates based on the leading weak-charge scalings for fixed orbital motion, $|\lambda_{\pm}|^2$ and $|\lambda_{\pm}|^4$, respectively. 
For the highly charged binary, approximately $99\%$ of the peak electromagnetic flux is carried by the $\ell=1$ sector, as expected from the time-dependent electric dipole generated by the orbital motion of the oppositely charged black holes. 
The highly charged configuration exhibits clearly visible differences from the neutral vacuum-GR baseline in the dominant gravitational waveform, particularly during the late inspiral and around merger. 
The weakly charged waveform remains close to this baseline. 

Section~\ref{subsec:Dependence_dilaton_coupling} examined the highly charged sequence with fixed $\lambda_{\pm}=\pm0.7$, comparing the Einstein--Maxwell reference $a=0$ with the EMD couplings $a=1$ and $a=\sqrt{3}$. 
The representative magnitudes of the individual EMD charges, $|Q_{\rm EMD}|$, remain nearly unchanged across this sequence, while the horizon-averaged scalar field becomes increasingly negative and the unweighted charge magnitude $|Q_e|$ decreases as $a$ increases. 
This behavior is quantitatively consistent with the charge-dressing relation $Q_e\simeq e^{2a\Phi_{\rm H}}Q_{\rm EMD}$ in Eq.~\eqref{eq:horizon_charge_dressing} and therefore reflects the increasing dilaton weighting of the electric flux rather than a loss of the conserved EMD charge. 
The interval from $D=10$ to the radiative merger increases monotonically with $a$, reaching approximately $10.7\%$ above the Einstein--Maxwell value for $a=\sqrt{3}$, as quantified in Eq.~\eqref{eq:merger_time_a_D10}. 
The electromagnetic-flux peak decreases for larger $a$, whereas the scalar-flux peak increases. 
With the dipole carrying approximately $99\%$ of the electromagnetic flux at its peak for all three couplings, this decrease primarily reflects a weaker dipolar peak rather than a substantial change in the multipolar composition at that time. 
This opposing behavior of the electromagnetic and scalar sectors indicates a coupling-dependent reorganization of the radiative response. 
The dominant gravitational waveforms retain similar overall shapes and comparable merger amplitudes after strain-peak alignment. 
Nevertheless, the strain exhibits visible differences between the EMD cases and the Einstein--Maxwell reference. 
After merger, the equal-magnitude opposite charges cancel on the common horizon and the scalar transient decays, so the remnant approaches a neutral black hole without stationary secondary dilaton hair. 

Section~\ref{subsec:like_charged} considered like-charged binaries with $\lambda_+=\lambda_-=+0.3$, for which the total EMD charge does not vanish. 
After merger, the common horizon inherits this nonzero charge, and the nonvanishing postmerger value of $\Phi_{\rm H}$ for $a\neq0$ is consistent with a charged remnant carrying secondary dilaton hair. 
As quantified in Eq.~\eqref{eq:sec6d_merger_intervals}, the interval from $D=10$ to the radiative merger is approximately $15\%$--$20\%$ longer than for the neutral vacuum-GR baseline, consistent with the repulsive electromagnetic interaction between like charges. 
Within the like-charged sequence, however, this interval decreases modestly with increasing $a$ and is approximately $4\%$ shorter for $a=\sqrt{3}$ than for the Einstein--Maxwell reference. 
A plausible interpretation is that the stronger scalar response at larger $a$ partially offsets the electromagnetic repulsion through an attractive dilaton-mediated interaction, while scalar radiation provides an additional channel for orbital-energy loss. 
The electromagnetic signal is strongly quadrupole dominated because the leading electric dipole approximately cancels for an equal-mass binary carrying equal charges. 
Unlike the dipole-dominated opposite-charge sequence, its peak flux changes little with $a$, even though the scalar radiation strengthens. 

Taken together, these results distinguish the roles of the charge magnitude, charge symmetry, and dilaton coupling in charged EMD binary mergers. 
Within the configurations sampled here, increasing the charge magnitude produces the largest changes in the late orbital evolution and the dominant gravitational waveform. 
The charge symmetry determines whether the electromagnetic interaction is attractive or repulsive, whether the leading electromagnetic radiation is dipolar or quadrupolar, and whether the remnant becomes neutral or retains a nonzero charge and, for $a\neq0$, secondary dilaton hair. 
The dilaton coupling modifies the scalar dressing and radiative response in a manner that depends on the charge configuration. 
In the weak-charge sequence of Sec.~\ref{subsec:main_results_low_charge}, the scalar response remains coupling dependent while the dominant gravitational waveforms are visually indistinguishable at the scale displayed. 
By contrast, the highly charged sequence exhibits visible differences in the dominant gravitational strain between the EMD cases and the Einstein--Maxwell reference. 
The reported merger-time intervals are coordinate-time diagnostics evaluated within the common gauge, numerical setup, and initial-data prescription adopted here. 
The present diagnostics do not provide systematic numerical error estimates, separate the conservative and dissipative contributions to the observed differences, or establish detector-level distinguishability between EMD and GR waveforms. 

An immediate next step is to construct charged EMD initial data with a self-consistent nontrivial dilaton profile, rather than imposing $\phi=K_\phi=0$ on the initial hypersurface. 
Such initial data should reduce the early scalar-field relaxation and improve control over the relaxed horizon masses, electric charges, and scalar profiles of the individual black holes. 
Combining this construction with an iterative eccentricity-reduction procedure such as \texttt{EccRed}~\cite{Habib:2020dba} will enable controlled comparisons among binaries with matched physical parameters and smaller residual eccentricity. 
In parallel, systematic resolution-convergence and waveform-extraction studies will quantify the numerical uncertainties. 
These developments will support accurate EMD waveform families for detector-weighted comparisons with vacuum-GR and charged Einstein--Maxwell models, accounting for degeneracies with the binary parameters. 
Injection studies across a range of signal-to-noise ratios can then determine when the dilaton-induced waveform differences reported in this paper are observationally distinguishable and guide tests of EMD theory using observed BBH signals. 

A complementary direction is to extend these simulations to magnetically or dyonically charged binaries, connecting the dynamical problem more directly to the broader GM--GHS family of charged black holes. 
These extensions would also motivate BBH merger studies in the broader class of $U(1)$ gauge-invariant scalar--vector--tensor theories~\cite{Heisenberg:2018acv}, which contains EMD theory as a special case. 
Static, spherically symmetric hairy BH solutions have been constructed in this class, including branches supported by magnetic charge~\cite{Heisenberg:2018vti,Taniguchi:2024ear,Zhang:2024cbw,Taniguchi:2025bmc,Kitagawa:2026tcl}. 

\section*{Acknowledgements}
We thank Gensei Tanaka for carefully checking the equations and providing helpful comments. 
K. Taniguchi is supported by the Grant-in-Aid for Research Activity Start-up of JSPS No.~26K24542.
R. K. is supported by the Grant-in-Aid for Scientific Research (C) of JSPS No.~23K03421. 

\appendix

\section{BSSN evolution and constraint monitoring}
\label{app:bssn_constraints}

\subsection{Motivation}
\label{app:bssn_motivation}

The charge- and coupling-dependent effects discussed in Sec.~\ref{sec:results}
must be distinguished from errors accumulated during the numerical evolution. This is particularly relevant for highly charged binaries, for which the
geometry, electromagnetic field, and dilaton evolve together through the
initial field relaxation and merger. Satisfying the constraints on the initial
slice does not ensure that their discretized counterparts remain satisfied
at later times. Truncation errors and numerical boundary treatments can
generate constraint violations during a free evolution.

Although the ADM equations in Sec.~\ref{sec:formulation} provide a direct
geometrical description of the system, their numerical behavior depends on
the formulation and gauge. The BSSN reformulation separates the conformal
geometry from the local volume element and evolves conformal connection
functions $\widetilde{\Gamma}^{i}$ defined in Eq.~\eqref{eq:app_bssn_connection} as additional variables~\cite{Shibata:1995nv,Baumgarte:1998rp}.
The momentum constraint is then used to replace the divergence
$\partial_j\widetilde{A}^{ij}$ of the conformal trace-free extrinsic
curvature defined in Eq.~\eqref{eq:app_bssn_curvature}. These changes reorganize the principal part of the gravitational
equations and have been found to improve numerical stability relative to
direct ADM evolution. Hyperbolicity nevertheless depends on the gauge and
the precise evolution system; it is not guaranteed by a conformal change
of variables alone~\cite{Sarbach:2002bt, Beyer:2004BSSN, Hilditch:2013sba}.
We therefore describe both the conformal evolution system and the constraint
diagnostics used to assess the evolutions.

\subsection{Evolution system and constraint control}
\label{app:bssn_method}

\subsubsection{Conformal gravitational variables}

In Cartesian coordinates, let $\gamma\equiv\det(\gamma_{ij})$ and define
\begin{align}
 \chi&\equiv\gamma^{-1/3}, &
 \widetilde\gamma_{ij}&\equiv\chi\gamma_{ij},
 \label{eq:app_bssn_metric}\\
 K&\equiv\gamma^{ij}K_{ij}, &
 \widetilde A_{ij}&\equiv\chi
 \left(K_{ij}-\frac13\gamma_{ij}K\right),
 \label{eq:app_bssn_curvature}\\
 \widetilde\Gamma^i&\equiv
 \widetilde\gamma^{jk}\widetilde\Gamma^i{}_{jk}
 =-\partial_j\widetilde\gamma^{ij}.
 \label{eq:app_bssn_connection}
\end{align}
Here $\widetilde{\Gamma}^{i}{}_{jk}$ denotes the Christoffel symbols
associated with the conformal metric $\widetilde{\gamma}_{ij}$, and
$\widetilde{D}_i$ denotes the corresponding covariant derivative. Indices on
conformal tensors are raised and lowered with $\widetilde\gamma_{ij}$.
The conformal factor $\chi$ is an auxiliary geometric variable introduced in the conformal decomposition, not an additional matter field such as the dilaton $\phi$.
The defining algebraic constraints are
\begin{equation}
 \det(\widetilde\gamma_{ij})=1,
 \qquad \widetilde\gamma^{ij}\widetilde A_{ij}=0.
 \label{eq:app_bssn_algebraic}
\end{equation}
The last equality in Eq.~\eqref{eq:app_bssn_connection} uses the unit
determinant condition. The physical variables are recovered from
$\gamma_{ij}=\chi^{-1}\widetilde\gamma_{ij}$ and
$K_{ij}=\chi^{-1}(\widetilde A_{ij}+\widetilde\gamma_{ij}K/3)$.

For compactness, define the componentwise advective derivative
$\partial_0\equiv\partial_t-\beta^k\partial_k$; it is not the tensor Lie
derivative. With the effective matter projections defined in Eqs.~\eqref{eq:rhoH_eff}, ~\eqref{eq:Ji_eff}, and~\eqref{eq:Sij_eff}, the standard BSSN continuum equations are
\begin{align}
 \partial_0\chi={}&\frac23\chi
       \left(\alpha K-\partial_k\beta^k\right),
 \label{eq:app_bssn_chi}\\
 \partial_0\widetilde\gamma_{ij}={}&-2\alpha\widetilde A_{ij}
       +2\widetilde\gamma_{k(i}\partial_{j)}\beta^k
       -\frac23\widetilde\gamma_{ij}\partial_k\beta^k,
 \label{eq:app_bssn_gammatilde}\\
 \partial_0K={}&-D_iD^i\alpha
       +\alpha\left(\widetilde A_{ij}\widetilde A^{ij}
                     +\frac13K^2\right)
       +4\pi\alpha\left(\rho_H^{\rm eff}+S^{\rm eff}\right),
 \label{eq:app_bssn_K}\\
 \partial_0\widetilde A_{ij}={}&
       \chi\left[-D_iD_j\alpha
          +\alpha\left(R^{(3)}_{ij}-8\pi S^{\rm eff}_{ij}\right)
       \right]^{\rm TF}
       +\alpha\left(K\widetilde A_{ij}
          -2\widetilde A_{ik}\widetilde A^k{}_j\right)
       \nonumber\\
       &+2\widetilde A_{k(i}\partial_{j)}\beta^k
       -\frac23\widetilde A_{ij}\partial_k\beta^k,
 \label{eq:app_bssn_A}\\
 \partial_0\widetilde\Gamma^i={}&
       -2\widetilde A^{ij}\partial_j\alpha
       +2\alpha\left(
           \widetilde\Gamma^i{}_{jk}\widetilde A^{jk}
           -\frac{3}{2\chi}\widetilde A^{ij}\partial_j\chi
           -\frac23\widetilde\gamma^{ij}\partial_jK
           -8\pi\widetilde\gamma^{ij}J_j^{\rm eff}
       \right)
       \nonumber\\
       &+\widetilde\gamma^{jk}\partial_j\partial_k\beta^i
       +\frac13\widetilde\gamma^{ij}\partial_j\partial_k\beta^k
       -\widetilde\Gamma^j\partial_j\beta^i
       +\frac23\widetilde\Gamma^i\partial_j\beta^j.
 \label{eq:app_bssn_Gamma}
\end{align}
The trace-free operation is
$X_{ij}^{\rm TF}=X_{ij}-\gamma_{ij}\gamma^{kl}X_{kl}/3$.
The Hamiltonian constraint has been used in Eq.~\eqref{eq:app_bssn_K},
and the momentum constraint in Eq.~\eqref{eq:app_bssn_Gamma}.
The lapse and shift are supplied by the gauge conditions described in
Sec.~\ref{sec:formulation}.

The change in the principal part is visible in the conformal Ricci tensor:
\begin{align}
 R^{(3)}_{ij}={}&\widetilde R_{ij}+R^\chi_{ij},
 \label{eq:app_ricci_split}\\
 R^\chi_{ij}={}&\frac{1}{2\chi}
   \left(\widetilde D_i\widetilde D_j\chi
    +\widetilde\gamma_{ij}\widetilde\gamma^{kl}
       \widetilde D_k\widetilde D_l\chi\right)
   -\frac{\partial_i\chi\partial_j\chi}{4\chi^2}
   -\frac{3\widetilde\gamma_{ij}}{4\chi^2}
      \widetilde\gamma^{kl}\partial_k\chi\partial_l\chi,
 \label{eq:app_ricci_chi}\\
 \widetilde R_{ij}={}&
   -\frac12\widetilde\gamma^{lm}
       \partial_l\partial_m\widetilde\gamma_{ij}
   +\widetilde\gamma_{k(i}\partial_{j)}\widetilde\Gamma^k
   +\widetilde\Gamma^k\widetilde\Gamma_{(ij)k}
   \nonumber\\
   &+\widetilde\gamma^{lm}
       \left(2\widetilde\Gamma^k{}_{l(i}
                    \widetilde\Gamma_{j)km}
              +\widetilde\Gamma^k{}_{im}\widetilde\Gamma_{klj}\right).
 \label{eq:app_ricci_tilde}
\end{align}
Here $\widetilde\Gamma_{ijk}\equiv
\widetilde\gamma_{il}\widetilde\Gamma^l{}_{jk}$.
The first term in Eq.~\eqref{eq:app_ricci_tilde} is a Laplace-type
operator acting on each conformal-metric component; other second
derivatives of the metric are represented through derivatives of the
independently evolved connection functions.

\subsubsection{Gauge conditions}
\label{app:gauge_conditions}

We close the BSSN evolution system with the advective moving-puncture
gauge~\cite{vanMeter:2006vi}.  Using the componentwise advective derivative
$\partial_0\equiv\partial_t-\beta^j\partial_j$ introduced above, the
lapse and shift satisfy
\begin{equation}
\begin{aligned}
  \partial_0\alpha
    &= -2\alpha K, \\
  \partial_0\beta^i
    &= \frac{3}{4}\,\mathcal{B}^i, \\
  \partial_0\mathcal{B}^i
    &= \partial_0\widetilde{\Gamma}^i-\eta\mathcal{B}^i.
\end{aligned}
\label{eq:app_moving_puncture_gauge}
\end{equation}
Here $\mathcal{B}^i$ is an auxiliary shift-driver variable; the
calligraphic symbol is used to distinguish it from the electromagnetic
magnetic field.  The first equation in
Eq.~\eqref{eq:app_moving_puncture_gauge} is the advective $1+\log$
slicing condition, while the second and third equations constitute the
advective hyperbolic Gamma-driver shift condition.  In terms of coordinate
time, the last equation is equivalently
$\partial_t\mathcal{B}^i=\partial_t\widetilde{\Gamma}^i
+\beta^j\partial_j(\mathcal{B}^i-\widetilde{\Gamma}^i)
-\eta\mathcal{B}^i$.
All simulations retain the advection terms displayed above and use $\eta M=1$. 
The shift-driver damping parameter $\eta$ is distinct from the electromagnetic
constraint-damping parameter $\kappa$ appearing in Eq.~\eqref{EMD_Zeta}.

\subsubsection{Electromagnetic and dilaton evolution in conformal variables}

We rewrite the $3+1$ evolution equations for the electromagnetic and dilaton fields in terms of the conformal metric variables introduced above. 
The conformal geometry enters these equations through the lapse, shift, spatial metric, extrinsic curvature, and spatial derivatives. 
The effective matter projections source the BSSN equations. 

Define the spatial Maxwell tensor and its conformally raised version by
\begin{equation}
 F_{ij}\equiv\partial_iA^\perp_j-\partial_jA^\perp_i,
 \qquad \widetilde F^{ij}\equiv
       \widetilde\gamma^{ik}\widetilde\gamma^{jl}F_{kl},
 \qquad F^{ij}=\chi^2\widetilde F^{ij}.
 \label{eq:app_spatial_F}
\end{equation}
The evolved spatial potential is the covector $A^\perp_i$, whereas the
electric field is the vector $E^i$. Define also
\begin{equation}
 E^2\equiv\chi^{-1}\widetilde\gamma_{ij}E^iE^j,
 \qquad B^2\equiv\frac{\chi^2}{2}
       \widetilde\gamma^{ik}\widetilde\gamma^{jl}F_{ij}F_{kl}.
 \label{eq:app_E2B2}
\end{equation}
These are the physical contractions $E_iE^i$ and $B_iB^i$.
Useful conformal identities are
\begin{align}
 D_iD^if={}&\chi\widetilde\gamma^{ij}
           \widetilde D_i\widetilde D_j f
       -\frac12\widetilde\gamma^{ij}\partial_i\chi\partial_jf,
 \label{eq:app_scalar_laplacian}\\
 D_iE^i={}&\partial_iE^i
       -\frac{3}{2\chi}E^i\partial_i\chi,
 \label{eq:app_vector_divergence}\\
 D_iA_\perp^i={}&\chi\widetilde\gamma^{ij}
           \widetilde D_iA^\perp_j
       -\frac12\widetilde\gamma^{ij}A^\perp_i\partial_j\chi.
 \label{eq:app_potential_divergence}
\end{align}
The divergence identities use $\det(\widetilde\gamma_{ij})=1$.
In this notation the matter evolution equations are
\begin{align}
 \partial_0\phi={}&-\alpha K_\phi,
 \label{eq:app_phi_rhs}\\
 \partial_0K_\phi={}&\alpha K K_\phi
       -\alpha\chi\widetilde\gamma^{ij}
             \widetilde D_i\widetilde D_j\phi
       +\frac\alpha2\widetilde\gamma^{ij}
             \partial_i\chi\partial_j\phi
       -\chi\widetilde\gamma^{ij}
             \partial_i\alpha\partial_j\phi
       +a\alpha e^{-2a\phi}(E^2-B^2),
 \label{eq:app_Kphi_rhs}\\
 \partial_0 A^\perp_i={}&
       A^\perp_j\partial_i\beta^j
       -\alpha\chi^{-1}\widetilde\gamma_{ij}E^j
       -\partial_i(\alpha A_\phi),
 \label{eq:app_A_rhs}\\
 \partial_0 A_\phi={}&\alpha K A_\phi-\alpha Z
       -\alpha\chi\widetilde\gamma^{ij}\widetilde D_iA^\perp_j
       +\frac\alpha2\widetilde\gamma^{ij}A^\perp_i\partial_j\chi
       -\chi\widetilde\gamma^{ij}A^\perp_i\partial_j\alpha,
 \label{eq:app_Aphi_rhs}\\
 \partial_0 E^i={}&-E^j\partial_j\beta^i
       +\alpha(K-2aK_\phi)E^i
       +\alpha\chi\widetilde\gamma^{ij}\partial_j Z
       \nonumber\\
       &+\chi^2\left[
           \alpha\widetilde D_j\widetilde F^{ij}
           +\left(\partial_j\alpha
                    +\frac{\alpha}{2\chi}\partial_j\chi
                    -2a\alpha\partial_j\phi\right)
                     \widetilde F^{ij}\right],
 \label{eq:app_E_rhs}\\
 \partial_0 Z={}&\alpha\left(
          \partial_iE^i-\frac{3}{2\chi}E^i\partial_i\chi
          -2aE^i\partial_i\phi-\kappa Z\right).
 \label{eq:app_Z_rhs}
\end{align}
The opposite signs of the shift-gradient terms in Eqs.~\eqref{eq:app_A_rhs} and \eqref{eq:app_E_rhs} reflect the covariant and contravariant index positions of the evolved fields. 
The dilaton equations~\eqref{eq:app_phi_rhs} and \eqref{eq:app_Kphi_rhs} use the momentum convention $K_\phi=-n^\mu\nabla_\mu\phi$ introduced in Sec.~\ref{sec:formulation}. 


\subsubsection{Constraint violation diagnostics and electromagnetic constraint damping}

To monitor whether the constraints remain satisfied during the evolution,
we substitute the numerically evolved fields into each constraint equation
and move all terms to the left-hand side. The resulting quantities are the
local constraint residuals: they vanish for an exact solution, whereas their
nonzero numerical values measure constraint violation. We define the
Hamiltonian, momentum, and EMD Gauss-constraint residuals by
\begin{align}
 \mathcal H&\equiv R^{(3)}+\frac23K^2
       -\widetilde A_{ij}\widetilde A^{ij}
       -16\pi\rho_H^{\rm eff},
 \label{eq:H_constraint_violation}\\
 \mathcal M^i&\equiv
       D_j(K^{ij}-\gamma^{ij}K)-8\pi J^i_{\rm eff},
 \label{eq:M_constraint_violation}\\
 \mathcal G&\equiv D_iE^i-2aE^iD_i\phi
       =e^{2a\phi}D_i(e^{-2a\phi}E^i).
 \label{eq:G_constraint_violation}
\end{align}
The Hamiltonian residual in Eq.~\eqref{eq:H_constraint_violation} is identical to
$R^{(3)}+K^2-K_{ij}K^{ij}-16\pi\rho_H^{\rm eff}$ in Eq.~\eqref{Ham}.
The conformal representation of the momentum residual is
\begin{equation}
 \mathcal M^i=\chi\left[
      \widetilde D_j\widetilde A^{ij}
      -\frac{3}{2\chi}\widetilde A^{ij}\partial_j\chi
      -\frac23\widetilde\gamma^{ij}\partial_jK
      -8\pi\widetilde\gamma^{ij}J_j^{\rm eff}\right].
 \label{eq:app_conformal_momentum_constraint}
\end{equation}
All three residuals vanish for an exact solution. The use of
$\rho_H^{\rm eff}$ and $J_i^{\rm eff}$ is essential: the scalar and
dilaton-weighted electromagnetic contributions must both be included.
The residual $\mathcal G$ tests the EMD Gauss law, rather than the
unweighted divergence $D_iE^i$ alone.

The auxiliary field $Z$ is distinct from $\mathcal G$.
Equation~\eqref{eq:app_Z_rhs} gives
\begin{equation}
 (\partial_t-\beta^i\partial_i)Z
       =\alpha(\mathcal G-\kappa Z).
 \label{eq:app_Z_constraint_relation}
\end{equation}
Together with the $+\alpha D^iZ$ term in the electric-field equation,
this supplies a propagation and damping mechanism for Gauss violations,
as in extended Maxwell systems~\cite{PhysRevD.81.084007,Hirschmann:2017psw}.
For example, about Minkowski spacetime with constant dilaton, unit lapse,
zero shift, and constant $\kappa>0$, the linearized constraint subsystem is
\begin{equation}
 \partial_t\mathcal G=\delta^{ij}\partial_i\partial_j Z,
 \qquad \partial_t Z=\mathcal G-\kappa Z,
 \qquad
 \partial_t^2Z+\kappa\partial_tZ
       -\delta^{ij}\partial_i\partial_jZ=0.
 \label{eq:app_cleaning_wave}
\end{equation}
This local model explains the purpose of the damping term; it does not
establish that every constraint violation in the dynamical binary
spacetime decreases monotonically. We monitor the auxiliary
field $Z$ as a diagnostic of the constraint-damping subsystem.

The conformal formulation also introduces the algebraic residuals
$\det(\widetilde\gamma_{ij})-1$ and
$\widetilde\gamma^{ij}\widetilde A_{ij}$, together with the connection
residual $\widetilde\Gamma^i+\partial_j\widetilde\gamma^{ij}$.
These distinguish consistency of the BSSN variable definitions from the
Hamiltonian, momentum, and Maxwell constraints. They are not displayed
in Fig.~\ref{fig:constraint_a1_charge07}.

\subsection{Behavior of the monitored constraints}
\label{app:bssn_results}

The spatial $L_2$ diagnostics are obtained over the full simulated
spatial domain using the standard \texttt{norm2} reduction of
\texttt{CarpetReduce}~\cite{CarpetReduce:documentation}, with output
provided by \texttt{CarpetIOScalar}~\cite{CarpetIOScalar:documentation}.
No additional mask is applied to exclude puncture neighborhoods or
apparent-horizon interiors.

\begin{figure}[!hbp]
 \centering
 \includegraphics[width=0.90\columnwidth]
 {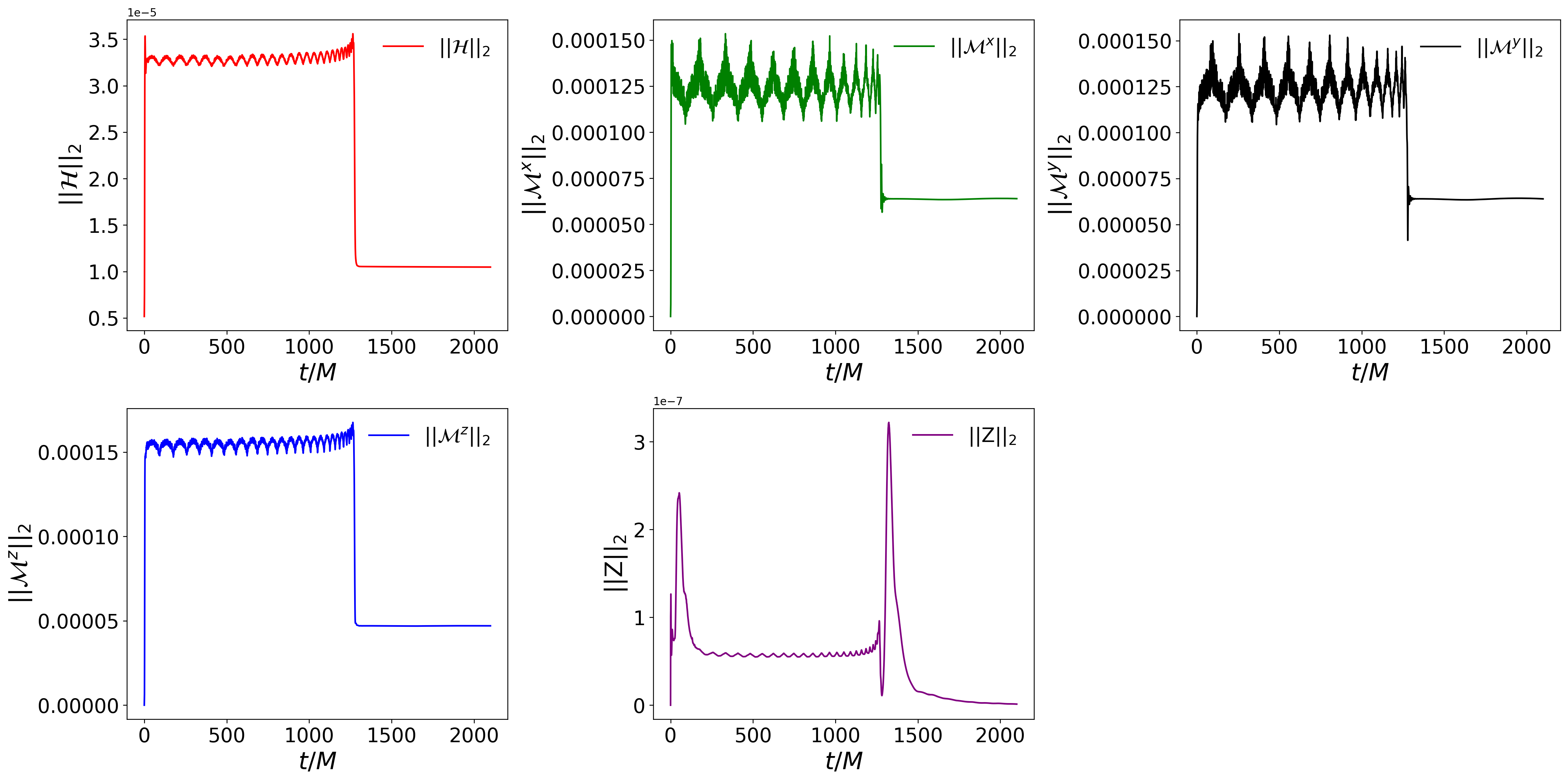}
 \caption{Constraint diagnostics for the highly charged binary with
 $a=1$ and $\lambda_{\pm}=\pm0.7$. The panels show the spatial $L_2$ reductions of the Hamiltonian
residual $\mathcal H$ and the individual coordinate components
$\mathcal M^x$, $\mathcal M^y$, and $\mathcal M^z$ of the momentum
residual, defined in Eqs.~\eqref{eq:H_constraint_violation}
and~\eqref{eq:M_constraint_violation}.
The remaining panel shows the auxiliary Gauss-constraint damping
field $Z$, whose evolution is given in Eq.~\eqref{eq:app_Z_rhs}.}
 \label{fig:constraint_a1_charge07}
\end{figure}

Figure~\ref{fig:constraint_a1_charge07} shows the spatial $L_2$ diagnostics
of $\mathcal H$, the three coordinate components of $\mathcal M^i$, and
$Z$ for $a=1$ and $\lambda_{\pm}=\pm0.7$.
The coordinate time $t/M$ is the simulation time, with no wave-propagation
correction. The Hamiltonian and momentum diagnostics oscillate about
slowly varying levels during the earlier evolution and settle to lower,
approximately stationary levels after the transition visible near
$t/M\simeq1200$. The auxiliary-field diagnostic exhibits an initial transient and
a later localized peak, followed by a marked decrease. None of the
displayed diagnostics shows a sustained runaway increase over the
plotted interval.

These observations provide a constraint-monitoring check over the duration
used for the radiation analysis. They do not, by themselves, establish
convergence to the continuum EMD solution or quantify the error in the
waveforms. In particular, the late-time Hamiltonian and momentum levels
are nonzero, and the absolute norm values depend on the spatial reduction
and its normalization. Demonstrating convergence requires comparisons
at several resolutions using a consistent reduction prescription.
The decrease around the late-time transition should likewise be
interpreted with the mesh hierarchy and the reduction domain specified.

\clearpage

%

\bibliographystyle{mybibstyle}
\bibliography{BBHsim_EMD}

\end{document}